\newif\ifdraftmode
\draftmodefalse

\documentclass[pdflatex,sn-nature]{sn-jnl}

\usepackage{graphicx}%
\usepackage{multirow}%
\usepackage{amsmath,amssymb,amsfonts}%
\usepackage{amsthm}%
\usepackage{mathrsfs}%
\usepackage[title]{appendix}%
\usepackage{xcolor}%
\usepackage{textcomp}%
\usepackage{manyfoot}%
\usepackage{booktabs}%
\usepackage{algorithm}%
\usepackage{algorithmicx}%
\usepackage{algpseudocode}%
\usepackage{listings}%
\usepackage{booktabs}

\theoremstyle{thmstyleone}%
\theoremstyle{thmstyletwo}%

\theoremstyle{thmstylethree}%

\usepackage{xcolor}
\usepackage[normalem]{ulem}

\ifdraftmode
    \newcommand{\delete}[1]{\textcolor{red}{\sout{#1}}}
    \newcommand{\add}[1]{\textcolor{red}{#1}}
    \newcommand{\notetoeditor}[1]{\textcolor{blue}{\emph{[#1]}}}
\else
    \newcommand{\delete}[1]{\ignorespaces}
    \newcommand{\add}[1]{#1}
    \newcommand{\notetoeditor}[1]{\ignorespaces}
\fi

\begin{document}

\newcommand\aj{AJ}
\newcommand\psj{PSJ}
\newcommand\araa{ARA\&A}
\newcommand\apj{ApJ}
\newcommand\apjl{ApJL}     
\newcommand\apjs{ApJS}
\newcommand\ao{ApOpt}
\newcommand\apss{Ap\&SS}
\newcommand\aap{A\&A}
\newcommand\aapr{A\&A~Rv}
\newcommand\aaps{A\&AS}
\newcommand\azh{AZh}
\newcommand\baas{BAAS}
\newcommand\icarus{Icarus}
\newcommand\jaavso{JAAVSO}  
\newcommand\jrasc{JRASC}
\newcommand\memras{MmRAS}
\newcommand\mnras{MNRAS}
\newcommand\pra{PhRvA}
\newcommand\prb{PhRvB}
\newcommand\prc{PhRvC}
\newcommand\prd{PhRvD}
\newcommand\pre{PhRvE}
\newcommand\prl{PhRvL}
\newcommand\pasp{PASP}
\newcommand\pasj{PASJ}
\newcommand\qjras{QJRAS}
\newcommand\skytel{S\&T}
\newcommand\solphys{SoPh}
\newcommand\sovast{Soviet~Ast.}
\newcommand\ssr{SSRv}
\newcommand\zap{ZA}
\newcommand\nat{Nature}
\newcommand\iaucirc{IAUC}
\newcommand\aplett{Astrophys.~Lett.}
\newcommand\apspr{Astrophys.~Space~Phys.~Res.}
\newcommand\bain{BAN}
\newcommand\fcp{FCPh}
\newcommand\gca{GeoCoA}
\newcommand\grl{Geophys.~Res.~Lett.}
\newcommand\jcp{JChPh}
\newcommand\jgr{J.~Geophys.~Res.}
\newcommand\jqsrt{JQSRT}
\newcommand\nphysa{NuPhA}
\newcommand\physrep{PhR}
\newcommand\physscr{PhyS}
\newcommand\planss{Planet.~Space~Sci.}
\newcommand\procspie{Proc.~SPIE}

\newcommand\actaa{AcA}
\newcommand\caa{ChA\&A}
\newcommand\cjaa{ChJA\&A}
\newcommand\jcap{JCAP}
\newcommand\na{NewA}
\newcommand\nar{NewAR}
\newcommand\pasa{PASA}
\newcommand\rmxaa{RMxAA}

\newcommand\maps{M\&PS}
\newcommand\aas{AAS Meeting Abstracts}
\newcommand\dps{AAS/DPS Meeting Abstracts}

\newcommand{\romanger}[1]{{#1}}

\hyphenation{BasicATLAS}
\let\astap=\aap 
\let\apjlett=\apjl 
\let\apjsupp=\apjs 
\let\applopt=\ao 

\newcommand{\nmone}{11}
\newcommand{\nmtwo}{9}
\newcommand{\nmtotal}{20}
\newcommand{\teffrangeone}{42}
\newcommand{\loggrangeone}{0.06}
\newcommand{\vtrangeone}{0.23}
\newcommand{\teffrangetwo}{21}
\newcommand{\loggrangetwo}{0.06}
\newcommand{\vtrangetwo}{0.18}
\newcommand{\teffrangemax}{42}
\newcommand{\loggrangemax}{0.06}
\newcommand{\vtrangemax}{0.23}

\newcommand{\Napopsep}{0.450}
\newcommand{\Naonegmu}{-0.382}
\newcommand{\Naonegmulowererr}{0.033}
\newcommand{\Naonegmuuppererr}{0.032}
\newcommand{\Naonegsigmalimit}{0.175}
\newcommand{\Natwogmu}{0.317}
\newcommand{\Natwogmulowererr}{0.051}
\newcommand{\Natwogmuuppererr}{0.051}
\newcommand{\Natwogsigma}{0.115}
\newcommand{\Natwogsigmalowererr}{0.036}
\newcommand{\Natwogsigmauppererr}{0.066}
\newcommand{\Namudiff}{0.699}
\newcommand{\Namudiffuppererr}{0.061}
\newcommand{\Namudifflowererr}{0.061}
\newcommand{\Mgpopsep}{-0.521}
\newcommand{\Mgonegmu}{0.121}
\newcommand{\Mgonegmulowererr}{0.013}
\newcommand{\Mgonegmuuppererr}{0.013}
\newcommand{\Mgonegsigmalimit}{0.059}
\newcommand{\Mgtwogmu}{-0.132}
\newcommand{\Mgtwogmulowererr}{0.114}
\newcommand{\Mgtwogmuuppererr}{0.115}
\newcommand{\Mgtwogsigma}{0.267}
\newcommand{\Mgtwogsigmalowererr}{0.077}
\newcommand{\Mgtwogsigmauppererr}{0.144}
\newcommand{\Mgmudiff}{-0.252}
\newcommand{\Mgmudiffuppererr}{0.116}
\newcommand{\Mgmudifflowererr}{0.114}
\newcommand{\Alpopsep}{0.835}
\newcommand{\Alonegmu}{-0.711}
\newcommand{\Alonegmulowererr}{0.048}
\newcommand{\Alonegmuuppererr}{0.048}
\newcommand{\Alonegsigmalimit}{0.211}
\newcommand{\Altwogmu}{0.589}
\newcommand{\Altwogmulowererr}{0.100}
\newcommand{\Altwogmuuppererr}{0.100}
\newcommand{\Altwogsigma}{0.227}
\newcommand{\Altwogsigmalowererr}{0.070}
\newcommand{\Altwogsigmauppererr}{0.129}
\newcommand{\Almudiff}{1.301}
\newcommand{\Almudiffuppererr}{0.111}
\newcommand{\Almudifflowererr}{0.111}
\newcommand{\Sipopsep}{-0.273}
\newcommand{\Sionegmu}{-0.056}
\newcommand{\Sionegmulowererr}{0.069}
\newcommand{\Sionegmuuppererr}{0.070}
\newcommand{\Sionegsigma}{0.138}
\newcommand{\Sionegsigmalowererr}{0.053}
\newcommand{\Sionegsigmauppererr}{0.105}
\newcommand{\Sitwogmu}{0.040}
\newcommand{\Sitwogmulowererr}{0.068}
\newcommand{\Sitwogmuuppererr}{0.070}
\newcommand{\Sitwogsigma}{0.152}
\newcommand{\Sitwogsigmalowererr}{0.048}
\newcommand{\Sitwogsigmauppererr}{0.088}
\newcommand{\Simudiff}{0.097}
\newcommand{\Simudiffuppererr}{0.099}
\newcommand{\Simudifflowererr}{0.097}
\newcommand{\Kpopsep}{-0.097}
\newcommand{\Konegmu}{-0.099}
\newcommand{\Konegmulowererr}{0.081}
\newcommand{\Konegmuuppererr}{0.082}
\newcommand{\Konegsigma}{0.167}
\newcommand{\Konegsigmalowererr}{0.058}
\newcommand{\Konegsigmauppererr}{0.118}
\newcommand{\Ktwogmu}{0.080}
\newcommand{\Ktwogmulowererr}{0.046}
\newcommand{\Ktwogmuuppererr}{0.047}
\newcommand{\Ktwogsigma}{0.102}
\newcommand{\Ktwogsigmalowererr}{0.034}
\newcommand{\Ktwogsigmauppererr}{0.060}
\newcommand{\Kmudiff}{0.179}
\newcommand{\Kmudiffuppererr}{0.094}
\newcommand{\Kmudifflowererr}{0.093}
\newcommand{\Capopsep}{0.039}
\newcommand{\Caonegmu}{-0.044}
\newcommand{\Caonegmulowererr}{0.012}
\newcommand{\Caonegmuuppererr}{0.012}
\newcommand{\Caonegsigmalimit}{0.065}
\newcommand{\Catwogmu}{0.036}
\newcommand{\Catwogmulowererr}{0.009}
\newcommand{\Catwogmuuppererr}{0.009}
\newcommand{\Catwogsigmalimit}{0.037}
\newcommand{\Camudiff}{0.080}
\newcommand{\Camudiffuppererr}{0.015}
\newcommand{\Camudifflowererr}{0.015}
\newcommand{\Scpopsep}{-0.006}
\newcommand{\Sconegmu}{-0.058}
\newcommand{\Sconegmulowererr}{0.019}
\newcommand{\Sconegmuuppererr}{0.019}
\newcommand{\Sconegsigmalimit}{0.068}
\newcommand{\Sctwogmu}{0.051}
\newcommand{\Sctwogmulowererr}{0.029}
\newcommand{\Sctwogmuuppererr}{0.029}
\newcommand{\Sctwogsigmalimit}{0.146}
\newcommand{\Scmudiff}{0.109}
\newcommand{\Scmudiffuppererr}{0.034}
\newcommand{\Scmudifflowererr}{0.034}
\newcommand{\Tipopsep}{0.038}
\newcommand{\Tionegmu}{-0.042}
\newcommand{\Tionegmulowererr}{0.014}
\newcommand{\Tionegmuuppererr}{0.014}
\newcommand{\Tionegsigmalimit}{0.057}
\newcommand{\Titwogmu}{0.032}
\newcommand{\Titwogmulowererr}{0.012}
\newcommand{\Titwogmuuppererr}{0.012}
\newcommand{\Titwogsigmalimit}{0.039}
\newcommand{\Timudiff}{0.074}
\newcommand{\Timudiffuppererr}{0.018}
\newcommand{\Timudifflowererr}{0.018}
\newcommand{\Vpopsep}{-0.028}
\newcommand{\Vonegmu}{-0.052}
\newcommand{\Vonegmulowererr}{0.043}
\newcommand{\Vonegmuuppererr}{0.043}
\newcommand{\Vonegsigma}{0.083}
\newcommand{\Vonegsigmalowererr}{0.033}
\newcommand{\Vonegsigmauppererr}{0.064}
\newcommand{\Vtwogmu}{0.037}
\newcommand{\Vtwogmulowererr}{0.017}
\newcommand{\Vtwogmuuppererr}{0.017}
\newcommand{\Vtwogsigmalimit}{0.073}
\newcommand{\Vmudiff}{0.089}
\newcommand{\Vmudiffuppererr}{0.046}
\newcommand{\Vmudifflowererr}{0.046}
\newcommand{\Crpopsep}{0.018}
\newcommand{\Cronegmu}{-0.040}
\newcommand{\Cronegmulowererr}{0.012}
\newcommand{\Cronegmuuppererr}{0.013}
\newcommand{\Cronegsigmalimit}{0.061}
\newcommand{\Crtwogmu}{0.036}
\newcommand{\Crtwogmulowererr}{0.016}
\newcommand{\Crtwogmuuppererr}{0.016}
\newcommand{\Crtwogsigmalimit}{0.081}
\newcommand{\Crmudiff}{0.075}
\newcommand{\Crmudiffuppererr}{0.020}
\newcommand{\Crmudifflowererr}{0.020}
\newcommand{\Mnpopsep}{0.008}
\newcommand{\Mnonegmu}{-0.031}
\newcommand{\Mnonegmulowererr}{0.017}
\newcommand{\Mnonegmuuppererr}{0.017}
\newcommand{\Mnonegsigmalimit}{0.068}
\newcommand{\Mntwogmu}{0.025}
\newcommand{\Mntwogmulowererr}{0.015}
\newcommand{\Mntwogmuuppererr}{0.014}
\newcommand{\Mntwogsigmalimit}{0.066}
\newcommand{\Mnmudiff}{0.056}
\newcommand{\Mnmudiffuppererr}{0.022}
\newcommand{\Mnmudifflowererr}{0.022}
\newcommand{\Fepopsep}{0.048}
\newcommand{\Feonegmu}{-0.047}
\newcommand{\Feonegmulowererr}{0.010}
\newcommand{\Feonegmuuppererr}{0.010}
\newcommand{\Feonegsigmalimit}{0.038}
\newcommand{\Fetwogmu}{0.035}
\newcommand{\Fetwogmulowererr}{0.011}
\newcommand{\Fetwogmuuppererr}{0.012}
\newcommand{\Fetwogsigmalimit}{0.051}
\newcommand{\Femudiff}{0.082}
\newcommand{\Femudiffuppererr}{0.016}
\newcommand{\Femudifflowererr}{0.016}
\newcommand{\Nipopsep}{0.013}
\newcommand{\Nionegmu}{-0.042}
\newcommand{\Nionegmulowererr}{0.014}
\newcommand{\Nionegmuuppererr}{0.014}
\newcommand{\Nionegsigmalimit}{0.065}
\newcommand{\Nitwogmu}{0.039}
\newcommand{\Nitwogmulowererr}{0.013}
\newcommand{\Nitwogmuuppererr}{0.013}
\newcommand{\Nitwogsigmalimit}{0.060}
\newcommand{\Nimudiff}{0.081}
\newcommand{\Nimudiffuppererr}{0.019}
\newcommand{\Nimudifflowererr}{0.019}
\newcommand{\Copopsep}{-0.018}
\newcommand{\Coonegmu}{-0.034}
\newcommand{\Coonegmulowererr}{0.018}
\newcommand{\Coonegmuuppererr}{0.018}
\newcommand{\Coonegsigmalimit}{0.077}
\newcommand{\Cotwogmu}{0.034}
\newcommand{\Cotwogmulowererr}{0.021}
\newcommand{\Cotwogmuuppererr}{0.020}
\newcommand{\Cotwogsigmalimit}{0.094}
\newcommand{\Comudiff}{0.068}
\newcommand{\Comudiffuppererr}{0.027}
\newcommand{\Comudifflowererr}{0.028}
\newcommand{\Znpopsep}{-0.053}
\newcommand{\Znonegmu}{-0.035}
\newcommand{\Znonegmulowererr}{0.027}
\newcommand{\Znonegmuuppererr}{0.027}
\newcommand{\Znonegsigmalimit}{0.121}
\newcommand{\Zntwogmu}{0.037}
\newcommand{\Zntwogmulowererr}{0.021}
\newcommand{\Zntwogmuuppererr}{0.021}
\newcommand{\Zntwogsigmalimit}{0.089}
\newcommand{\Znmudiff}{0.072}
\newcommand{\Znmudiffuppererr}{0.034}
\newcommand{\Znmudifflowererr}{0.035}
\newcommand{\Srpopsep}{-0.298}
\newcommand{\Sronegmu}{-0.034}
\newcommand{\Sronegmulowererr}{0.058}
\newcommand{\Sronegmuuppererr}{0.058}
\newcommand{\Sronegsigmalimit}{0.317}
\newcommand{\Srtwogmu}{0.038}
\newcommand{\Srtwogmulowererr}{0.070}
\newcommand{\Srtwogmuuppererr}{0.068}
\newcommand{\Srtwogsigma}{0.154}
\newcommand{\Srtwogsigmalowererr}{0.051}
\newcommand{\Srtwogsigmauppererr}{0.092}
\newcommand{\Srmudiff}{0.072}
\newcommand{\Srmudiffuppererr}{0.089}
\newcommand{\Srmudifflowererr}{0.091}
\newcommand{\Ypopsep}{-0.071}
\newcommand{\Yonegmu}{-0.020}
\newcommand{\Yonegmulowererr}{0.040}
\newcommand{\Yonegmuuppererr}{0.040}
\newcommand{\Yonegsigmalimit}{0.213}
\newcommand{\Ytwogmu}{0.020}
\newcommand{\Ytwogmulowererr}{0.018}
\newcommand{\Ytwogmuuppererr}{0.018}
\newcommand{\Ytwogsigmalimit}{0.071}
\newcommand{\Ymudiff}{0.040}
\newcommand{\Ymudiffuppererr}{0.044}
\newcommand{\Ymudifflowererr}{0.043}
\newcommand{\Zrpopsep}{-0.189}
\newcommand{\Zronegmu}{-0.008}
\newcommand{\Zronegmulowererr}{0.062}
\newcommand{\Zronegmuuppererr}{0.060}
\newcommand{\Zronegsigmalimit}{0.335}
\newcommand{\Zrtwogmu}{0.005}
\newcommand{\Zrtwogmulowererr}{0.039}
\newcommand{\Zrtwogmuuppererr}{0.040}
\newcommand{\Zrtwogsigmalimit}{0.184}
\newcommand{\Zrmudiff}{0.013}
\newcommand{\Zrmudiffuppererr}{0.072}
\newcommand{\Zrmudifflowererr}{0.073}
\newcommand{\Bapopsep}{-0.246}
\newcommand{\Baonegmu}{0.004}
\newcommand{\Baonegmulowererr}{0.090}
\newcommand{\Baonegmuuppererr}{0.091}
\newcommand{\Baonegsigma}{0.185}
\newcommand{\Baonegsigmalowererr}{0.064}
\newcommand{\Baonegsigmauppererr}{0.132}
\newcommand{\Batwogmu}{-0.007}
\newcommand{\Batwogmulowererr}{0.022}
\newcommand{\Batwogmuuppererr}{0.023}
\newcommand{\Batwogsigmalimit}{0.085}
\newcommand{\Bamudiff}{-0.011}
\newcommand{\Bamudiffuppererr}{0.093}
\newcommand{\Bamudifflowererr}{0.093}
\newcommand{\Lapopsep}{-0.329}
\newcommand{\Laonegmu}{0.029}
\newcommand{\Laonegmulowererr}{0.100}
\newcommand{\Laonegmuuppererr}{0.095}
\newcommand{\Laonegsigma}{0.193}
\newcommand{\Laonegsigmalowererr}{0.069}
\newcommand{\Laonegsigmauppererr}{0.141}
\newcommand{\Latwogmu}{0.016}
\newcommand{\Latwogmulowererr}{0.034}
\newcommand{\Latwogmuuppererr}{0.033}
\newcommand{\Latwogsigmalimit}{0.161}
\newcommand{\Lamudiff}{-0.013}
\newcommand{\Lamudiffuppererr}{0.101}
\newcommand{\Lamudifflowererr}{0.106}
\newcommand{\Eupopsep}{-0.274}
\newcommand{\Euonegmu}{0.013}
\newcommand{\Euonegmulowererr}{0.107}
\newcommand{\Euonegmuuppererr}{0.105}
\newcommand{\Euonegsigma}{0.219}
\newcommand{\Euonegsigmalowererr}{0.074}
\newcommand{\Euonegsigmauppererr}{0.153}
\newcommand{\Eutwogmu}{-0.002}
\newcommand{\Eutwogmulowererr}{0.022}
\newcommand{\Eutwogmuuppererr}{0.021}
\newcommand{\Eutwogsigmalimit}{0.097}
\newcommand{\Eumudiff}{-0.014}
\newcommand{\Eumudiffuppererr}{0.107}
\newcommand{\Eumudifflowererr}{0.110}

\title[\delete{Globular Cluster Metal Retention} \add{Supernova in Globular Cluster}]{\delete{Retention of Supernova Ejecta in a Typical Globular Cluster} \add{Evidence of Supernova Between Formation of Stellar Populations in a Globular Cluster}}


\author*[1]{\fnm{Evan} N.\ \sur{Kirby}}\email{ekirby@nd.edu}

\author[1]{\fnm{Roman} \sur{Gerasimov}}\email{rgerasim@nd.edu}

\author[1,2,3]{\fnm{Alice} \sur{Cai}}\email{acai@u.northwestern.edu}

\author[1,4,5]{\fnm{Benjamin} \sur{Coco}}\email{bcoco@fordham.edu}

\author[1]{\fnm{Pranav} \sur{Nalamwar}}\email{pnalamwa@nd.edu}

\author[1]{\fnm{Lauren} \sur{Henderson}}\email{lhender6@nd.edu}

\affil*[1]{\orgdiv{Department of Physics and Astronomy}, \orgname{University of Notre Dame}, \orgaddress{\street{225 Nieuwland Science Hall}, \city{Notre Dame}, \postcode{46637}, \state{Indiana}, \country{USA}}}

\affil[2]{\orgdiv{Department of Physics and Astronomy}, \orgname{Northwestern University}, \orgaddress{\city{Evanston}, \postcode{60208}, \state{Illinois}, \country{USA}}}

\affil[3]{\orgdiv{Center for Interdisciplinary Exploration and Research in Astronomy}, \orgname{Northwestern University}, \orgaddress{\street{1800 Sherman Avenue}, \city{Evanston}, \postcode{60201}, \state{Illinois}, \country{USA}}}

\affil[4]{\orgdiv{Department of Physics and Engineering Physics}, \orgname{Fordham University}, \orgaddress{\street{Freeman Hall 208}, \city{Bronx}, \postcode{10458}, \state{New York}, \country{USA}}}

\affil[5]{\orgdiv{School of Professional Studies}, \orgname{City University of New York}, \orgaddress{\street{119 W 31st St}, \city{New York}, \postcode{10001}, \state{New York}, \country{USA}}}

\keywords{globular clusters, nucleosynthesis, element abundance correlations, differential line-by-line analysis}



\maketitle

\textbf{\delete{Most globular clusters do not retain supernova ejecta. As a result,} Globular clusters do not undergo conventional chemical evolution \add{driven by supernova enrichment.} \delete{and} Instead, \add{they} exhibit unique abundance patterns of the light elements, which cannot be fully explained by any of the proposed enrichment mechanisms \cite{gra19,bas18}.  ``Normal'' stars of low sodium abundances comprise the first \delete{generation} \add{population} of cluster stars, and ``enriched'' stars of high sodium abundances, which are found only in globular clusters \citep{car10}, comprise the second \delete{generation} \add{population}. 
Here we show from a differential line-by-line analysis of stars that span a small range of effective temperature that the globular cluster M92 has higher Fe abundances in second-\delete{generation}\add{population} (sodium-enhanced) stars than first-\delete{generation}\add{population} stars.  \delete{This is the first example of a cluster with a correlation between light and heavy element abundances.}  \add{The two populations are well separated in Na, Al, and Fe abundances.}  The rise in \delete{metallicity} \add{Fe abundance} between the first and second stellar \delete{generations} \add{populations} suggests that M92 was able to retain at least some \delete{of the} supernova ejecta, \add{all of which exploded after the first population finished forming.} \delete{for the first time providing a clock for globular cluster formation} \add{This result provides a lower limit for the time delay between populations}.  \delete{We show that the stellar populations are a sequence in time; that the first population formed in a near-instantaneous burst; that the second population formed after a pause; and that the observed abundance dispersions imply the former existence of a stellar population that predates the ``first'' population.}
}

\notetoeditor{For readability, subsequent revisions of ``generation'' to ``population,'' ``1G'' to ``1P,'' and ``2G'' to ``2P'' are not indicated.  Changes to punctuation are also not indicated.}

\add{Until recently,} the globular cluster (GC) M92 seemed to conform to the expectation that GCs do not retain appreciable amounts of supernova ejecta \citep{bai09,bai18}.  Like the majority of GCs, it has a dispersion in Fe abundances less than 0.1~dex \citep{coh11b,wil12}, and it is not chemically anomalous in its $s$-process elements \citep{mil17}.  Like all globular clusters, it exhibits anti-correlations in light element abundances, such as the Na--O anti-correlation \citep{car09a}.

\delete{There is no definitive evidence that the Fe-peak abundances in any GC correlate with the light element abundances.  Even} Anomalous GCs, which are defined by star-to-star variation in Fe-peak abundances and represent about 15\% of all GCs \citep{mil17}, rarely show correlation between the absolute abundances of light and heavy elements.  The different metallicity populations in anomalous GCs sometimes show different strengths of light element anti-correlations \citep{joh10,joh17}\delete{, but not absolute abundances}.  
\add{Differential line-by-line analyses in a few GCs show correlations between Na and Fe-peak elements, such as in the ``RGB bump'' sample of NGC~6752 \citep{yon13_differential} and in M22 \citep{mck22}.  A similar analysis revealed statistically significant dispersion in heavy elements, including $s$-process elements in NGC~288 and NGC~362 \citep{mon23a}.  M22 and NGC~362 are classified as anomalous from space-based photometry, but NGC~6752 and NGC~288 are not \citep{mil17}.  The implied question is whether even ``ordinary'' GCs have heavy-element abundance variations at some level.}

The apparent lack of correlation between light and heavy elements \delete{makes it difficult to construct a self-consistent theory of cluster formation.} \add{in most clusters means that the light element abundance patterns are made before supernovae explode or that the clusters do not retain supernova ejecta.  Proposed sources of the light element abundance patterns include asymptotic giant branch (AGB) stars \citep{dan01}, fast-rotating massive stars (FRMSs) \citep{dec07}, and extremely massive stars (EMSs) \citep{gie25}.}  Nucleosynthesis of Fe-peak elements (from core collapse supernovae) is expected to happen on a shorter timescale than \delete{light elements (e.g., from asymptotic giant branch, or AGB, stars).} \add{AGB stars.  Therefore, explaining the abundance patterns with AGB stars requires that the GCs do not retain supernova ejecta but do retain AGB ejecta.}  \delete{Fast-rotating massive stars} FRMSs can make the Na--O anti-correlation, and they have much shorter timescales than AGB stars, but they cannot produce the Mg--Al anti-correlation seen in many clusters, nor can they explain the small He abundance variations seen in smaller clusters \citep{cha16}.  \add{\delete{Extremely massive stars} EMSs can explain the light element abundance patterns \citep{gie25}, but they might not be able to explain a different Fe abundance in the second population of stars than in the first population.}

\delete{How can a cluster enrich itself in light elements, requiring a long time to form, without enriching itself in Fe, which needs only a few Myr?}  Many theories \add{of GC formation} have been proposed, but \delete{all} \add{most} of them violate multiple observational constraints \citep{bas18}.  The most likely path to identifying the source of the light element abundance patterns is to measure the timescale of production because AGB stars, FRMSs, \add{and EMSs} have such different timescales.  Tying the timescale of light element production to Fe, which is produced on the well-known timescale of core collapse supernovae, would be illuminating.  \delete{However, the lack of correlation between Fe and light elements means that Fe abundance cannot be used to derive timescales for cluster formation and thus identify the source of the light element variations.}

Most GCs have uniform compositions of neutron-capture elements \add{formed in the rapid process ($r$-process)}.  \add{However, M5, M15, M92, and NGC~3201 have a dispersion in $r$-process elements \citep{sne97,roe11b,roe11a}.}  M92's first-population (1P) stars show a measurable dispersion in the $r$-process elements, whereas the second-population (2P) stars do not \citep{kir23}.  M15 and NGC~2298 have also recently been shown to display star-to-star $r$-process variations in 1P but not 2P \citep{cab24,ban25,hen25}.  This pattern is subtle, and it is easier to find in metal-poor clusters, so it is not clear whether these clusters are unusual or simply the first to be shown to have $r$-process variations \citep{nal25}.  The patterns of neutron-capture abundances in these clusters are distinct from those in \add{the} anomalous clusters \delete{which} \add{that} show $s$-process enrichment\add{, like NGC~288 and NGC~362 \citep{car13_ngc362,mon23a}}.  The neutron-capture abundances in \add{M5}, M15, M92, \delete{and} NGC~2298 \add{and NGC~3201} are from $r$-process enrichment based on their [Ba/Eu] ratios, as discussed more in the Supplementary Discussion.

Although prior spectroscopic studies have not shown a variation of Fe-peak elements in M92, it has been photometrically classified as a ``metal-complex'' cluster \citep{lee24}.  Narrow-band photometry in the Ca~HK spectral region hints that the stars in M92 clump into two groups separated in metallicity by about 0.12~dex.  However, this spectral region is sensitive to the abundances of many elements in addition to Fe.  Spectroscopy can definitively identify the elements with varying abundances.

\begin{figure}[b!]
\centering
\includegraphics[width=0.9\textwidth]{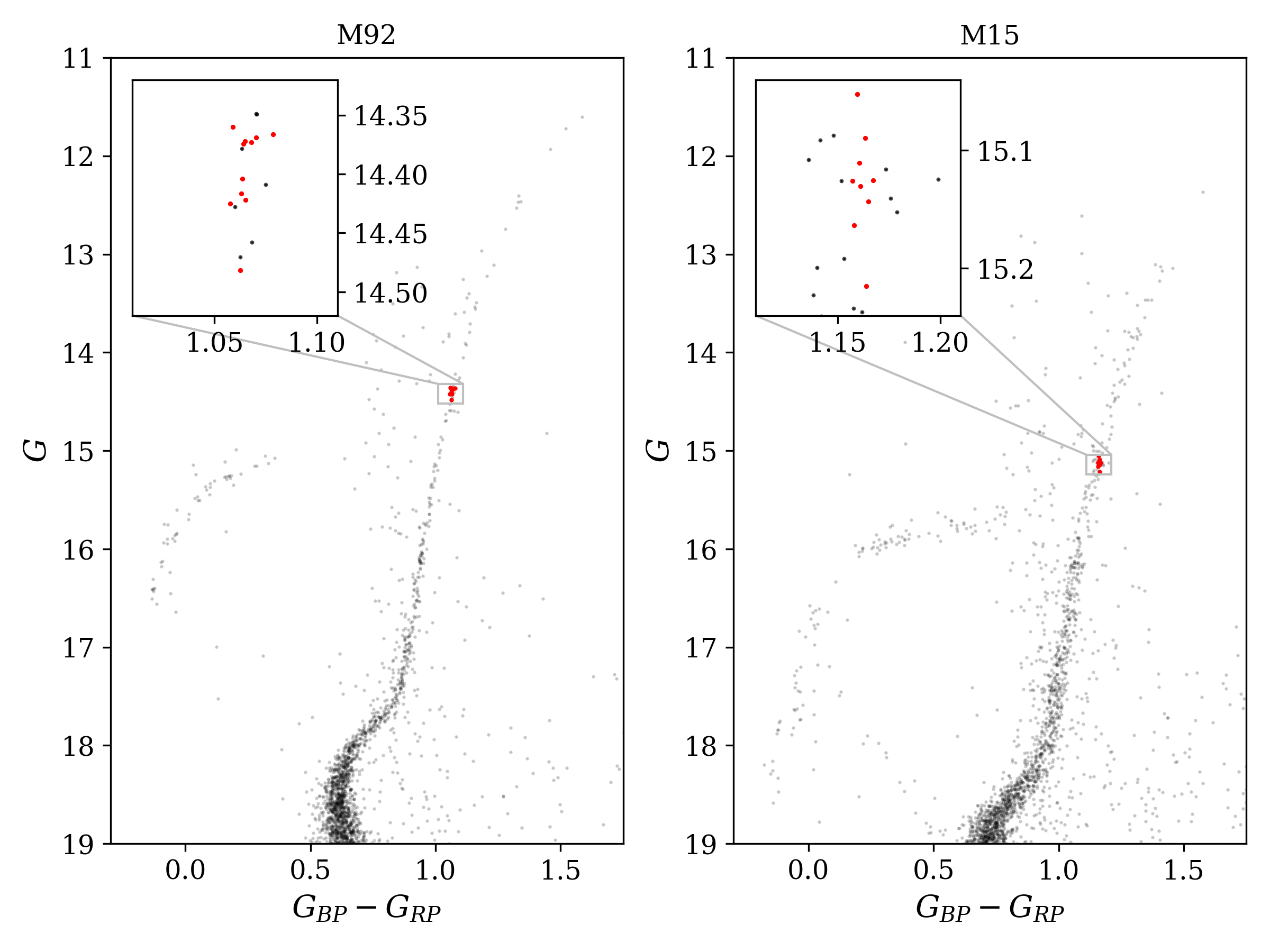}
\caption{\textit{Gaia} apparent color--magnitude diagrams of M92 and M15.  The red points show stars observed with Keck/HIRES, shown in detail in the insets.  The black points are stars in the \textit{Gaia} catalog at least 3~arcmin from the center of the cluster (to avoid photometry errors due to crowding) and with proper motions within 3~mas~yr$^{-1}$ of the mean proper motion.}\label{fig:cmd}
\end{figure}

\begin{figure}[b]
\centering
\includegraphics[width=0.9\textwidth]{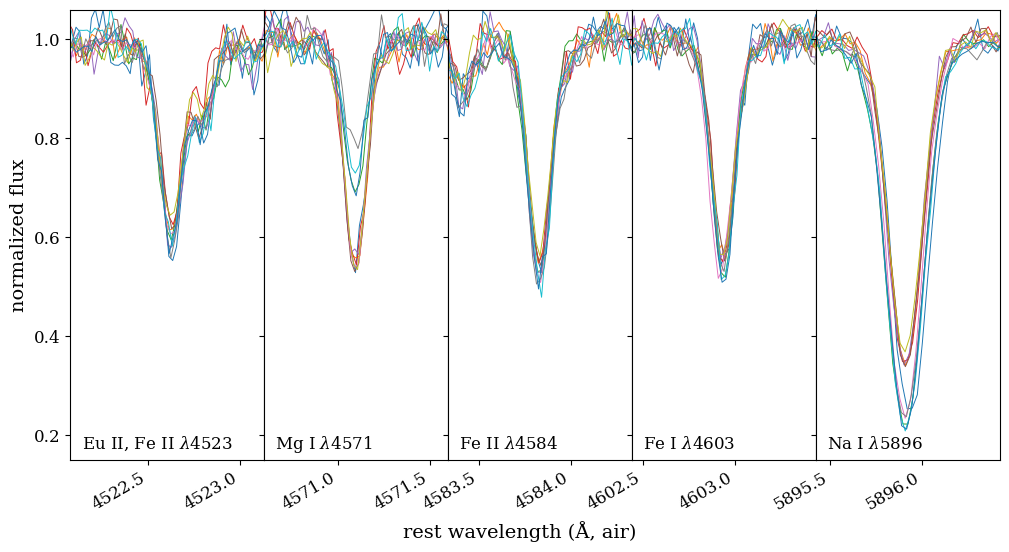}
\caption{Select absorption lines in all stars observed in M92.  A single star is represented by the same colored line in each panel.  Stars with strong Na absorption also have strong Fe absorption but weak Mg absorption.  The Eu absorption strength is variable, but it does not correlate with the other elements.}\label{fig:spectra}
\end{figure}

\subsection*{Measurements}

In order to further investigate the origin of the abundance patterns in detail, we obtained high-resolution spectra of \nmone\ stars in M92 and \nmtwo\ stars in M15.  We focus our discussion on M92.  M15 is used as a contrasting case.  For example, the fact that we do not see \delete{Fe retention in M15 that correlates with stellar population} \add{an obvious difference in Fe abundance between the stellar populations of M15} shows that GCs evolve differently from each other.  It is also evidence (in addition to evidence presented \delete{below an} in the Supplementary Discussion) that our results for M92 do not result from a systematic error.

We chose the sample from a very narrow range of color and magnitude within each cluster.  This approach minimizes systematic differences in the abundance measurements between stars.  Color and magnitude \add{at fixed distance} roughly correspond to effective temperature ($T_{\rm eff}$) and surface gravity ($\log g$), respectively.  For stars of identical $T_{\rm eff}$ and $\log g$, the abundance of an element almost solely determines the strength of its absorption lines.  Figure~\ref{fig:cmd} shows the {\it Gaia} DR3 \citep{gaiadr3} color--magnitude diagrams of \delete{both clusters} \add{M92 and M15}.  Figure~\ref{fig:spectra} shows select absorption lines in the spectra of M92 stars.  Their small temperature range allows line strengths to be interpreted as abundances.

We observed all but two of the stars with Keck/HIRES \citep{vog94} on 2022 Aug 13--14 UT\@.  We used a 1.148'' slit for a resolving power of \delete{37,500} \add{45,000 (determined by the smoothing required to match a synthetic spectrum to the observed line profiles)}.  For M92, we obtained two exposures of 1080~s for each star in 2022.  We observed M92-star-13 on 2024 Aug 14 with two exposures of 1200~s and one exposure of 1320~s.  M15 is farther and fainter, so we obtained two exposures of 1320--1380~s in 2022.  We observed M15-star-4 on 2025 May 9 with two exposures of 810~s each.  \delete{The} Methods \delete{section} describes the data reduction.

\begin{sidewaystable}
\caption{Star coordinates, photometry, and atmospheric parameters.}\label{tab:starcoords}
\begin{tabular}{llcccccccc}
\toprule
{\it Gaia} DR3 ID & Nickname & RA & Dec & $G_0$ & $(G_{BP} - G_{RP})_0$ & \add{S/N (pix$^{-1}$)} & $T_{\rm eff}$ (K)\footnotemark[1] & $\log g$ (cm s$^{-2}$) & $\xi$ (km s$^{-1}$) \\ \midrule
1360404232151729024 & M92-star-1  & 17h16m56.0s & +43d04m47.9s & 14.312 & 1.038 & 49 & $5001 \pm 3$ & $2.03 \pm 0.10$ & $2.03 \pm 0.10$ \\ 
1360407358887922560 & M92-star-2  & 17h17m29.5s & +43d12m14.7s & 14.316 & 1.036 & 56 & $5006 \pm 3$ & $2.03 \pm 0.10$ & $1.98 \pm 0.07$ \\ 
1360216353098847232 & M92-star-3  & 17h17m12.2s & +43d02m20.9s & 14.315 & 1.033 & 62 & $5012 \pm 2$ & $2.03 \pm 0.10$ & $2.06 \pm 0.08$ \\ 
1360216181300170240 & M92-star-4  & 17h17m03.9s & +43d02m03.0s & 14.317 & 1.032 & 70 & $5014 \pm 3$ & $2.03 \pm 0.10$ & $2.04 \pm 0.07$ \\ 
1360405091145275904 & M92-star-6  & 17h16m59.4s & +43d07m09.4s & 14.347 & 1.031 & 69 & $5015 \pm 4$ & $2.05 \pm 0.10$ & $1.96 \pm 0.07$ \\ 
1360405670963035648\footnotemark[2] & M92-star-8  & 17h17m04.2s & +43d08m55.0s & 14.309 & 1.046 & 62 & $4985 \pm 12$ & $2.02 \pm 0.10$ & $1.83 \pm 0.06$ \\ 
1360405774042282112 & M92-star-9  & 17h17m10.0s & +43d10m17.3s & 14.365 & 1.033 & 55 & $5012 \pm 5$ & $2.05 \pm 0.10$ & $1.99 \pm 0.08$ \\ 
1360405572181946112 & M92-star-10 & 17h17m18.9s & +43d09m29.7s & 14.359 & 1.031 & 56 & $5016 \pm 5$ & $2.05 \pm 0.10$ & $1.94 \pm 0.08$ \\ 
1360405258646144896 & M92-star-11 & 17h16m56.6s & +43d07m23.0s & 14.303 & 1.027 & 52 & $5024 \pm 6$ & $2.03 \pm 0.10$ & $2.05 \pm 0.10$ \\ 
1360405018127998848 & M92-star-12 & 17h17m03.0s & +43d06m36.5s & 14.368 & 1.025 & 42 & $5027 \pm 10$ & $2.06 \pm 0.10$ & $2.00 \pm 0.10$ \\ 
1360381451645627904 & M92-star-13 & 17h16m43.3s & +43d04m16.1s & 14.425 & 1.030 & 84 & $5017 \pm 3$ & $2.08 \pm 0.10$ & $2.02 \pm 0.10$ \\ 
1745972173687459456 & M15-star-4  & 21h30m10.6s & +12d14m11.4s & 14.883 & 1.029 & 37 & $5018 \pm 4$ & $2.07 \pm 0.10$ & $2.06 \pm 0.11$ \\ 
1745977362007982592 & M15-star-5  & 21h29m52.3s & +12d19m39.6s & 14.901 & 1.027 & 42 & $5023 \pm 4$ & $2.08 \pm 0.10$ & $1.93 \pm 0.10$ \\ 
1745934000017148928 & M15-star-6  & 21h29m57.1s & +12d04m21.9s & 14.973 & 1.026 & 52 & $5025 \pm 4$ & $2.11 \pm 0.10$ & $1.96 \pm 0.10$ \\ 
1745948461173090432 & M15-star-7  & 21h30m03.9s & +12d10m52.5s & 14.847 & 1.026 & 49 & $5026 \pm 8$ & $2.06 \pm 0.10$ & $2.05 \pm 0.09$ \\ 
1745971864449368064 & M15-star-8  & 21h30m21.0s & +12d13m00.7s & 14.888 & 1.023 & 66 & $5031 \pm 3$ & $2.08 \pm 0.10$ & $1.90 \pm 0.08$ \\ 
1745948461173092480 & M15-star-9  & 21h30m04.2s & +12d11m27.8s & 14.868 & 1.023 & 69 & $5032 \pm 7$ & $2.07 \pm 0.10$ & $1.89 \pm 0.08$ \\ 
1745948534187384704 & M15-star-10 & 21h30m00.7s & +12d11m48.9s & 14.810 & 1.022 & 65 & $5034 \pm 8$ & $2.05 \pm 0.10$ & $1.91 \pm 0.08$ \\ 
1745947808337966592 & M15-star-11 & 21h29m49.4s & +12d08m26.9s & 14.921 & 1.020 & 57 & $5037 \pm 6$ & $2.10 \pm 0.10$ & $1.96 \pm 0.09$ \\ 
1745948190585815680\footnotemark[2] & M15-star-12 & 21h30m07.5s & +12d10m11.5s & 14.883 & 1.019 & 51 & $5039 \pm 38$ & $2.08 \pm 0.10$ & $2.02 \pm 0.08$ \\ 
\bottomrule
\end{tabular}
\footnotetext[1]{The uncertainty is the random component propagated from the $G_{BP}-G_{RP}$ color.  The color--$T_{\rm eff}$ relation \citep{muc21} gives an additional systematic uncertainty of 83~K\@.  The Methods section discusses our separate treatment of these errors.}\footnotetext[2]{\add{The \textit{Gaia} magnitudes for these stars are possibly affected by crowding.}}\end{sidewaystable}

To maximize the benefit of choosing stars of similar temperature, we determined $T_{\rm eff}$ photometrically.  We used a calibration based on the infrared flux method \citep{muc21} between {\it Gaia} $G_{BP}-G_{RP}$ colors and $T_{\rm eff}$.  \add{(Methods discusses the effects of photometric crowding.)}  The surface gravity was determined from the relation between luminosity, temperature, and gravity (see Equation~1 of Kirby et al.\ 2023 \cite{kir23}).  We adopted apparent distance moduli of 14.74 and 15.42 for M92 and M15, respectively \citep{van16}.  We corrected for reddening assuming $E(B-V) = 0.023$ and $0.10$ \citep{van16}, respectively, \delete{and} using the reddening and extinction formulae appropriate for {\it Gaia} photometry \citep{gaia18}.  We inferred the luminosity from the absolute, extinction-corrected $G$ magnitude after applying a bolometric correction \citep{and18}.  Table~\ref{tab:starcoords} gives the names, coordinates, extinction- and reddening-corrected photometry, \add{signal-to-noise ratio (S/N),} $T_{\rm eff}$, $\log g$, and microturbulent velocity $\xi$ for each star.  \add{The S/N is measured as 0.6745 times the inverse median absolute deviation from 1 of continuum-normalized flux values in the range 5715--5795~\AA, which is relatively free from absorption lines.  (The factor of 0.6745 scales the median absolute deviation to the standard deviation for a normal distribution.)}  The ranges of $T_{\rm eff}$ and $\log g$ are \teffrangeone~K and \loggrangeone\ in M92 and \teffrangetwo~K and \loggrangetwo\ in M15, respectively.

Measuring absolute stellar abundances is complicated by uncertainties and assumptions.  The largest sources of uncertainty are the atmospheric parameters, especially $T_{\rm eff}$, and the oscillator strengths of the absorption lines.  One of the biggest often-used assumptions is local thermodynamic equilibrium (LTE)\@.  These effects can be mitigated by measuring differential abundances \citep{ram09,yon13_differential,mck22,mon23a,ber26}, 
\notetoeditor{references added}
where stars are compared to a reference star of similar parameters.  Instead of averaging the absolute abundances of all the lines of a species, like Fe$\,${\sc i}, the differences in abundance are computed line by line, and the differences \add{in abundance for all lines of the same species} are averaged.  For stars that span a small range of $T_{\rm eff}$, $\log g$, and metallicity, systematic errors from stellar parameters, oscillator strengths, and the assumption of LTE are minimized because they affect the same line by about the same amount in different stars.  Therefore, the correlated errors are subtracted away.

\begin{sidewaystable}
\caption{Differential abundances for C through Si.}\label{tab:abund1}
\begin{tabular*}{\textheight}{@{\extracolsep\fill}l c c c c c c@{}}
\toprule
Star & \add{C} & \add{N} & Na & Mg & Al & Si \\ \midrule
M92-star-1 & $+0.161 \pm 0.180$ & --- & $-0.379 \pm 0.039$ & $+0.113 \pm 0.021$ & $-0.771 \pm 0.091$ & $-0.221 \pm 0.050$ \\ 
M92-star-2 & $-0.279 \pm 0.188$ & $+0.394 \pm 0.249$ & $+0.419 \pm 0.035$ & $-0.298 \pm 0.021$ & $+0.693 \pm 0.057$ & $+0.168 \pm 0.061$ \\ 
M92-star-3 & $-0.155 \pm 0.186$ & $+0.350 \pm 0.248$ & $+0.380 \pm 0.032$ & $-0.242 \pm 0.019$ & $+0.730 \pm 0.051$ & $+0.114 \pm 0.053$ \\ 
M92-star-4 & $-0.155 \pm 0.190$ & $+0.277 \pm 0.254$ & $+0.369 \pm 0.032$ & $-0.141 \pm 0.019$ & $+0.644 \pm 0.048$ & $+0.045 \pm 0.040$ \\ 
M92-star-6 & $+0.127 \pm 0.183$ & --- & $-0.391 \pm 0.037$ & $+0.105 \pm 0.019$ & $-0.666 \pm 0.073$ & $-0.107 \pm 0.038$ \\ 
M92-star-8 & $-0.319 \pm 0.196$ & $+0.342 \pm 0.249$ & $+0.333 \pm 0.036$ & $-0.377 \pm 0.019$ & $+0.659 \pm 0.052$ & $+0.108 \pm 0.052$ \\ 
M92-star-9 & $-0.230 \pm 0.162$ & --- & $-0.377 \pm 0.038$ & $+0.104 \pm 0.020$ & $-0.628 \pm 0.093$ & $-0.058 \pm 0.050$ \\ 
M92-star-10 & $+0.381 \pm 0.173$ & $-1.011 \pm 0.305$ & $-0.468 \pm 0.042$ & $+0.144 \pm 0.020$ & $-0.694 \pm 0.068$ & $+0.033 \pm 0.046$ \\ 
M92-star-11 & $+0.153 \pm 0.181$ & --- & $-0.300 \pm 0.040$ & $+0.142 \pm 0.022$ & $-0.800 \pm 0.079$ & $+0.076 \pm 0.054$ \\ 
M92-star-12 & $+0.096 \pm 0.180$ & $-0.037 \pm 0.253$ & $+0.250 \pm 0.036$ & $+0.127 \pm 0.025$ & $+0.596 \pm 0.076$ & $+0.026 \pm 0.059$ \\ 
M92-star-13 & $+0.219 \pm 0.182$ & $-0.314 \pm 0.266$ & $+0.150 \pm 0.032$ & $+0.142 \pm 0.018$ & $+0.207 \pm 0.055$ & $-0.197 \pm 0.037$ \\ 
M15-star-4 & $-0.024 \pm 0.185$ & --- & $-0.363 \pm 0.041$ & $-0.040 \pm 0.026$ & --- & $-0.226 \pm 0.091$ \\ 
M15-star-5 & $+0.006 \pm 0.181$ & $+0.257 \pm 0.237$ & $+0.386 \pm 0.035$ & $-0.020 \pm 0.025$ & $+0.525 \pm 0.090$ & $+0.087 \pm 0.076$ \\ 
M15-star-6 & $+0.230 \pm 0.178$ & $-0.776 \pm 0.259$ & $-0.160 \pm 0.036$ & $+0.113 \pm 0.023$ & $-0.436 \pm 0.077$ & $-0.085 \pm 0.052$ \\ 
M15-star-7 & $+0.100 \pm 0.178$ & --- & $-0.282 \pm 0.039$ & $+0.186 \pm 0.025$ & $-0.507 \pm 0.082$ & $-0.016 \pm 0.062$ \\ 
M15-star-8 & $-0.323 \pm 0.188$ & $+0.711 \pm 0.255$ & $+0.292 \pm 0.034$ & $-0.558 \pm 0.022$ & $+0.388 \pm 0.059$ & $+0.257 \pm 0.054$ \\ 
M15-star-9 & $-0.064 \pm 0.187$ & $+0.153 \pm 0.264$ & $+0.358 \pm 0.033$ & $+0.027 \pm 0.020$ & $+0.316 \pm 0.055$ & $-0.049 \pm 0.052$ \\ 
M15-star-10 & $+0.262 \pm 0.181$ & $-0.517 \pm 0.276$ & $-0.136 \pm 0.035$ & $+0.095 \pm 0.020$ & $-0.319 \pm 0.075$ & $-0.019 \pm 0.050$ \\ 
M15-star-11 & $+0.011 \pm 0.184$ & --- & $-0.250 \pm 0.038$ & $+0.120 \pm 0.024$ & $-0.398 \pm 0.088$ & $-0.095 \pm 0.058$ \\ 
M15-star-12 & $-0.197 \pm 0.202$ & $+0.171 \pm 0.260$ & $+0.138 \pm 0.057$ & $-0.058 \pm 0.030$ & $+0.441 \pm 0.077$ & $+0.032 \pm 0.077$ \\ 
\botrule
\end{tabular*}
\end{sidewaystable}
\begin{sidewaystable}
\caption{Differential abundances for K through Fe.}\label{tab:abund2}
\begin{tabular*}{\textheight}{@{\extracolsep\fill}l c c c c c c@{}}
\toprule
Star & K & Ca & Sc & Ti & V & Cr \\ \midrule
M92-star-1 & $-0.160 \pm 0.038$ & $-0.021 \pm 0.017$ & $-0.048 \pm 0.035$ & $-0.034 \pm 0.024$ & $-0.040 \pm 0.035$ & $-0.015 \pm 0.018$ \\ 
M92-star-2 & $+0.121 \pm 0.039$ & $+0.048 \pm 0.018$ & $+0.136 \pm 0.039$ & $+0.040 \pm 0.025$ & $+0.061 \pm 0.033$ & $+0.094 \pm 0.018$ \\ 
M92-star-3 & $+0.181 \pm 0.038$ & $+0.060 \pm 0.017$ & $+0.081 \pm 0.035$ & $+0.051 \pm 0.026$ & $+0.065 \pm 0.030$ & $+0.051 \pm 0.017$ \\ 
M92-star-4 & $-0.046 \pm 0.031$ & $+0.018 \pm 0.015$ & $+0.032 \pm 0.035$ & $+0.015 \pm 0.024$ & $+0.021 \pm 0.031$ & $+0.016 \pm 0.016$ \\ 
M92-star-6 & $-0.077 \pm 0.034$ & $-0.030 \pm 0.014$ & $-0.056 \pm 0.033$ & $-0.048 \pm 0.023$ & $+0.022 \pm 0.031$ & $-0.069 \pm 0.016$ \\ 
M92-star-8 & $+0.167 \pm 0.037$ & $+0.024 \pm 0.017$ & $+0.060 \pm 0.035$ & $+0.032 \pm 0.025$ & $+0.064 \pm 0.030$ & $+0.026 \pm 0.019$ \\ 
M92-star-9 & $+0.051 \pm 0.038$ & $-0.047 \pm 0.015$ & $-0.045 \pm 0.033$ & $-0.035 \pm 0.024$ & $-0.040 \pm 0.033$ & $-0.032 \pm 0.018$ \\ 
M92-star-10 & $-0.291 \pm 0.037$ & $-0.077 \pm 0.015$ & $-0.077 \pm 0.035$ & $-0.069 \pm 0.024$ & $-0.167 \pm 0.033$ & $-0.045 \pm 0.019$ \\ 
M92-star-11 & $-0.014 \pm 0.043$ & $-0.043 \pm 0.018$ & $-0.064 \pm 0.035$ & $-0.023 \pm 0.024$ & $-0.029 \pm 0.046$ & $-0.033 \pm 0.019$ \\ 
M92-star-12 & $+0.038 \pm 0.052$ & $+0.038 \pm 0.020$ & $-0.051 \pm 0.037$ & $+0.035 \pm 0.027$ & $+0.026 \pm 0.042$ & $+0.003 \pm 0.022$ \\ 
M92-star-13 & $+0.029 \pm 0.028$ & $+0.033 \pm 0.015$ & $+0.051 \pm 0.032$ & $+0.024 \pm 0.024$ & $-0.006 \pm 0.027$ & $+0.021 \pm 0.015$ \\ 
M15-star-4 & $-0.212 \pm 0.050$ & $-0.099 \pm 0.021$ & $-0.128 \pm 0.043$ & $-0.089 \pm 0.026$ & $-0.377 \pm 0.133$ & $-0.033 \pm 0.032$ \\ 
M15-star-5 & $+0.036 \pm 0.049$ & $+0.001 \pm 0.018$ & $-0.016 \pm 0.039$ & $+0.005 \pm 0.026$ & $+0.092 \pm 0.060$ & $-0.003 \pm 0.025$ \\ 
M15-star-6 & $-0.012 \pm 0.044$ & $-0.002 \pm 0.016$ & $-0.005 \pm 0.039$ & $+0.002 \pm 0.025$ & $-0.022 \pm 0.043$ & $-0.005 \pm 0.023$ \\ 
M15-star-7 & $-0.039 \pm 0.041$ & $-0.021 \pm 0.017$ & $+0.018 \pm 0.039$ & $+0.009 \pm 0.025$ & $+0.106 \pm 0.050$ & $+0.004 \pm 0.020$ \\ 
M15-star-8 & $+0.444 \pm 0.041$ & $+0.133 \pm 0.016$ & $+0.176 \pm 0.037$ & $+0.053 \pm 0.024$ & $+0.046 \pm 0.031$ & $+0.054 \pm 0.017$ \\ 
M15-star-9 & $+0.010 \pm 0.036$ & $-0.013 \pm 0.015$ & $+0.000 \pm 0.034$ & $+0.004 \pm 0.024$ & $-0.029 \pm 0.034$ & $+0.003 \pm 0.017$ \\ 
M15-star-10 & $-0.034 \pm 0.034$ & $+0.004 \pm 0.015$ & $+0.019 \pm 0.034$ & $-0.002 \pm 0.023$ & $+0.018 \pm 0.040$ & $-0.006 \pm 0.018$ \\ 
M15-star-11 & $+0.077 \pm 0.042$ & $+0.010 \pm 0.016$ & $-0.035 \pm 0.032$ & $+0.004 \pm 0.024$ & $+0.018 \pm 0.038$ & $+0.011 \pm 0.020$ \\ 
M15-star-12 & $-0.267 \pm 0.046$ & $-0.036 \pm 0.029$ & $-0.052 \pm 0.043$ & $-0.020 \pm 0.033$ & $-0.018 \pm 0.045$ & $-0.028 \pm 0.032$ \\ 
\botrule
\end{tabular*}
\end{sidewaystable}
\begin{sidewaystable}
\caption{Differential abundances for Co through Zr.}\label{tab:abund3}
\begin{tabular*}{\textheight}{@{\extracolsep\fill}l c c c c c c@{}}
\toprule
Star & Mn & Fe & Co & Ni & Zn & Sr \\ \midrule
M92-star-1 & $-0.068 \pm 0.032$ & $-0.038 \pm 0.019$ & $-0.015 \pm 0.033$ & $-0.007 \pm 0.024$ & $+0.013 \pm 0.045$ & $-0.131 \pm 0.051$ \\ 
M92-star-2 & $+0.054 \pm 0.028$ & $+0.072 \pm 0.021$ & $+0.037 \pm 0.039$ & $+0.066 \pm 0.021$ & $-0.040 \pm 0.049$ & $+0.095 \pm 0.050$ \\ 
M92-star-3 & $+0.067 \pm 0.024$ & $+0.050 \pm 0.021$ & $+0.081 \pm 0.034$ & $+0.033 \pm 0.020$ & $+0.031 \pm 0.040$ & $+0.125 \pm 0.047$ \\ 
M92-star-4 & $+0.017 \pm 0.024$ & $+0.019 \pm 0.019$ & $+0.028 \pm 0.030$ & $+0.006 \pm 0.018$ & $+0.081 \pm 0.035$ & $-0.057 \pm 0.044$ \\ 
M92-star-6 & $-0.026 \pm 0.023$ & $-0.053 \pm 0.018$ & $-0.028 \pm 0.027$ & $-0.055 \pm 0.020$ & $-0.053 \pm 0.037$ & $+0.086 \pm 0.046$ \\ 
M92-star-8 & $+0.002 \pm 0.025$ & $+0.031 \pm 0.021$ & $-0.010 \pm 0.034$ & $+0.023 \pm 0.021$ & $+0.036 \pm 0.037$ & $+0.133 \pm 0.046$ \\ 
M92-star-9 & $-0.017 \pm 0.029$ & $-0.041 \pm 0.019$ & $-0.010 \pm 0.032$ & $-0.035 \pm 0.022$ & $-0.011 \pm 0.042$ & $-0.057 \pm 0.049$ \\ 
M92-star-10 & $-0.028 \pm 0.027$ & $-0.049 \pm 0.019$ & $-0.055 \pm 0.034$ & $-0.060 \pm 0.027$ & $-0.040 \pm 0.050$ & $-0.125 \pm 0.052$ \\ 
M92-star-11 & $-0.021 \pm 0.038$ & $-0.055 \pm 0.020$ & $-0.066 \pm 0.035$ & $-0.054 \pm 0.024$ & $-0.096 \pm 0.054$ & $+0.045 \pm 0.049$ \\ 
M92-star-12 & $-0.010 \pm 0.036$ & $+0.034 \pm 0.022$ & $-0.028 \pm 0.045$ & $+0.046 \pm 0.031$ & $+0.065 \pm 0.059$ & $-0.212 \pm 0.062$ \\ 
M92-star-13 & $+0.013 \pm 0.020$ & $+0.010 \pm 0.018$ & $+0.069 \pm 0.028$ & $+0.066 \pm 0.018$ & $+0.032 \pm 0.031$ & $+0.119 \pm 0.040$ \\ 
M15-star-4 & $-0.036 \pm 0.091$ & $-0.080 \pm 0.022$ & $-0.021 \pm 0.044$ & $+0.006 \pm 0.040$ & $-0.038 \pm 0.072$ & $-0.189 \pm 0.065$ \\ 
M15-star-5 & $-0.046 \pm 0.048$ & $+0.019 \pm 0.021$ & $+0.022 \pm 0.041$ & $-0.042 \pm 0.036$ & $+0.072 \pm 0.072$ & $+0.119 \pm 0.055$ \\ 
M15-star-6 & $-0.068 \pm 0.033$ & $-0.002 \pm 0.021$ & $-0.070 \pm 0.035$ & $+0.008 \pm 0.026$ & $+0.102 \pm 0.063$ & $+0.063 \pm 0.051$ \\ 
M15-star-7 & $+0.030 \pm 0.037$ & $+0.007 \pm 0.020$ & $+0.069 \pm 0.039$ & $-0.018 \pm 0.028$ & $-0.062 \pm 0.058$ & $+0.093 \pm 0.053$ \\ 
M15-star-8 & $+0.020 \pm 0.026$ & $+0.038 \pm 0.019$ & $+0.001 \pm 0.029$ & $+0.033 \pm 0.020$ & $+0.094 \pm 0.036$ & $-0.040 \pm 0.047$ \\ 
M15-star-9 & $+0.049 \pm 0.026$ & $+0.002 \pm 0.019$ & $-0.007 \pm 0.030$ & $+0.012 \pm 0.022$ & $-0.065 \pm 0.042$ & $+0.048 \pm 0.047$ \\ 
M15-star-10 & $-0.036 \pm 0.026$ & $-0.011 \pm 0.018$ & $+0.019 \pm 0.030$ & $-0.003 \pm 0.023$ & $+0.013 \pm 0.039$ & $+0.117 \pm 0.044$ \\ 
M15-star-11 & $+0.023 \pm 0.032$ & $+0.007 \pm 0.020$ & $-0.014 \pm 0.033$ & $-0.015 \pm 0.026$ & $-0.035 \pm 0.047$ & $-0.028 \pm 0.051$ \\ 
M15-star-12 & $+0.030 \pm 0.046$ & $-0.011 \pm 0.032$ & $+0.004 \pm 0.056$ & $+0.020 \pm 0.041$ & $-0.042 \pm 0.057$ & $-0.186 \pm 0.063$ \\ 
\botrule
\end{tabular*}
\end{sidewaystable}
\begin{sidewaystable}
\caption{Differential abundances for Ba through Dy.}\label{tab:abund4}
\begin{tabular*}{\textheight}{@{\extracolsep\fill}l c c c c c c c@{}}
\toprule
Star & Y & Zr & Ba & La & Nd & Eu & Dy \\ \midrule
M92-star-1 & $-0.055 \pm 0.036$ & $-0.189 \pm 0.065$ & $-0.171 \pm 0.042$ & $-0.170 \pm 0.070$ & --- & $-0.150 \pm 0.063$ & $-0.062 \pm 0.072$ \\ 
M92-star-2 & $+0.059 \pm 0.035$ & $+0.064 \pm 0.066$ & $-0.012 \pm 0.045$ & $+0.062 \pm 0.046$ & --- & $-0.041 \pm 0.046$ & $+0.033 \pm 0.160$ \\ 
M92-star-3 & $+0.041 \pm 0.032$ & $+0.041 \pm 0.075$ & $-0.040 \pm 0.045$ & $+0.054 \pm 0.045$ & $-0.252 \pm 0.072$ & $-0.013 \pm 0.035$ & $-0.122 \pm 0.076$ \\ 
M92-star-4 & $-0.009 \pm 0.035$ & $+0.053 \pm 0.053$ & $+0.016 \pm 0.043$ & $+0.065 \pm 0.043$ & $+0.031 \pm 0.075$ & $+0.068 \pm 0.031$ & $+0.017 \pm 0.057$ \\ 
M92-star-6 & $+0.058 \pm 0.031$ & $+0.008 \pm 0.049$ & $+0.202 \pm 0.044$ & $+0.238 \pm 0.037$ & $+0.127 \pm 0.060$ & $+0.234 \pm 0.028$ & $+0.216 \pm 0.074$ \\ 
M92-star-8 & $+0.036 \pm 0.032$ & $-0.107 \pm 0.052$ & $+0.000 \pm 0.045$ & $-0.030 \pm 0.050$ & $+0.003 \pm 0.081$ & $-0.011 \pm 0.035$ & $-0.112 \pm 0.065$ \\ 
M92-star-9 & $-0.074 \pm 0.034$ & $+0.078 \pm 0.059$ & $-0.076 \pm 0.043$ & $-0.014 \pm 0.091$ & $-0.206 \pm 0.075$ & $-0.120 \pm 0.043$ & $-0.077 \pm 0.067$ \\ 
M92-star-10 & $-0.076 \pm 0.035$ & $-0.025 \pm 0.061$ & $-0.034 \pm 0.043$ & $-0.044 \pm 0.069$ & --- & $-0.071 \pm 0.043$ & $-0.059 \pm 0.068$ \\ 
M92-star-11 & $+0.048 \pm 0.042$ & $+0.082 \pm 0.066$ & $+0.103 \pm 0.046$ & $+0.093 \pm 0.058$ & $+0.255 \pm 0.083$ & $+0.149 \pm 0.037$ & $+0.172 \pm 0.103$ \\ 
M92-star-12 & $+0.013 \pm 0.044$ & $-0.067 \pm 0.081$ & $+0.052 \pm 0.049$ & $+0.038 \pm 0.089$ & --- & $-0.001 \pm 0.057$ & $+0.068 \pm 0.136$ \\ 
M92-star-13 & $-0.013 \pm 0.028$ & $+0.063 \pm 0.070$ & $-0.044 \pm 0.040$ & $-0.092 \pm 0.046$ & $+0.042 \pm 0.075$ & $-0.026 \pm 0.027$ & $-0.102 \pm 0.091$ \\ 
M15-star-4 & $-0.266 \pm 0.071$ & $+0.788 \pm 0.073$ & $-0.463 \pm 0.048$ & $-0.364 \pm 0.092$ & $-0.185 \pm 0.159$ & $-0.385 \pm 0.080$ & $-0.294 \pm 0.301$ \\ 
M15-star-5 & $+0.119 \pm 0.047$ & $-0.077 \pm 0.082$ & $+0.087 \pm 0.057$ & $+0.037 \pm 0.054$ & $+0.099 \pm 0.066$ & $+0.140 \pm 0.033$ & $+0.142 \pm 0.153$ \\ 
M15-star-6 & $+0.054 \pm 0.036$ & $-0.110 \pm 0.075$ & $+0.133 \pm 0.057$ & $+0.115 \pm 0.040$ & $+0.072 \pm 0.048$ & $+0.141 \pm 0.030$ & $+0.104 \pm 0.077$ \\ 
M15-star-7 & $+0.026 \pm 0.034$ & $-0.063 \pm 0.101$ & $+0.069 \pm 0.055$ & $+0.058 \pm 0.039$ & $-0.008 \pm 0.048$ & $+0.092 \pm 0.030$ & $+0.022 \pm 0.115$ \\ 
M15-star-8 & $-0.040 \pm 0.030$ & $-0.081 \pm 0.052$ & $-0.003 \pm 0.050$ & $-0.038 \pm 0.034$ & $-0.123 \pm 0.036$ & $-0.075 \pm 0.028$ & $-0.073 \pm 0.075$ \\ 
M15-star-9 & $+0.035 \pm 0.033$ & $-0.055 \pm 0.050$ & $+0.146 \pm 0.055$ & $+0.076 \pm 0.031$ & $+0.029 \pm 0.034$ & $+0.133 \pm 0.026$ & $+0.070 \pm 0.083$ \\ 
M15-star-10 & $+0.089 \pm 0.033$ & $-0.073 \pm 0.072$ & $+0.228 \pm 0.054$ & $+0.175 \pm 0.101$ & $+0.106 \pm 0.031$ & $+0.185 \pm 0.025$ & $+0.153 \pm 0.119$ \\ 
M15-star-11 & $+0.030 \pm 0.034$ & $-0.087 \pm 0.064$ & $+0.100 \pm 0.057$ & $+0.063 \pm 0.036$ & $+0.014 \pm 0.044$ & $+0.083 \pm 0.029$ & $+0.098 \pm 0.068$ \\ 
M15-star-12 & $-0.111 \pm 0.051$ & $-0.244 \pm 0.079$ & $-0.312 \pm 0.057$ & $-0.277 \pm 0.056$ & $-0.272 \pm 0.087$ & $-0.307 \pm 0.047$ & $-0.264 \pm 0.144$ \\ 
\botrule
\end{tabular*}
\end{sidewaystable}

We computed the differential abundance of each line by computing the difference between the abundance of that line and the average abundance of that line in all stars in the cluster.  (Differential abundances are often computed with respect to a reference star\delete{, in which case the reference star is not included in the sample of differential abundances}.  Taking the difference with respect to the average \delete{allows us to include all the stars in the sample} \add{lessens the effect of noise in a single reference star}.)  We use only those absorption lines measured in at least five stars within each cluster.  Tables~\ref{tab:abund1}--\ref{tab:abund4} report the differential abundances.  \delete{The} Methods \delete{section} gives more details on the abundance measurements.

\subsection*{Results}

\begin{figure}[t]
\centering
\includegraphics[width=0.6\textwidth]{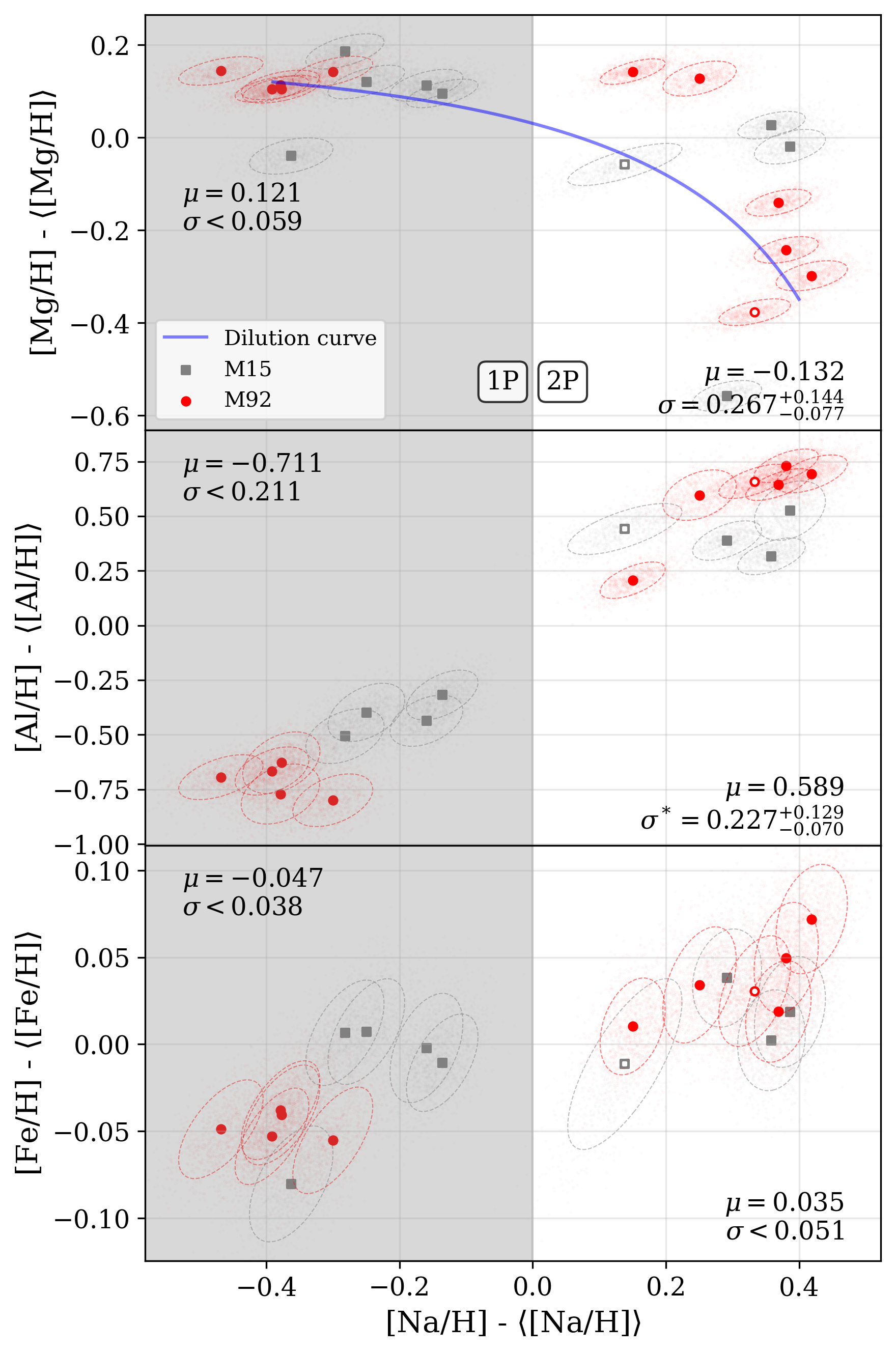}
\caption{Differential Mg, \add{Al,} and Fe abundances vs.\ differential Na abundances.  Each point is surrounded by a cloud of MC trials, representing the error distribution.  The ellipses enclose 68\% of these trials and thus represent the $1\sigma$ covariance.  \add{Hollow symbols indicate that crowding possibly affects the photometry used to derive temperature and surface gravity (see Table~\ref{tab:starcoords}).}  The gray shading identifies 1P, defined by Na \add{(or Al)} abundance.  The statistics show the mean ($\mu$) and standard deviation ($\sigma$).  Detections of $\sigma$ are given with 68\% C.I., and upper limits are specified at 95\% C.L.\@  {\it Top:} The blue curve represents the dilution curve, a linear combination of the two compositions represented by the endpoints of the curve.  (The line appears curved because the axes are logarithmic.)  \delete{The black arrows show the approximate trajectory of AGB nucleosynthesis \citep{der10}.  After about 30~Myr, the most massive AGB stars have generated Mg-poor, Na-rich material.  The composition jumps discontinuously along the dashed arrow.  Then, the lower-mass AGB stars make less extreme material, following the curve of the solid arrow.}  {\it Bottom:} 2P is enhanced in Fe compared to 1P, indicating retention of supernova ejecta.}\label{fig:mgalfe}
\end{figure}

The light element abundance patterns define the stellar populations.  \add{The populations in M92 are particularly well defined in Al abundance \citep{mas19}.  The Na--Al plane (Figure~\ref{fig:mgalfe}) shows two distinct populations.  The upper left and lower right quadrants, defined by ${\rm [Na/H]} - \langle {\rm [Na/H]} \rangle = 0.0$ and ${\rm [Al/H]} - \langle {\rm [Al/H]} \rangle = 0.0$ are devoid of stars, and there is gap of \Napopsep~dex in Na abundance and \Alpopsep~dex in Al abundance.  We define the Na-poor, Al-poor population in the lower left quadrant as 1P and the Na-rich, Al-rich population in the upper right quadrant as 2P\@.}

The top panel of Figure~\ref{fig:mgalfe} shows the anti-correlation of Na and Mg abundances.  This pattern reflects high-temperature (``advanced'') hydrogen burning.  Na is enhanced in the Ne--Na cycle, and Mg is depleted and Al enhanced in the Mg--Al cycle.  \delete{Each of these cycles has a different characteristic temperature, so they are generated in different stars at different times in the formation of the GC\@.}  The observed pattern of Mg vs.\ Na in M92 shows that Mg depletion and Na enhancement occur \delete{at different times} \add{in stars with a range of properties, such as burning temperature}.  If the abundances were explained solely by a linear combination of a single primordial (1P) source and a single enriched (2P) source, then the abundances would follow the dilution curve \citep{pra06,bas18} in Figure~\ref{fig:mgalfe} (assuming that the diluting gas had uniform composition).  Instead, Mg abundances exceed the dilution curve: \add{two stars have higher Na abundance than the curve by $4\sigma$}.  Therefore, \delete{there are multiple sources of advanced hydrogen burning.  In some, the temperature is high enough to enhance Na but not enough to deplete Mg.} \add{the Na--Mg pattern in M92 cannot be explained by two sources each with a single yield of Na and Mg.}  It is only in sources where Na is enriched to its highest value that Mg is depleted.

The wide range of Mg abundances in 2P indicates that it was polluted with nucleosynthetic sources with timescales that vary by more than the gas mixing time.  Otherwise, the 2P gas would have uniform composition (or at least fall on the dilution curve).  The cluster gas homogenizes over a short time ($\sim 1$~Myr).  The reason is that the gas turbulently mixes on a timescale comparable to the cluster orbital or crossing time \citep{mur90}.  The crossing time today is less than 1~Myr, but it might be \delete{different} \add{shorter} in the past, when the cluster was more massive and possibly more compact than today ($t_{\rm cross} \propto R^3/M$).  The cluster undoubtedly has lost gas and stellar mass \citep{lam10,kra12}.  It has also expanded due to tidal interaction with the Milky Way \citep{gie11}.  Therefore, the crossing/mixing time of the cluster was not likely to be much longer than $\sim 1$~Myr.  As a result, the nucleosynthetic sources that span the range of the Na--Mg relation evolve continuously over many Myr.

Unlike most GCs---even anomalous GCs---M92 shows evidence that its 2P retained core collapse supernova ejecta.  The bottom panel of Figure~\ref{fig:mgalfe} shows the abundances of Fe vs.\ Na.  Extended Data Figure~\ref{fig:fepeak_ext} additionally shows Ca and all of the other \delete{Fe-peak} \add{Fe-group} elements.  2P is more enhanced in all of these elements compared to 1P\@.  \add{An exact permutation test with Monte Carlo resampling (see Methods) shows that the mean Fe abundance of 2P is higher than 1P, even accounting for the sample size, at a significance of $p = 0.002$.}  Fe in 2P is enhanced by $\Femudiff \pm \Femudiffuppererr$~dex (68\% C.I.) compared to 1P\@.  If 2P formed after 1P, then 2P must have acquired Fe-peak elements from supernova explosions.  \delete{The only other cluster to have} \add{Another GC that has} evidence of retaining supernova ejecta is Terzan~5 \citep{ori11}, but the [$\alpha$/Fe] ratio in its 2P is lower than in 1P, meaning that it retained Type~Ia supernova ejecta.  The timescale for Type~Ia enrichment is so long that Massari et al.\ (2014) \citep{massari14} concluded that Terzan~5 is not a GC at all.

On its own, a dispersion in Fe would not be a highly significant finding.  \add{Other GCs also have Fe-peak dispersions, like NGC~6852 \citep{yon13_differential} and M22 \citep{mck22}.}  Small dispersions in Fe can be explained by the initial heterogeneity of the gas \citep{bai18}.  \add{Space-based photometric measurements of abundance have shown metallicity dispersion even within the first-generation stars \citep{mar19,lar22,lar23,leg22,leg24,lat25}.}  The fact that Fe is \delete{correlated with light elements, like Na, proves} \add{is higher in every 2P star than every 1P star suggests} that the Fe was generated by 1P and retained by 2P\@.

The most massive supernovae take \delete{$\gtrsim 4$} \add{$\gtrsim 3$}~Myr \add{\citep{hig23}} to explode.  Therefore, 2P started forming at least \delete{$\gtrsim 4$} \add{$\gtrsim 3$}~Myr after 1P\@.  The supernovae would produce $\alpha$ elements, such as Mg and Ca, in addition to Fe-peak elements.  Figure~\ref{fig:mgalfe} shows that Mg is depleted in 2P, whereas Fe is enhanced.  Therefore, Mg nucleosynthesis is dominated by advanced hydrogen burning, whereas Fe nucleosynthesis is done by supernovae.

Some GCs, like $\omega$~Centauri \citep{nor95,joh10}, show discrete populations.  The vast majority of GCs show a continuum along the Na--O anticorrelation \citep{car09a}.  M92 \delete{uniquely} shows \delete{discrete} \add{highly distinct} low-Na and high-Na populations, \add{a distinction previously seen in Al abundances \citep{mas19}}.  M92's 1P shows an immeasurably small scatter in \delete{nearly all} \add{most} elements with atomic number less than 30 \add{(see ``Uncertainty estimation'' in Methods)}.  Fe is the most precisely measured element by virtue of its many absorption lines spanning two ionization states and many excitation potentials.  The upper limit on Fe abundance dispersion in 1P is $\sigma < \Feonegsigmalimit$ (95\% C.L.).  With the exception of \delete{K} \add{Si, K, and V} mentioned in the Supplementary Discussion (Extended Data Figure~\ref{fig:lightelements_ext}), all the other elements with atomic number less than 30 have \delete{(less stringent)} upper limits on dispersion in 1P\@.  Furthermore, 1P is cleanly separated from 2P by \delete{0.2} \add{\Napopsep}~dex in Na abundance.  \delete{It could be that M92 is unique, or it might be that the precision of our abundance measurements permitted us to detect this separation, although we did not detect a clean separation in M15.}

\add{The pattern of Fe in M92 places timescales on the formation of 1P and 2P\@.}  The small scatter in the light elements in 1P indicates that M92 had an initial starburst.  1P completed its formation before the first high-temperature hydrogen burning sources and before the first supernovae.  \add{2P did not begin forming until supernovae yielded Fe and high-temperature hydrogen-burning sources yielded Na and Al.}  

\begin{figure}[t]
\centering
\includegraphics[width=0.45\textwidth]{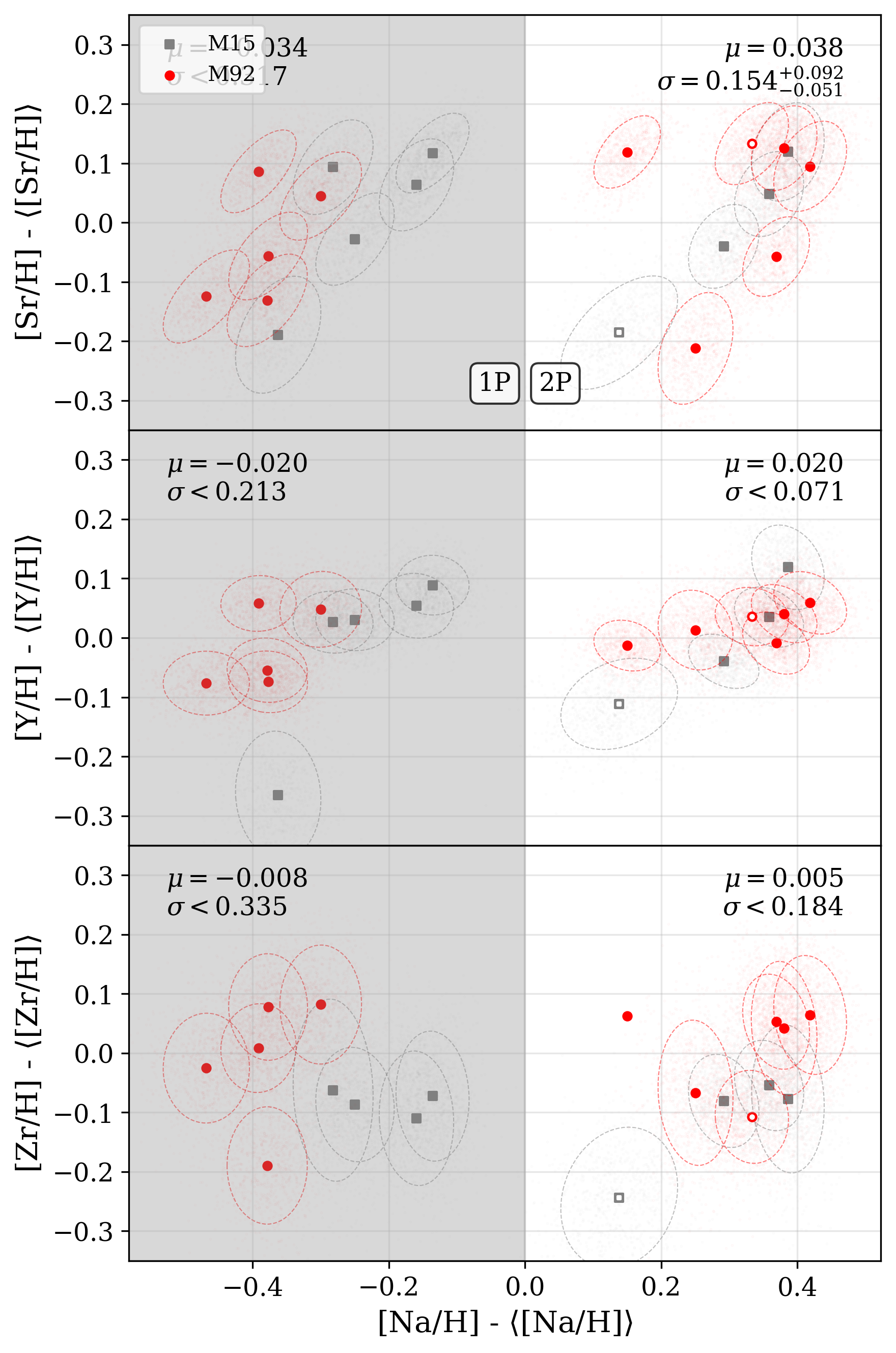}
\includegraphics[width=0.45\textwidth]{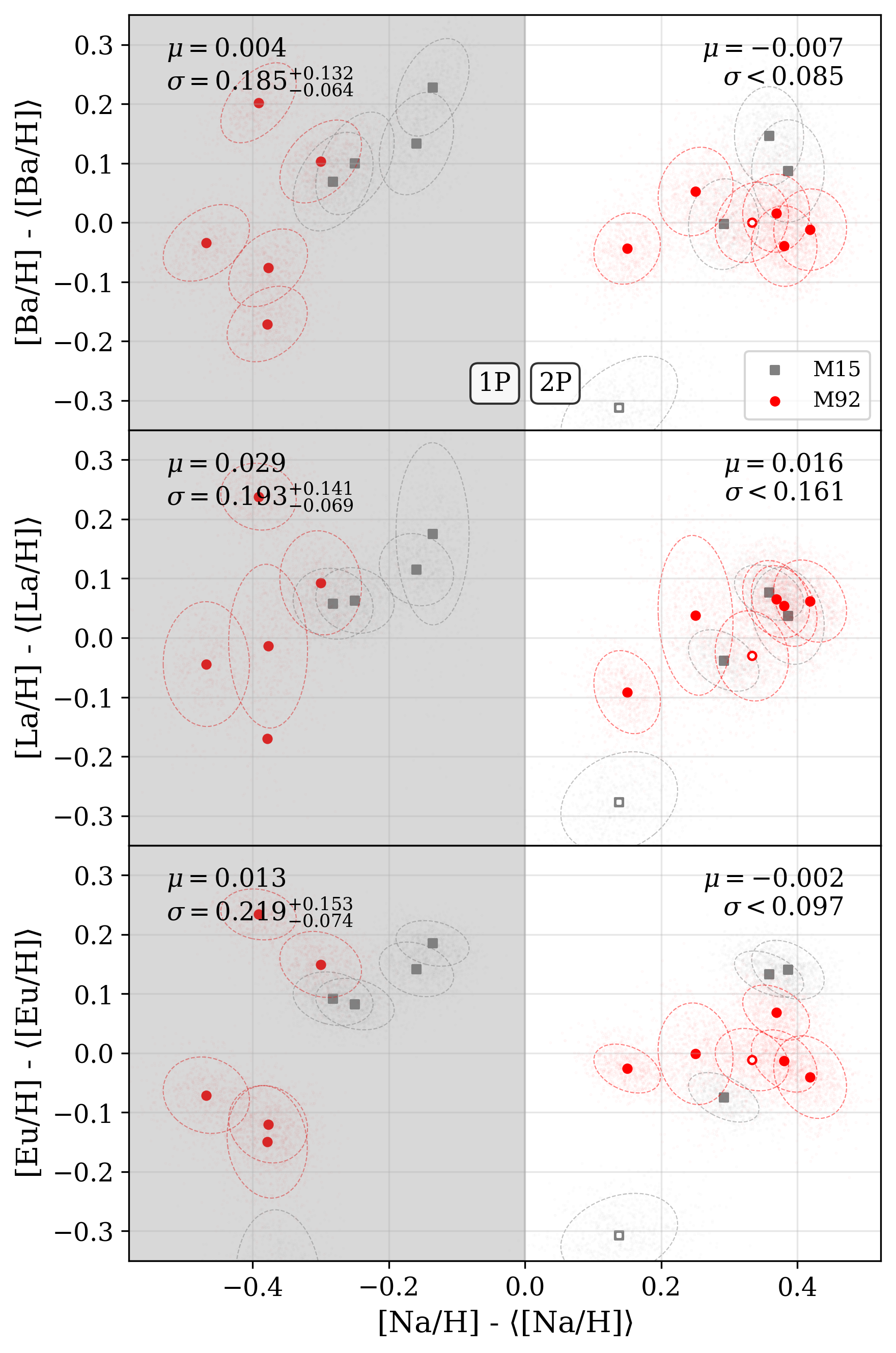}
\caption{\textit{Left:} Differential abundances of first-peak (``weak'') $r$-process elements vs.\ differential abundances of Na.  \textit{Right:} Differential abundances of Ba and two lanthanides (La and Eu), representative of the main $r$-process.  The statistical methods are the same as in Figure~\ref{fig:mgalfe}.}\label{fig:rprocess}
\end{figure}

\notetoeditor{The following two paragraphs were moved to this location from above.}

FRMSs are disfavored \add{sources of high-temperature hydrogen burning in M92} because they produce the light element anti-correlations on the same timescale as Fe.  Therefore, FRMS nucleosynthesis is inconsistent with a large \add{2P} dispersion in \delete{Na and Mg} \add{light elements} but a small dispersion in Fe.  \add{EMSs are disfavored because they make all the light-element abundance variations in less than 3~Myr \citep{gie25}.  In M92, the Mg abundances continued to evolve over the course of 2P, in which all the stars formed after supernovae exploded.}

AGB stars could plausibly explain \add{some (but not all) of} these patterns \citep{dan01,kar07,der10,kar14,doh14}.  The most massive AGB stars, around $8~M_{\odot}$, produce Mg-depleted, Na-enhanced material after about 30~Myr.  Then, after 70~Myr, the Mg returns to a near-primordial enhancement, and the Na returns to near its initial value.  In this scenario, the evolution of the cluster proceeds from the top left of the Mg--Na relation to the bottom right without forming any stars along the way.  Then, the evolution proceeds to higher Mg, then lower Na, forming 2P stars along the way.  Although the existing AGB yield tracks \delete{do} reproduce the qualitative abundance patterns that we observe, they do not reproduce the trends in quantitative detail \citep{der10}.  In any case, AGB stars are still inconsistent with other GC observations, such as Li abundance \citep{muc11} \add{without fine-tuning of Li production by the Cameron--Fowler mechanism \citep{cam71,nor24}}.

The heaviest elements in most GCs are created in the $r$-process.  The $r$-process has a main component, which generates the full range of neutron-capture elements, and a ``weak'' component, which produces only the first $r$-process peak (Sr, Y, and Zr) \citep{tra04}.  The main $r$-process must be generated in rare sources, such as neutron star mergers \citep{lat74} or magnetorotational supernovae \citep{nis15}.  The weak $r$-process, which is about ten times as abundant as the main $r$-process \citep{pra20}, could be produced more commonly in ordinary core collapse supernovae \citep{fro06}.  Figure~\ref{fig:rprocess} shows the weak and main $r$-process elements in M92 and M15.  We corroborate the discovery of a larger scatter for the main $r$-process in M92's 1P than in 2P \citep{kir23}.

We conclude that the proto-cluster experienced a main $r$-process event prior to the formation of 1P.  The gas had not mixed evenly before 1P finished forming.  The smaller dispersion in the weak $r$-process supports that those events (putatively ordinary core collapse supernovae) could be more common than main $r$-process events because the primordial gas cloud(s) that formed M92 \delete{were} \add{was} enriched with many sources that had time to mix well before the 1P stars formed.  Scenarios where an $r$-process event happened after 1P formed would predict different abundances for stars below the first dredge-up than above it, which has been shown not to be the case for at least one GC, M15 \citep{kir20}.

The short gas mixing timescale ($\sim 1$~Myr) of the GC indicates that 1P formed in a nearly instantaneous burst.  Otherwise, 1P would have homogenized.  The dispersion in the main $r$-process contrasts sharply with the very low dispersion in Fe in 1P.  Somehow, the proto-cluster mixed evenly in Fe without mixing in the $r$-process, or the proto-cluster formed from separate clouds of gas that had identical Fe content but disparate $r$-process content.

\subsection*{Conclusions and Interpretation}

Our conclusions about the formation of M92 are as follows:

\begin{itemize}
    \item 1P formed in a nearly instantaneous burst, lasting $\lesssim 1$~Myr.
    \item High-temperature hydrogen burning was not active during the formation of 1P\@.
    \item The gas that formed 1P was not well-mixed in \add{main} $r$-process elements, but it was well-mixed in other elements, including Fe.
    \item 2P began forming at least \delete{$\gtrsim 4$} \add{$\gtrsim 3$}~Myr later, having retained a small amount of core collapse supernova ejecta.  \delete{Uniquely among GCs,} The Fe-peak elements are enhanced only in 2P stars.
    \item 2P formed over an extended duration, sampling hydrogen burning at a range of temperatures.
\end{itemize}

The $r$-process dispersion in 1P, coupled with the homogeneity of Fe, indicates that the $r$-process event happened in the proto-cluster gas cloud.  It has been proposed \citep{kir23} that M92 could have assembled from multiple gas clouds, only one of which experienced an $r$-process event.  However, the Fe abundances of those clouds would be distinct, and the resulting 1P would have some Fe abundance dispersion.  Similarly, the $r$-process could not have come from an external ``interloper'' source that fortuitously happened near the GC\@.  Otherwise, there would also be more common interloper sources of nucleosynthesis, like core collapse supernovae, that would have produced Fe dispersion.  An $r$-process source in the proto-cloud suggests that there was a population of proto-cluster stars that pre-dated 1P\@.  This population also would have come with its own core collapse supernovae, but they would have had to be numerous enough to pollute the cluster gas evenly with Fe.  However, these stars are not seen today.  Either the cluster lost them, or they had a top-heavy initial mass function, such that no stars from this ``zeroth'' population are still burning today.

The discovery of supernova ejecta retention \add{in 2P but not 1P} was possible through choosing a narrow $T_{\rm eff}$ range of the stars and through differential, line-by-line abundance analysis.  Without this level of precision, M92 appears like most other ordinary GCs.  Therefore, the abundance patterns in M92 might be \delete{typical, not unusual} \add{found in other GCs observed in the same way}.  This approach has promise for new discoveries about chemical evolution of other GCs.

Supplementary Information is available for this paper.  Correspondence and requests for materials should be addressed to Evan Kirby (ekirby@nd.edu).

\backmatter

\section*{Methods}

\add{\bmhead{Photometry and atmospheric parameters} We determined atmospheric parameters from \textit{Gaia} photometry.  \textit{Gaia} magnitudes are susceptible to photometric crowding in dense fields.  The \textit{Gaia} archive provides the flux excess, $C$, in the $G_{BP}$ and $G_{RP}$ bands, which is often inflated by crowding sources.  A corrected flux excess, $C_*$, can be computed based on the star's color.  We computed $C_*$ for all our stars.  It exceeds 0.05 for two stars, M92-star-8 and M15-star-12, which indicates that the \textit{Gaia} magnitudes for these stars could be suspect.  These stars are indicated with footnotes in Table~\ref{tab:starcoords} and with hollow points in Figures~\ref{fig:mgalfe}, \ref{fig:rprocess}, and \ref{fig:lightelements_ext}--\ref{fig:rprocess_ext}.}

\add{To estimate the effect of crowding, we also computed color temperatures based on 2MASS infrared magnitudes.  We found all of the stars in our sample in the 2MASS catalog \citep{skr06}.  Crowding also affects 2MASS photometry, and it can be diagnosed by the ``C'' flag and by proximity to another bright source less than 5~arcsec.  Two stars had nearest-neighbor separations of 4.6 and 4.7 arcsec (M92-star-8 and M15-star-7).  The C flag is raised for the $K_s$ filter in both of these cases.  The $K_s$ magnitude of one star (M92-star-9) was given only an upper limit.}

\add{We diagnosed the effect of crowding by comparing $T_{\rm eff}$ derived from the $G_{BP}-G_{RP}$ and $G_{BP}-K_s$ colors \citep{muc21}.  The absolute deviation between the temperatures ranges from less than 1 K to 113 K.  The crowded stars (M92-star-8, M15-star-7, and M15-star-12) have deviations of $61$~K, $-18$~K, and $9$~K\@.  We conclude that crowding probably does not add significant error to the photometric temperatures in excess of the scatter in the color--temperature relations (83~K for $G_{BP}-G_{RP}$ and 49~K for $G_{BP}-K_s$).}

\bmhead{Data reduction} We reduced the HIRES spectra with the software package \texttt{makee} \citep{bar24}.  This set of programs performs flat-fielding, wavelength calibration, sky subtraction, and echelle order extraction.  The final product is a one-dimensional spectrum for each order.  We stitched all of the echelle orders into a single, continuous spectrum, as described below.

The first step in that process is continuum normalization.  We constructed an approximate model of the spectrum in order to divide out the absorption lines.  We used the same model spectrum for each star because all the stars in the sample have similar stellar parameters.  \texttt{MOOG} was used in conjunction with a model atmosphere with $T_{\rm eff} = 5009$~K, $\log g = 2.11$, ${\rm [Fe/H]} = -2.40$, and $\xi = 1.65$~km~s$^{-1}$.  It is not necessary for these parameters to match the spectrum exactly.  Each echelle order was shifted to the rest frame and divided by the model spectrum.  Then, the quotient was fit with a spline with a breakpoint spacing of 500 pixels coupled with $\pm 1.5\sigma$ clipping.  The continuum-normalized spectrum is the original echelle order divided by the spline fit.  Each echelle order was interpolated onto a common, logarithmically-spaced wavelength array.  Some of the blue orders overlap in wavelength.  The overlapping pixels were averaged with inverse variance weighting.

We used only one absorption line for potassium: K$\,${\sc i}~$\lambda 7699$.  It is contaminated by telluric ${\rm O}_2$ absorption.  We removed the telluric absorption following a procedure based on \delete{measuring} \add{fitting} an adjacent ${\rm O}_2$ line, scaling \delete{it} \add{the fit} by the ratio of oscillator strengths, and subtracting the ${\rm O}_2$ line that overlaps K$\,${\sc i} \citep{kir23}.  We did not estimate the uncertainty introduced by this subtraction.  As a result, our K measurements could have error that is not represented by our error estimates.

\begin{figure}[t]
\centering
\includegraphics[width=0.8\textwidth]{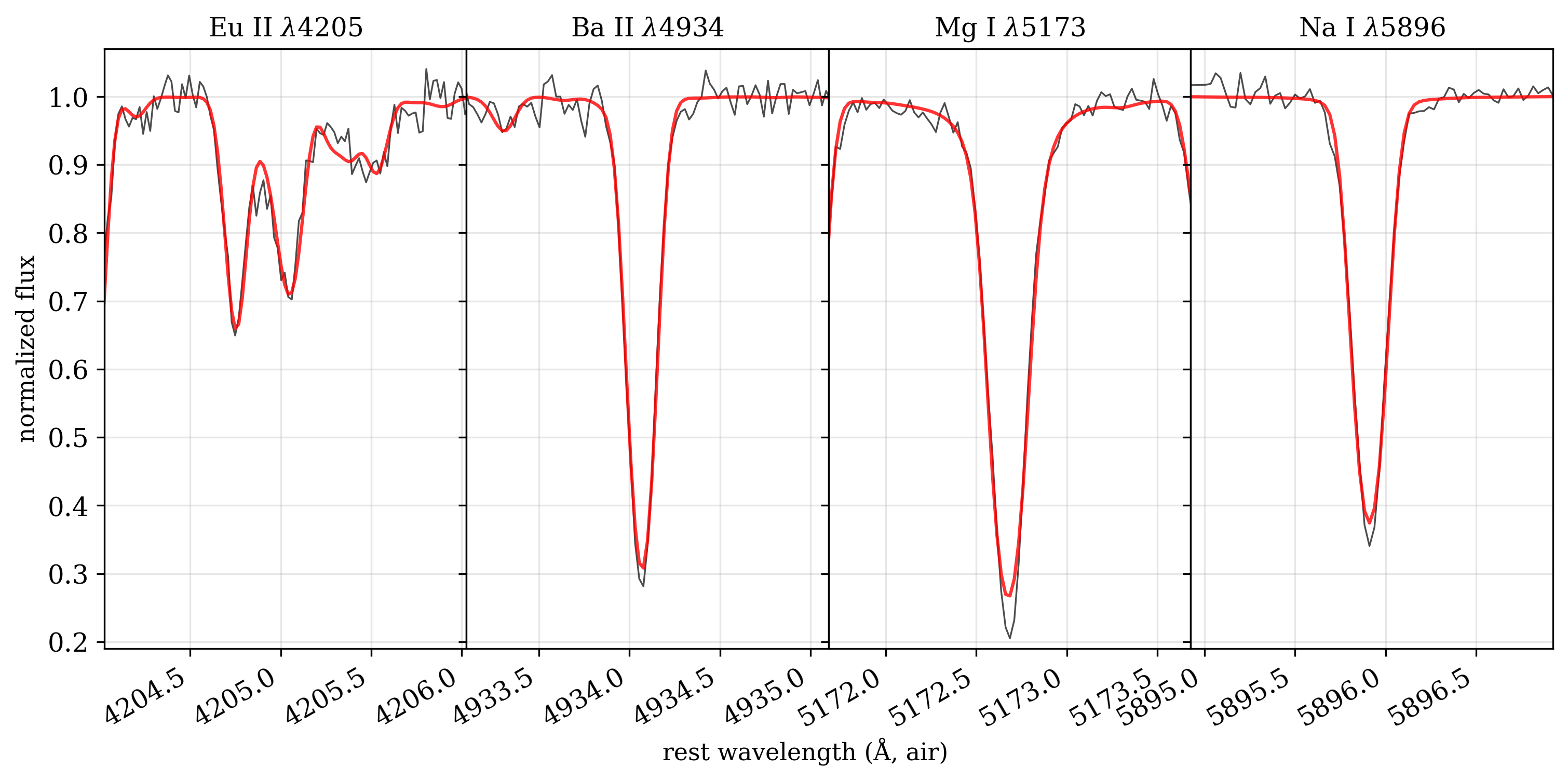}
\caption{\add{Example syntheses for four lines in M92-star-6.  The Eu and Ba lines incorporate hyperfine structure from isotopic splitting.  The line cores in the Mg~b and Na~D lines are too weak, which is typical for 1D LTE syntheses.}\label{fig:synth_ext}}
\end{figure}

\bmhead{Abundance measurements} We measured \delete{equivalent widths (EWs)} \add{abundances of lines} from the line list of Ji et al.\ (2020) \citep{ji20}.  \add{We omitted some weak or badly blended lines, including Na$\,${\sc i}~$\lambda 5683$, Na$\,${\sc i}~$\lambda 5688$, Y$\,${\sc i}~$\lambda 5200$, Y$\,${\sc ii}~$\lambda 5206$, Eu$\,${\sc ii}~$\lambda 4436$, Eu$\,${\sc ii}~$\lambda 4523$, Eu$\,${\sc ii}~$\lambda 6645$, and Dy$\,${\sc ii}~$\lambda 4450$.  We measured abundances from a combination of equivalent widths (EWs) and spectral synthesis.}  We \delete{used} \add{measured EWs with} a custom GUI written in IDL \citep{kir23}.  Gaussians were fit to \delete{most of} the absorption lines using a local determination of the continuum.  \add{We computed abundances with the LTE radiative transfer code \texttt{MOOG} \citep{sne73}.  \texttt{MOOG}'s \texttt{abfind} driver was used to compute abundances from most EWs.  The \texttt{blends} driver was used for Fe-group odd-numbered elements with hyperfine splitting (Sc, V, Mn, and Co).}  \delete{The two Mg~b triplet lines in the line list are strong enough that Gaussians are not good fits.  For those two lines, we fit Voigt profiles.  In principle, Voigt profiles could be fit to even the weak lines.  However, when we did so, the damping term is almost always zero, in which case the Voigt profile reduces to a Gaussian.  Omitting the damping term for most lines allows more reliable fits because the fitter, \texttt{MPFIT} \citep{mar12}, has a smaller parameter space to explore.  We computed abundances with a combination of EWs and spectral synthesis.}  For \add{the strong Na D lines, the Mg~b lines at 5173~\AA\ and 5184~\AA, the Al~$\,${\sc i}$\lambda\lambda 3944,3962$ doublet, and} neutron-capture elements with significant hyperfine splitting or blends (Sr, Y, Zr, Ba, La, Nd, \delete{and} Eu, \add{and Dy}), we used spectral synthesis.  \add{(Mg~b~$\lambda 5167$ is excluded from the line list due to blending.)  We use the solar $r$-process pattern for the hyperfine splitting \citep{sne08}.  We measured carbon and nitrogen abundances from synthesis of the CH 4300~\AA\ and NH 3875~\AA\ molecular bands.}  \delete{For all other elements, we used EWs.}  \add{\texttt{MOOG}'s \texttt{synth} driver was used to compute synthetic spectra in 10~\AA\ windows (60~\AA\ for CH and 20~\AA\ for NH) around the lines.  The line lists for the syntheses were sourced from \texttt{linemake} \citep{pla21a,pla21b}.  \texttt{MPFIT} found the best fit abundance by minimizing the $\chi^2$ of the fit.  The local continuum was re-determined in each iteration of \texttt{MPFIT} by fitting a low-order spline to the quotient of the observed and synthetic spectra.  Extended Data Figure~\ref{fig:synth_ext} shows an example of spectral syntheses for one star in M92.  The line cores for the strong Mg~b triplet and Na~D doublet are underestimated, which is a common problem for 1D LTE models \citep{lin11,oso15,lin22}.}

\delete{We computed abundances with the LTE radiative transfer code \texttt{MOOG} \citep{sne73} coupled with} \add{We computed abundances with \texttt{MOOG} and} \texttt{ATLAS9} model atmospheres \citep{kur93,kir11d}.  The abundances were computed in two iterations.  In the first iteration, we used a model atmosphere with the photometric values of temperature and gravity and with the metallicity fixed to ${\rm [Fe/H]} = -2.39$ and ${\rm [\alpha/Fe]} = +0.41$.  The abundances other than the $\alpha$ elements (O, Ne, Si, S, Ar, Ca) obeyed the solar abundance pattern \citep{asp09} scaled by [Fe/H]\@.  The microturbulent velocity ($\xi$) was iterated with \texttt{MPFIT} \citep{mar12} until the slope of the abundances of Fe$\,${\sc i} lines with reduced width (EW divided by wavelength) was minimized.  In the second iteration, we computed an \texttt{ATLAS9} model atmosphere with an opacity distribution function (ODF) determined by the abundances computed in the first iteration.  \add{We could not measure O abundances, so we assumed that the O abundance follows the same trend as NGC~2808 \citep{car15}:}

\add{
\begin{equation}
    A({\rm O}) = \begin{cases}
    -2.375 A({\rm Na}) + 0.625 & A({\rm Na}) > 0.2 \\
    -0.666 A({\rm Na}) + 0.283 & A({\rm Na}) \le 0.2
    \end{cases}
\end{equation}
}

\noindent
We used \texttt{BasicATLAS} \citep{BasicATLAS} to generate the ODFs and model atmospheres.  The rest of the iteration proceeded in the same way.  Computing a model atmosphere with the measured abundance pattern reduces uncertainty due to the change in continuum opacity caused by the light element abundance variations, especially the electron donors Na and Mg.  In other words, it allows us to be more confident that our procedure does not introduce spurious correlations between the abundances of light elements and other elements.

\add{The error, $\sigma_{\rm tot}$, on each line measurement is based on uncertainty in EW or spectral synthesis and uncertainty in atmospheric parameters:}

\add{\begin{equation}
    \sigma_{\rm tot}^2 = \sigma_{\rm noise}^2 + \sigma_{T_{\rm eff}}^2 + \sigma_{\log g} + \sigma_{\xi}^2 \label{eq:sigmatot}
\end{equation}}

\noindent
\add{For lines measured with EW, $\sigma_{\rm noise}$ is computed by taking the average deviation in abundance when the EW is increased and decreased by the error in EW, as computed in the Gaussian fit.  For lines measured with synthesis, $\sigma_{\rm noise}$ is the $1\sigma$ error returned by \texttt{MPFIT}\@.  The remaining error terms are errors in abundance estimated by increasing $T_{\rm eff}$, $\log g$, or $\xi$ by $1\sigma$.  Equation~\ref{eq:sigmatot} does not account for covariance in line measurement.  Covariance is considered for final abundance measurements (see ``Uncertainty estimation'' below).}

We report abundances relative to a reference abundance, which we chose to be the average of the abundances in the cluster.  For each absorption line, we computed the unweighted average abundance within a GC\@.  We subtracted that average abundance from the absorption line's abundance in each star.  Then, for each star, we computed the \delete{un}weighted average abundance of all the lines of that element, regardless of ionization state.  \add{The weight is the inverse variance, $\sigma_{\rm tot}^{-2}$, of the line measurement.}  These abundances are essentially an arbitrary zeropoint.  As a result, the exact value of the reference abundance does not meaningfully influence the measurements.

\add{Most previous differential abundance analyses \citep{mel09,yon13_differential,ram14,mck22,mon23a} compute differential abundances relative to a single reference star.  An absorption line's abundance is subtracted from the abundance of the same line in a reference star.  This procedure works best when the reference star has line measurements of higher quality than all the other stars.  In our case, we chose the stars in our sample to be from a very tight region of the color--magnitude diagram so that all stars would have very similar atmospheric parameters.  The brightness our selection is limited by the need to go far enough down the luminosity function of the red giant branch to include enough stars in a tight selection box.  As a result, no one star is of especially high spectral quality.  In essence, we maximized the accuracy of the differential technique at the expense of some precision.  Using the average abundance of a line as the reference rather the abundance from a single star lessens the effect of spectral noise.}

\bmhead{Uncertainty estimation} To estimate \add{abundance} uncertainties, we recomputed the abundances using 1,000 Monte Carlo (MC) trials.  In each trial, we sampled the atmospheric parameters and the EWs from their error distributions.  It is computationally infeasible to recompute a stellar atmosphere model for the exact temperature and composition of the trial, so we used a precomputed grid \citep{kir11d} of \texttt{ATLAS9} model atmospheres.

To determine $T_{\rm eff}$ for a given MC trial, we isolated the random component of the \textit{Gaia} color--$T_{\rm eff}$ relation \citep{muc21} from the systematic component.  The random component is the propagation of the $G_{BP}-G_{RP}$ color through the fitted polynomial relation.  The systematic component determined by Mucciarelli et al.\ (2021) is 83~K\@.  We assumed that the systematic component is highly correlated from star to star.  Therefore, we sampled the random component of the $T_{\rm eff}$ uncertainty using a different random number for each star, but we sampled the systematic component of the uncertainty using the same chain of 1,000 random numbers (i.e., same random number generator ``seed'') for each star.  \add{Below, we show that the assumption of correlation of the systematic error from star to star does not lead to an underestimate of the uncertainty.}

We determined the surface gravity $\log g$ by assuming a mass of $M = 0.9~M_{\odot}$ and a radius from the luminosity and $T_{\rm eff}$ via the Stefan--Boltzmann law.  We propagated the uncertainty from the \textit{Gaia} $G$ magnitude to calculate the luminosity uncertainty.  The luminosity also requires a bolometric correction \citep{and18}, which depends on $T_{\rm eff}$.  The $T_{\rm eff}$ used is consistent within the same MC trial, which properly preserves the correlation between $T_{\rm eff}$ and $\log g$.  We further propagated an assumed error of $0.1~M_{\odot}$ on the stellar mass.

We fit for $\xi$ for the abundance measurements, but that would have been computationally prohibitive for MC trials.  Therefore, we used an empirical relation between $\xi$ and $\log g$ \citep{kir09}.  The MC trials account for correlation between these two atmospheric parameters \add{as well as a random 0.05~km~s$^{-1}$ scatter in the original fit between $\xi$ and $\log g$}.

The atmospheric composition in the \texttt{ATLAS9} atmosphere grid is characterized by [Fe/H] and [$\alpha$/Fe].  We took these values from the abundances measured as described above, where $\alpha$ is the average of Mg, Si, and Ca abundances.  We sampled from uncertainties on [Fe/H] and [$\alpha$/Fe], which are estimated as the standard error on the mean of the abundances from the lines of those elements.

The Gaussian \delete{and Voigt} profiles fit to the absorption lines also have uncertainties estimated from \texttt{MPFIT}\@.  Each MC trial samples the EW of each line from a normal distribution centered on the EW measurement with a standard deviation given by the \texttt{MPFIT} uncertainty.  Although the atmospheric parameters are correlated with each other within each MC trial, the EW errors are not.

\add{We test whether the resulting abundance uncertainties accurately represent the data using the ANOVA $F$-test.  The $F$-statistic is the ratio of variance among the population to the average variance of the Monte Carlo samples for each star.  A value of $F \sim 1$ indicates that the measurements have no intrinsic variance beyond the uncertainties.  A value of $F \gg 1$ indicates that there is an intrinsic variance or that the uncertainties are underestimated.  A value of $F \ll 1$ indicates that the uncertainties are overestimated.  Extended Data Table~\ref{tab:fstats} gives the $F$-statistics in M92 for different elements where the stars are grouped by population (1P and 2P)\@.  Each $F$-statistic is accompanied by a $p$-value, which quantifies the significance of detection of variance in the population.  Small $p$-values indicate highly significant detections of dispersion.}

\add{The $F$-test simultaneously tests for intrinsic population variance and the reasonableness of error bars.  $F$-statistics should not be much less than one unless the uncertainties are overestimated.  The $F$-statistic for nearly all Fe-group elements (Sc, Ti, Mn, Fe, Co, Ni, and Zn) in 1P is less than one.  Therefore, the uncertainties are at least as large as the population variance.  The only elements in 1P with detected variances at $>95\%$ significance ($p < 0.05$) are Si, K, V, and neutron-capture elements.  These are also the elements reported to have a dispersion ($\sigma$) in Extended Data Figures~\ref{fig:lightelements_ext} and \ref{fig:fepeak_ext}.  The K abundances could be affected by telluric contamination, as discussed in the Supplementary Discussion.  The Si and V abundances either have a systematic error that we have not accounted for, or they have a genuine 1P dispersion.}

\add{The $F$-statistic for most Fe-group elements is higher in 2P than in 1P\@.  This could indicate a genuine dispersion in those elements in 2P, but the significance exceeds 95\% only for Sc and Cr, whose dispersion is driven by one star.  The neutron-capture elements other than Sr also have no detected dispersion in 2P\@.  On the other hand, the light elements (Na, Mg, Al, Si, and K) have highly significant dispersions in 1P\@.}

\add{\bmhead{Tests of significance of abundance differences between populations} We tested the significance of our finding that the abundances of some elements, like Fe, are different in 2P than in 1P in M92.  First, we computed the mean value of the abundance of an element in 1P for all 1,000 MC trials.  We separately computed the mean of 2P in the same way.  Then, we calculated the fraction of those trials had a mean abundance in 1P greater than in 2P\@.  This number is reported as $f({\rm 1P}>\rm{2P})$ in Extended Data Table \ref{tab:fstats}.}

\add{Second, we performed an exact permutation test.  We assigned five stars (the sample size of 1P) to ``Set A'' and the six remaining stars (the sample size of 2P) to ``Set B.''  Then, we computed the mean abundance of each set.  We repeated this for all 462 combinations of A and B, including the combination where A and B correspond to 1P and 2P\@.  The significance of a detection of a difference is the fraction of combinations for which the difference in means between A and B meets or exceeds the difference in means between 1P and 2P\@.  This value is reported as $p_{\rm perm}$ in Extended Data Table~\ref{tab:fstats}.  It cannot be less than $1/462 = 0.002$ because the actual 1P and 2P separation are part of the 462 combinations.}

\add{All elements less massive than Zn (including Zn) except Mg and Si are found to have a lower mean abundance in 1P than in 2P at $p < 0.05$.  Mg is the only element to have a higher mean abundance in 1P than in 2P\@.  Although the mean 1P abundance of Si was lower than 2P for all the MC trials, the significance from the exact permutation test was weak.  The mean abundances of neutron-capture elements are not significantly higher or lower in either population.}

\bmhead{Quantification of abundance dispersion} We quantified the intrinsic dispersion---accounting for measurement uncertainty---in abundances separately for the 1P and 2P populations.  We modeled the distributions as Gaussian.  The log-likelihood of the distribution is

\begin{equation}
    \ln \mathscr{L} = -\frac{1}{2} \sum_i \left(\ln (\delta_i^2 + \sigma^2) + \frac{(A_i - \mu)^2}{\delta_i^2 + \sigma^2}\right)
\end{equation}

\noindent where $A_i$ is the differential abundance measurement for star $i$, $\delta_i$ is its uncertainty estimated from the standard deviation of the MC trials, $\mu$ is the mean, and $\sigma$ is the intrinsic dispersion.  We used the Monte Carlo Markov chain code \texttt{emcee} \citep{for13} to estimate the posteriors on $\mu$ and $\sigma$.  We used uniform, uninformative priors and 32 walkers with $10^5$ steps each, including $10^4$ burn-in steps.

We determined whether we detected an intrinsic dispersion by evaluating $\ln \mathscr{L}$ at $\sigma = 0$ and $\mu$ set to the median value determined by the walkers.  We considered $\sigma$ significantly detected if $\mathscr{L}(\sigma = 0)/\mathscr{L}_{\rm best} < 0.05$ where $\mathscr{L}_{\rm best}$ is evaluated at the median values of $\mu$ and $\sigma$.  In other words, we report a detection of $\sigma$ only if it is different from zero at 95\% confidence or better.  In the case of a detection of $\sigma$ we report upper and lower error bars at the 16$^{\rm th}$ and 84$^{\rm th}$ percentiles (68\% C.I.)\@.  Otherwise, we report an upper limit at the 95$^{\rm th}$ percentile (95\% C.L.)\@.

\section*{Supplementary Discussion}

\begin{figure}[t]
\centering
\includegraphics[width=0.45\textwidth]{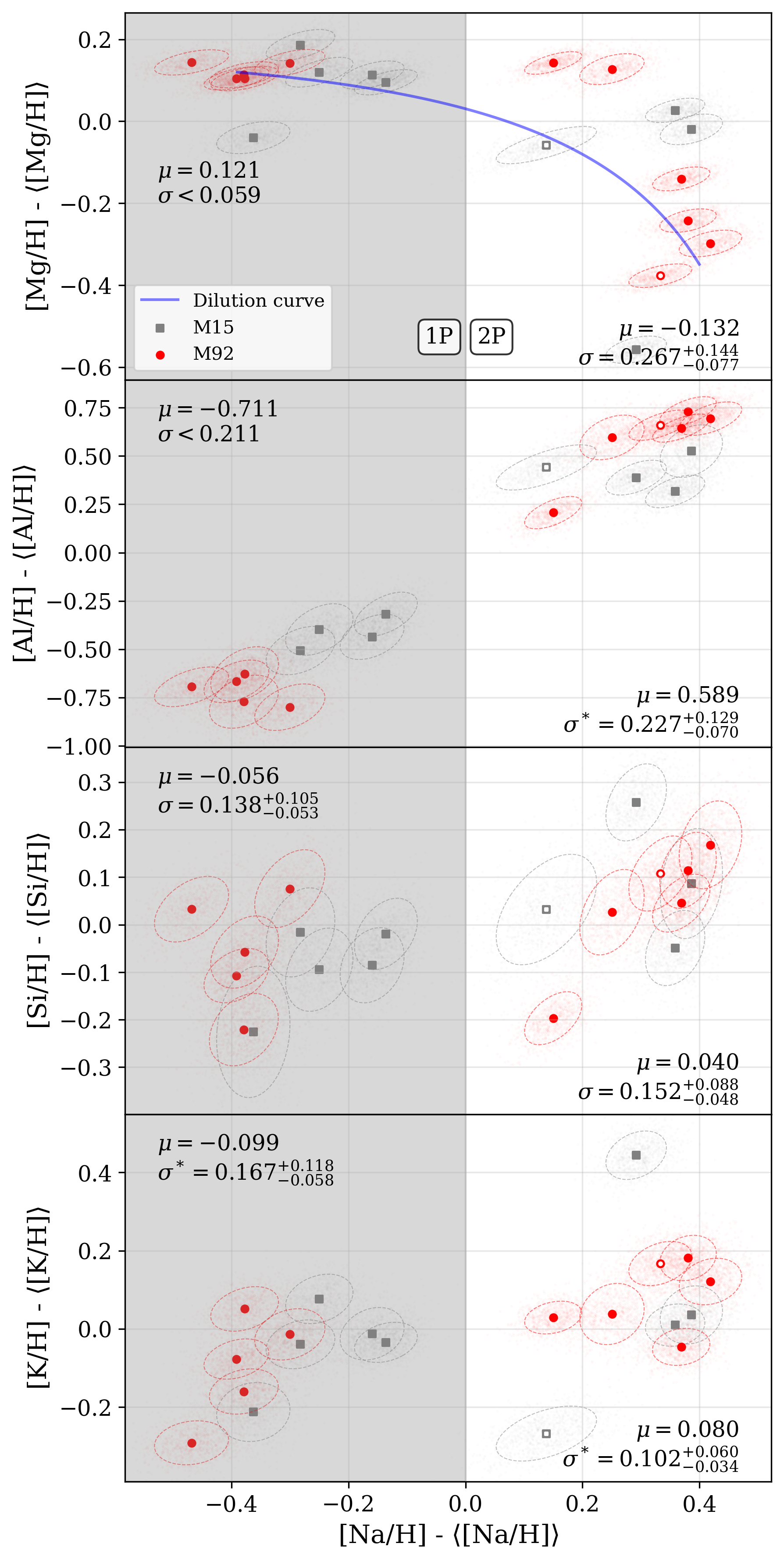}
\includegraphics[width=0.45\textwidth]{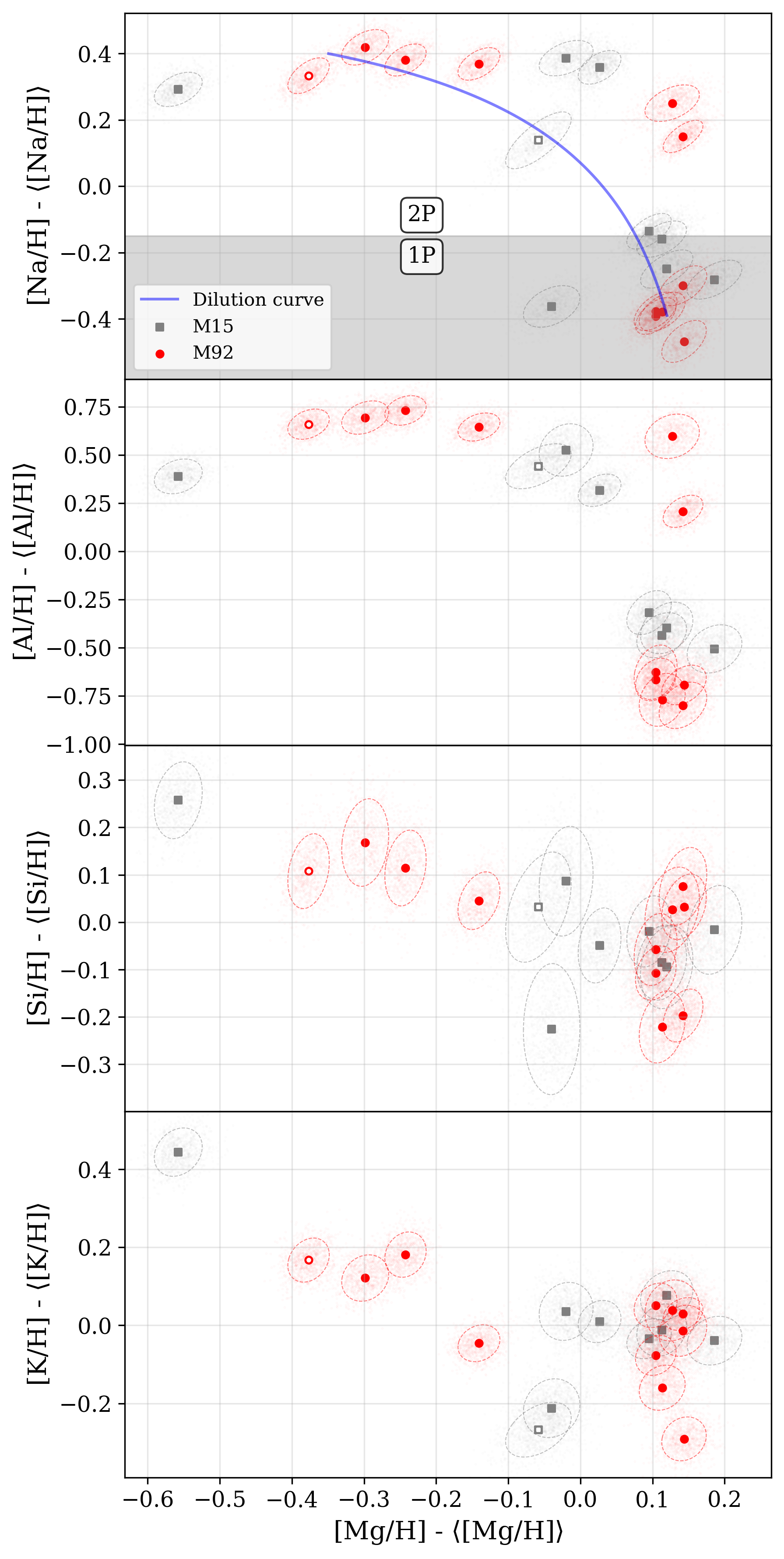}
\caption{The light element abundance patterns in M92 and M15.  The five elements shown here participate in advanced hydrogen burning.  The left column shows differential abundances vs.\ Na, and the right column shows differential abundances vs.\ Mg.  The Na column gives the mean ($\mu$) and dispersion ($\sigma$) for each population, as described in ``Quantification of abundance dispersion'' in Methods.   Upper limits on $\sigma$ are given as 95\% C.L.\ and detections given as 68\% C.I.\@  The asterisks on $\sigma$ for the K measurements indicates that K could have systematic errors due to telluric decontamination that are not accounted for.\label{fig:lightelements_ext}}
\end{figure}

\begin{figure}[t]
\centering
\includegraphics[width=0.45\textwidth]{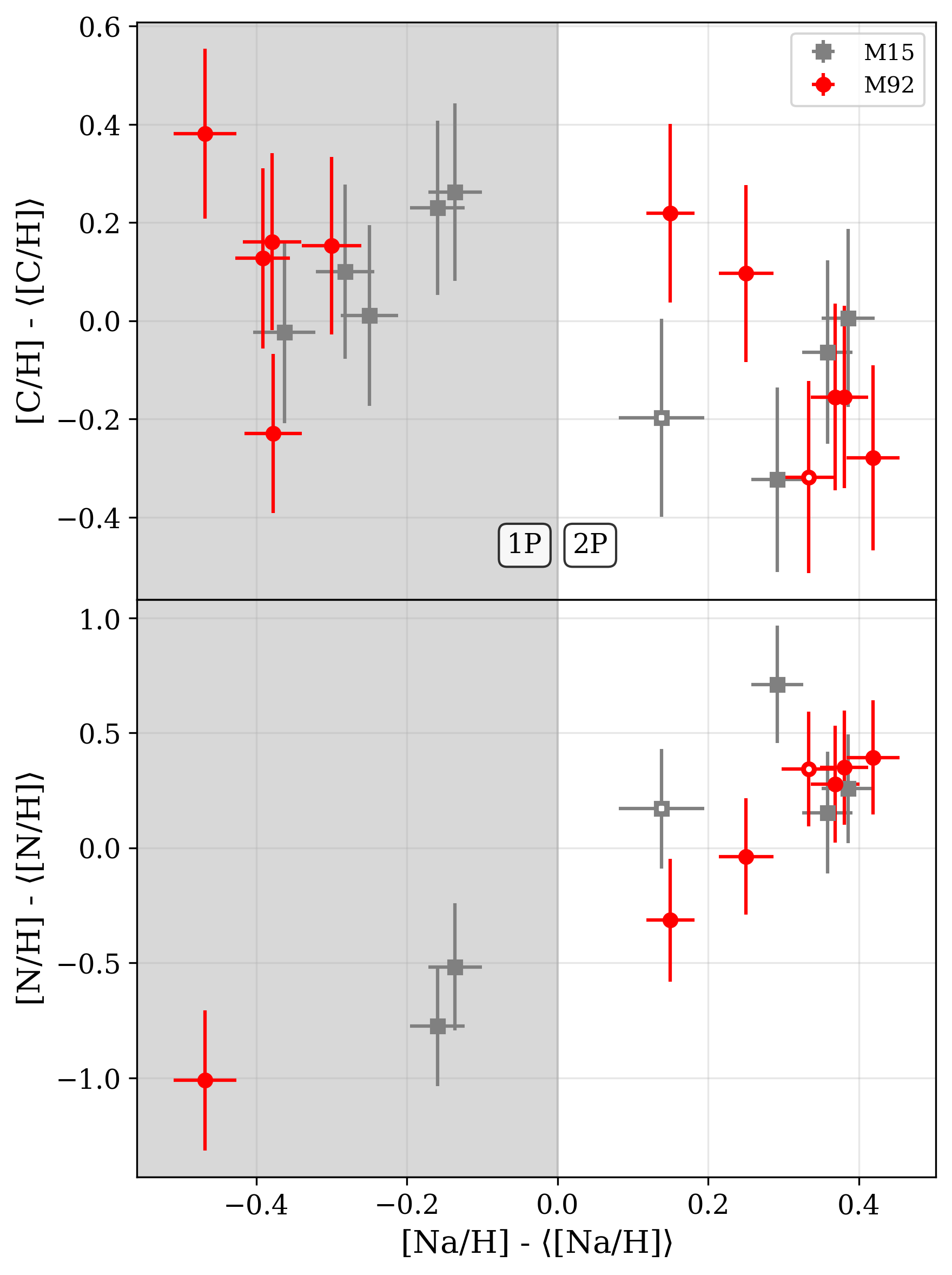}
\caption{\add{Differential abundances of C and N vs.\ differential abundances of Na.  N could not be measured in all 1P stars.  Error bars are calculated from Equation~\ref{eq:sigmatot}, which is a different method than the Monte Carlo error ellipses shown in other figures.}}\label{fig:cn}
\end{figure}

\begin{figure}[t]
\centering
\includegraphics[width=0.45\textwidth]{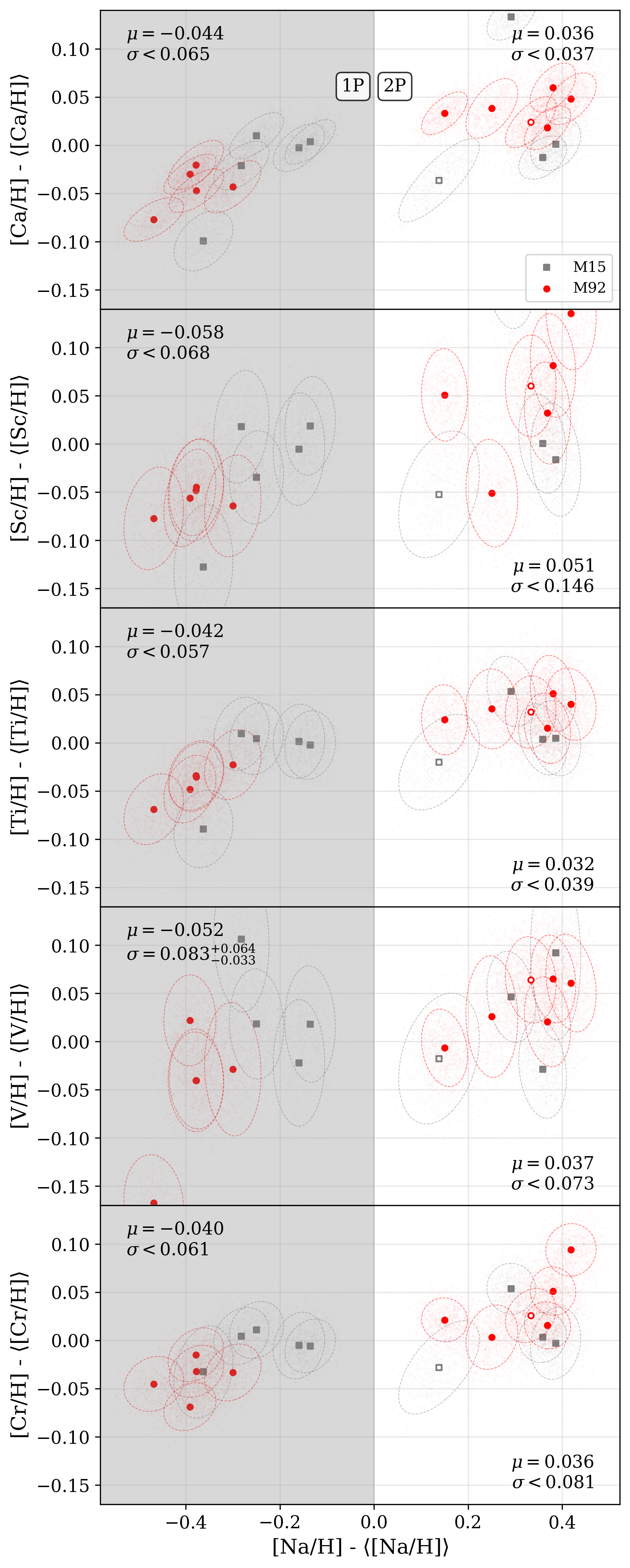}
\includegraphics[width=0.45\textwidth]{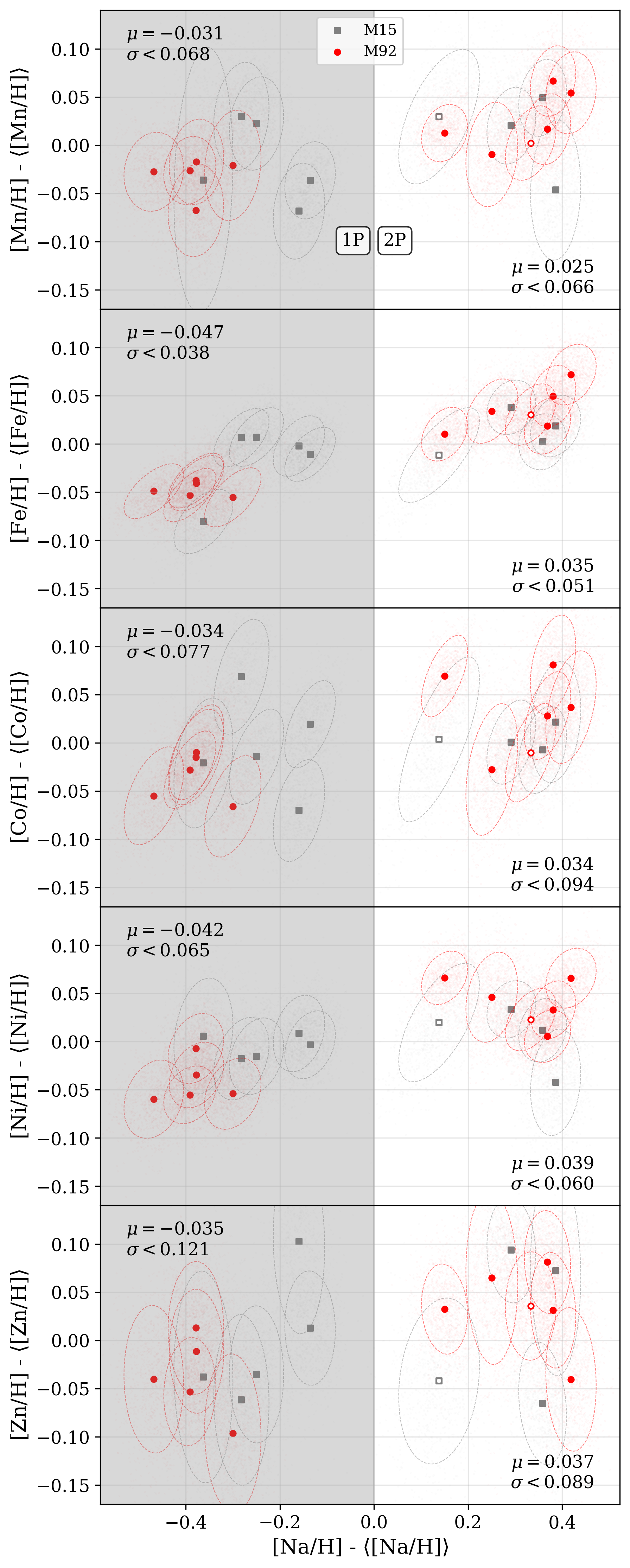}
\caption{Ca and additional \delete{Fe-peak} \add{Fe-group} elements to supplement Fe shown in Figure~\ref{fig:mgalfe}.  \add{The $y$-axis range is the same in each panel.}  \delete{The blue line is a guide, not a fit.  It is the same in each panel to demonstrate the result if all these elements originated in the same source(s).}\label{fig:fepeak_ext}}
\end{figure}

\begin{figure}[t]
\centering
\includegraphics[width=0.6\textwidth]{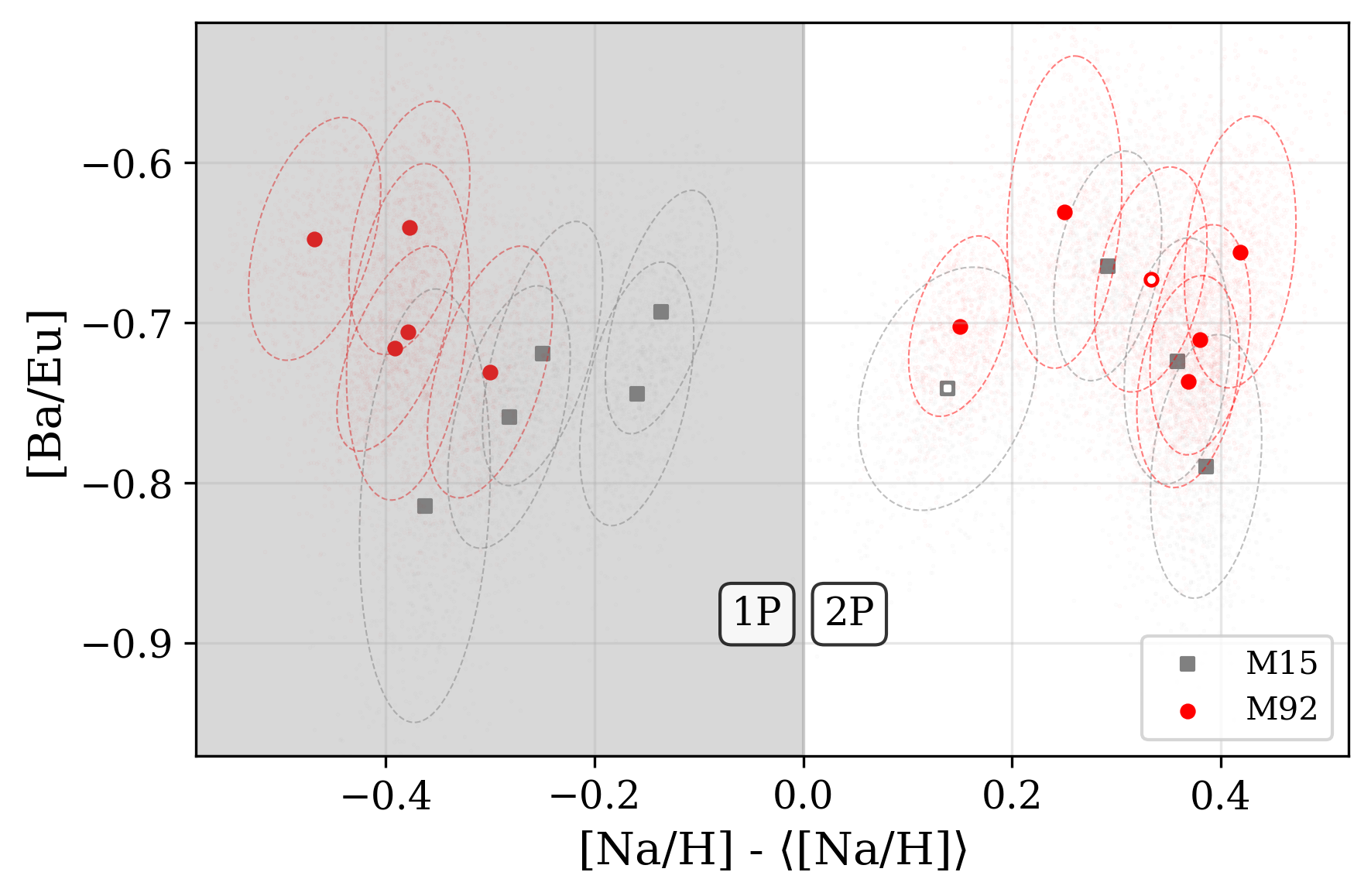}
\caption{The [Ba/Eu] ratio, a diagnostic of the $s$- vs.\ $r$-process.  A value of ${\rm [Ba/Eu]} \approx -0.7$ indicates that all of the stars were enriched exclusively in the $r$-process.\label{fig:rprocess_ext}}
\end{figure}

\begin{figure}[t]
\centering
\includegraphics[width=0.6\textwidth]{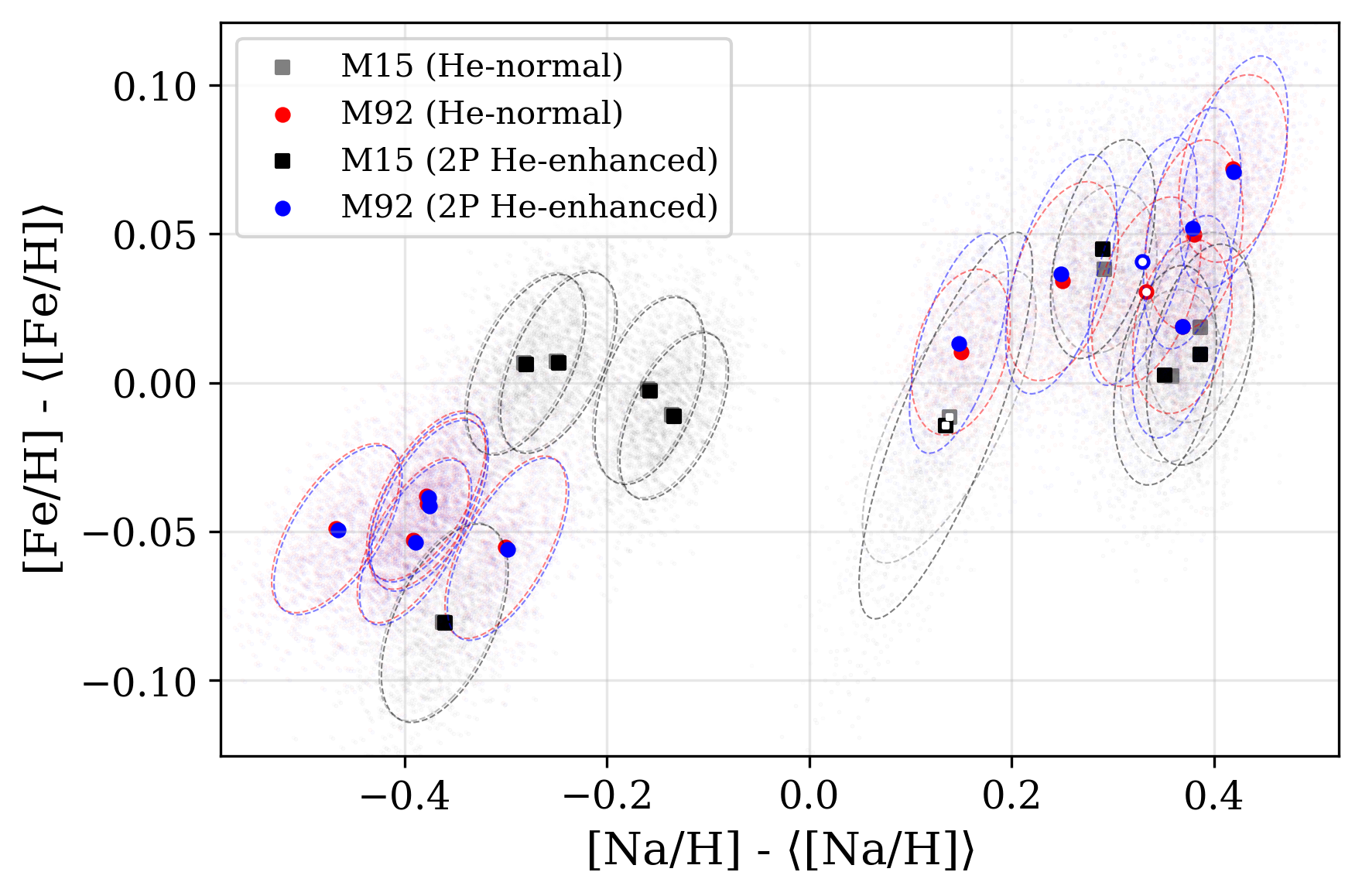}
\add{\caption{Differential abundances of Fe vs.\ Na assuming $Y=0.25$ (``He-normal'') and $Y=0.28$ (``He-enhanced'') for 2P stars.}\label{fig:helium}}
\end{figure}

\bmhead{Note on relative abundances} We report abundances relative to their averages within the cluster.  Extended Data Table~\ref{tab:avgabund} gives these average abundances.  The absolute abundances may be computed from the relative abundances (Tables~\ref{tab:abund1}--\ref{tab:abund4} and Figures~\ref{fig:mgalfe}--\ref{fig:rprocess} and Extended Data Figures~\ref{fig:lightelements_ext}--\ref{fig:helium}) by adding the relative abundances to the average abundances in Extended Data Table~\ref{tab:avgabund}.  However, the absolute abundances are subject to all the systematic uncertainties (oscillator strengths, the assumption of LTE, and more) that we mitigated through the differential analysis.

\bmhead{Further discussion of abundance patterns} Figure~\ref{fig:mgalfe} shows only \delete{two} \add{three} of the five light elements that we measured (Na, Mg, Al, Si, and K).  Extended Data Figure~\ref{fig:lightelements_ext} shows these abundance patterns in two ways: vs.\ Na abundance and vs.\ Mg abundance.  All of these elements participate in advanced hydrogen burning.  At higher temperatures, Si can be created by proton capture on Al \citep{pra17}.  It might even be possible at temperatures exceeding $10^8$~K to convert Ar into K \citep{ventura12,pra17}.  In agreement with many previous studies on GC light element abundances, we find that Mg and Al are anti-correlated.  Si and Mg are also anti-correlated, as expected.  Even K seems to be anti-correlated with Mg.  Ventura et al.\ (2012) \citep{ventura12} speculated that Ca could be created in extremely high-temperature hydrogen burning, but our results show that it is created in the same ratio as Fe.  Therefore, the source of Ca in M92's 2P is supernovae, not hydrogen burning.

The abundances of Al and K should be treated with some suspicion.  We measured Al exclusively from the Al$\,${\sc i}~$\lambda\lambda 3944,3962$ resonance doublet, which is extremely strong.  Furthermore, both lines lie in the wings of the even stronger Ca~HK doublet.  The strength of the lines and their unfortunate proximity to some of the strongest lines in the optical spectrum make it difficult to place the continuum in order to measure EWs.  Our error analysis does not account for possibly large errors in continuum placement.  Likewise, we measured K exclusively from K$\,${\sc i}~$\lambda 7699$.  As described in Methods, it is contaminated by telluric ${\rm O}_2$.  Our error analysis does not account for the uncertainty introduced by telluric removal.

We formally detected a dispersion in K abundances, even in 1P\@.  If this dispersion is real, it could mean that extremely high-temperature hydrogen burning was in place in the earliest stages of cluster formation, yet it somehow did not affect the abundances of Na, Mg, and possibly not even Al, though our upper limit on Al dispersion in 1P is not very restrictive.  Given the possibly large systematic uncertainty in K abundances, we do not consider the K dispersion significant enough to alter our conclusions.

The infrared spectrum is better suited to measuring Al abundances.  In fact, APOGEE spectra show a distinct separation between Al-normal and Al-enhanced stars in M92 \citep{mas19}.  This result supports our discovery of an abundance gap between 1P and 2P based on Na abundances.

\add{Extended Data Figure~\ref{fig:cn} shows differential abundances of C and N\@.  Both of them are measured with spectral synthesis of molecular absorption.  We could not apply the Monte Carlo technique to estimate measurement uncertainty.  The error bars are those calculated by Equation~\ref{eq:sigmatot}, which does not account for covariance between the atmospheric parameter error terms.}

Extended Data Figure~\ref{fig:fepeak_ext} shows Ca and nine \delete{Fe-peak} \add{Fe-group} elements to supplement Fe, shown in Figure~\ref{fig:mgalfe}.  All of their abundances are enhanced in 2P compared to 1P, albeit with varying levels of significance.  None of them \add{except V, whose dispersion is driven by a single star,} show a dispersion in 1P detected at better than 95\% confidence.  \delete{The blue lines have the same slopes in all panels.  The lines are not fits but just a guide.  The abundances of all Fe-peak elements follow this slope, indicating that they were all enhanced by the same factor between 1P and 2P.  Therefore, they originate from the same source (supernovae).  Furthermore, the fact that all Fe-peak elements follow this trend and not just Fe validates the quality of the measurements.}

Neutron-capture elements can be generated in the $s$-process, $r$-process, or other sub-dominant processes.  An $s$-process vs.\ $r$-process origin can be diagnosed by the ratios of neutron-capture elements.  The most common diagnostic is [Ba/Eu] because Eu is made nearly exclusively in the $r$-process.  Extended Data Figure~\ref{fig:rprocess_ext} shows the absolute abundances of [Ba/Eu], based on adding the relative abundances to Table~\ref{tab:avgabund} and subtracting the solar abundance pattern \citep{asp09}.  The pure $r$-process abundance ratio is ${\rm [Ba/Eu]} = -0.94$ \citep{pra20}, based on the solar system abundance pattern coupled with a chemical evolution model.  However, metal-poor stars that are extremely $r$-process-enhanced have slightly higher values (e.g., 2MASS J22132050–5137385, ${\rm [Ba/Eu]} = -0.73$ \citep{roe24}).  We measured ${\rm [Ba/Eu]} \approx -0.6$ in all stars in M92, regardless of stellar population, whereas an $s$-process origin would have ${\rm [Ba/Eu]} > 0$.  Regardless, the strength of our study is in relative abundances, whereas [Ba/Eu] must be placed on an absolute scale to interpret it in the context of $s$- vs.\ $r$-process.

\add{\bmhead{Effect of He abundance} Helium variation in GCs has been suspected for a long time as the possible solution to the diversity of horizontal branch morpohologies \citep[e.g.,][]{mil18}.  Spectroscopic measurements \citep{dup13} and photometric inferences \citep{mil15,val16,zen19} show that He abundances in 2P are higher than in 1P in some clusters.}

\add{He abundance can affect the spectroscopic abundance measurements of all elements in multiple ways.  Some of these effects might mimic the small abundance differences between 1P and 2P that we found for some elements, like Fe.  First, He is the second most abundant element in most stars.  Increasing the He fraction necessarily decreases the H fraction.  This effect causes [X/H] to increase when He abundance is increased.  Second, less H means less continuum opacity due to the H$^-$ ion.  To first order, line strengths are set by the ratio of line opacity to continuum opacity.  Therefore, increasing He abundance (and decreasing H abundance) increases the strengths of modeled absorption lines.  The result would be to decrease abundance to match the observed line strength.  Third, the damping wings of some strong lines are caused by collisions with hydrogen.  Increasing He abundance (and decreasing H abundance) will decrease the strengths of strong lines and cause inferred abundances to increase.  Fourth, the atmospheric structure changes, generally becoming more transparent with increasing He abundance.  Most lines will weaken, causing the inferred abundances to increase.}

\add{The effect of changing He mass fraction on differential abundances in NGC~6752 has been considered extensively \citep{yon13_differential}.  We repeat some of that analysis for M15 and M92.  Whereas the baseline He mass fraction is $Y=0.25$, we recomputed abundances for all stars following the procedures above after changing it to $Y=0.28$ in the \texttt{ATLAS9}/\texttt{BasicATLAS} model atmospheres.  This value is propagated to \texttt{MOOG} so that both the atmosphere and the synthetic spectrum use consistent values of H and He abundance.  The number density of all species relative to H is adjusted accordingly.  Therefore, this procedure accounts for all four effects discussed in the previous paragraph.}

\add{Extended Data Figure~\ref{fig:helium} compares the differential Na and Fe abundances for the $Y=0.25$ and $Y=0.28$ analyses.  In the former case, all stars have $Y=0.25$.  In the latter case, 1P stars have $Y=0.25$ and 2P stars have $Y=0.28$.  The intent of this choice is to investigate whether He abundance differences between 1P and 2P could spuriously cause the observed differences in Fe and other element abundances.  As found previously \citep{yon13_differential,mon23a}, the effect of varying He is nearly negligible.  For example, the changes in differential Fe abundances between the two helium treatments range from $-0.010$ to $+0.009$, much smaller than the observed difference between the populations.}

\bmhead{Independent validation of abundance measurements} To unambiguously demonstrate that the [Fe/H] discrepancy between 1P and 2P stars of M92 is genuine and not an artifact of our analysis, we re-measured [Fe/H] and [Na/H] for all M92 and M15 stars in this study using an independent method described in this section.

\begin{figure}[t]
\centering
\includegraphics[width=0.7\textwidth]{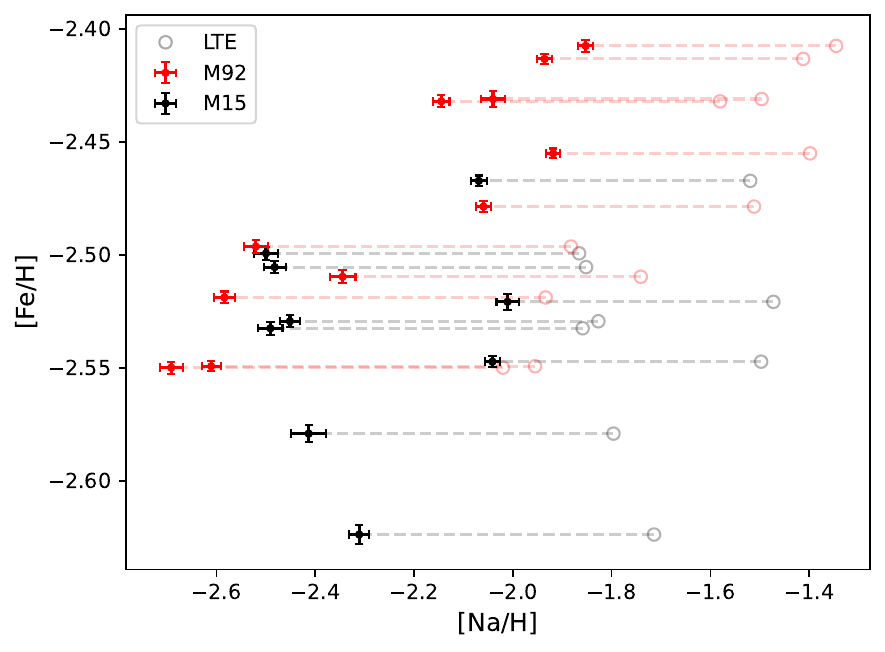}
\caption{[Na/H] and [Fe/H] abundances for the M92 and M15 stars considered in this study, inferred from direct forward modeling of the observed spectra. These measurements independently confirm the correlation between light element (Na) and Fe-peak abundances in M92. The open markers show the original measurements made under the assumption of LTE\@. The final values were corrected for NLTE effects as described in text. The error bars \delete{only} include \add{only} random measurement errors. The scatter in [Fe/H] within the individual populations is likely driven by much larger systematic errors.\label{fig:independent_verification}}
\end{figure}

We determined the abundances by fitting forward models to the entire spectrum of each star redward of 4000~\AA\@. The model spectra were synthesized at each iteration of the fitting process similarly to the second stage of the method of Henderson et al.\ (2025) \citep{hen25}. The atmospheric pressure--temperature profiles used in the synthesis were generated by interpolating a pre-computed 5D grid of \texttt{ATLAS9} \citep{kur93} atmospheres that spans wide ranges of $T_\mathrm{eff}$, $\log g$, metallicity, enhancement of $\alpha$-elements, and carbon-to-oxygen ratio. The synthetic spectra were then generated with \texttt{BasicATLAS}/\texttt{SYNTHE} \citep{BasicATLAS,SYNTHE}, downsampled to the HIRES instrument resolution with Gaussian convolution and corrected for the flux normalization using a third-order spline. In addition to the five free parameters of the \texttt{ATLAS9} grid, the synthetic spectra were also varied in Fe abundance (independently of the overall metallicity), Na abundance, radial velocity and instrument resolution. Since purely spectroscopic $T_\mathrm{eff}$ estimates are sensitive to NLTE effects, we also included the difference between synthetic photometry and \textit{Gaia} $G_{BP}-G_{RP}$ colors of the stars in our goodness-of-fit parameter. The results of our measurements are shown in Extended Data Figure~\ref{fig:independent_verification} with open markers.

The [Na/H] measurements shown in the figure are largely driven by the resonance Na~D doublet at $\lambda\lambda 5890,5896$, which is significantly stronger than all other Na lines in our spectral range. However, the strength and profile shapes of the Na~D doublet are strongly affected by NLTE corrections, especially at the low metallicities of M92 and M15 \citep{Baumueller_Na_NLTE,Lind_Na_NLTE}. Excluding the Na~\add{D} doublet from the fit would improve the accuracy of the measurements, but yield very low precision due to the relative weakness of other Na lines in the spectrum. We analyzed the spectrum of each star twice: with the Na~\add{D} doublet included and excluded from the fit. We then fit a linear relationship to the obtained [Na/H] values, which can be determined accurately due to the much larger number of stars in the sample (20 stars across both M92 and M15) compared to the small number of free parameters (2 for a linear relationship). We then used this relationship as an empirical NLTE correction for our [Na/H] measurements with all Na lines included. These NLTE-corrected values are shown in Extended Data Figure~\ref{fig:independent_verification} with error bars.

Our measurements in Extended Data Figure~\ref{fig:independent_verification} are in qualitative agreement with the lower panel of Figure~\ref{fig:mgalfe}, and clearly show that in \delete{the globular cluster} M92, \delete{the second (Na-rich) stellar population} \add{2P} has a higher Fe content than \delete{the first (Na-poor) population} \add{1P}\@. In contrast, M15 does not display a comparable trend.

\bmhead{Comment on color temperatures} Lee (2024) \citep{lee24} found a difference between the color temperature scale of Mucciarelli et al.\ (2021) \citep{muc21} and evolutionary temperatures, based on stellar luminosities.  Lee (2024) argued that this effect spuriously introduced the apparent spread in $r$-process abundances detected in 1P \citep{kir23}.  However, the difference between the two temperature scales in our present sample is only about 25~K at the luminosity of the stars in our sample.  Such a small difference would not translate to a large enough abundance change to explain our results.  Furthermore, the differential nature of our analysis coupled with the small range of luminosities renders negligible the effect on differential abundances of systematic difference between temperature scales.

Lee (2024) also surmised that M92 is the result of a merger between two sub-clusters of different metallicity.  Our measurements do not support this hypothesis because the low-[Fe/H] population is exclusively 1P from the light element abundances, and the high-[Fe/H] population is exclusively 2P\@.  The sub-clusters would have had to evolve independently, with one lacking 1P stars, the other lacking 2P stars.  We rule out the merger hypothesis on the basis that there is no known cluster that is exclusively 2P\@.

\add{\bmhead{Acknowledgements}
We thank Alexander Ji for a helpful discussion.
}

\section*{Declarations}

\noindent {\bf Funding.} The authors acknowledge funding from the University of Notre Dame.  PN acknowledges financial support from the Arthur J. Schmitt Foundation to complete this project. AC and BC acknowledge financial support from the National Science Foundation grant No.\ 2050527 through the Research Experience for Undergraduates program at the University of Notre Dame.

\noindent {\bf Competing interests.} The authors declare no competing interests.

\noindent {\bf Ethics.} Generative AI was not used in the writing of this article.  Claude Sonnet 4.5 was used to generate some Python code and improve the readability of all code.

\noindent {\bf Consent for publication.} Not applicable.

\noindent {\bf Data availability.} Most of the raw spectra are available at the Keck Observatory Archive.  The \delete{2024 and 2025 spectra are} \add{2025 spectrum of M15-star-4 is} still subject to an 18-month proprietary period.  \add{It will become public in November 2026.}  \delete{Nonetheless,} All the merged, continuum-normalized spectra are available at the following Github repository: \url{https://github.com/enkirby/M15_M92}.  The repository also contains our measurements of equivalent widths and abundances for each line in each star as binary FITS tables.

\noindent {\bf Materials availability.} Not applicable.

\noindent {\bf Code availability.}  The custom code that we wrote to analyze the spectra is available at the following Github repository: \url{https://github.com/enkirby/M15_M92}.  It will be made into a Zenodo repository with its own permanent DOI if this article is accepted for publication.  \texttt{BasicATLAS} \citep{BasicATLAS} is available at \url{https://github.com/Roman-UCSD/BasicATLAS} and \url{https://zenodo.org/records/7145514}.  The version of \texttt{MOOG} \citep{sne73,sob11} used in this study is available at \url{https://github.com/alexji/MOOG17scat}.

\noindent {\bf Author contributions.} ENK led the project, acquired the data, performed most of the analysis, and wrote the manuscript.  RG consulted closely with ENK on the interpretation of the results and led the independent validation of the measurements \add{described in the Supplementary Discussion}.  AC wrote some of the initial code for data reduction.  BC conducted the first draft of equivalent width measurements.  PN provided some code for statistics computation.  LH gave input on interpretation of the results.  All authors participated in editing the manuscript.
\clearpage

\section*{Extended data figures and tables}

\begin{table}[h!]
\caption{Average abundances ($12 + \log n({\mathrm X})/n({\mathrm H})$) in each cluster.}\label{tab:avgabund}
\begin{tabular}{lcc}
\toprule
Element & M15 & M92 \\ \midrule
Na & $+4.415$ & $+4.440$ \\ 
Mg & $+5.398$ & $+5.408$ \\ 
Al & $+3.793$ & $+3.861$ \\ 
Si & $+5.522$ & $+5.612$ \\ 
K & $+3.241$ & $+3.216$ \\ 
Ca & $+4.205$ & $+4.235$ \\ 
Sc & $+0.840$ & $+0.826$ \\ 
Ti & $+2.866$ & $+2.873$ \\ 
V & $+1.647$ & $+1.690$ \\ 
Cr & $+3.041$ & $+3.101$ \\ 
Mn & $+2.500$ & $+2.570$ \\ 
Fe & $+5.007$ & $+5.052$ \\ 
Co & $+2.735$ & $+2.735$ \\ 
Ni & $+3.756$ & $+3.813$ \\ 
Zn & $+2.284$ & $+2.283$ \\ 
Sr & $+0.286$ & $+0.166$ \\ 
Y & $-0.311$ & $-0.439$ \\ 
Zr & $+0.492$ & $+0.289$ \\ 
Ba & $-0.213$ & $-0.577$ \\ 
La & $-0.927$ & $-1.282$ \\ 
Nd & $-0.300$ & $-0.608$ \\ 
Eu & $-1.137$ & $-1.552$ \\ 
Dy & $-0.509$ & $-0.922$ \\ 
\bottomrule
\end{tabular}
\end{table}

\begin{table}
\centering
\caption{\add{Tests of intrinsic dispersion and population differences in M92.}}
\label{tab:fstats}
\begin{tabular}{lrrrrrr}
\hline
        & \multicolumn{2}{c}{$1P$} & \multicolumn{2}{c}{$2P$} \\ \cmidrule(lr){2-3} \cmidrule(lr){4-5}
Element & $F$ & $p$ & $F$ & $p$ & $f({\rm 1P}>\rm{2P})$ & $p_{\rm perm}$ \\
\hline
Na & 2.332 & 0.054 & 8.590 & $<$0.001 & 0.000 & 0.002 \\
Mg & 0.950 & 0.434 & 116.347 & $<$0.001 & 1.000 & 0.965 \\
Al & 0.781 & 0.537 & 11.281 & $<$0.001 & 0.000 & 0.002 \\
Si & 5.984 & $<$0.001 & 6.315 & $<$0.001 & 0.000 & 0.110 \\
K & 12.159 & $<$0.001 & 5.408 & $<$0.001 & 0.000 & 0.015 \\
Ca & 1.856 & 0.115 & 0.814 & 0.539 & 0.000 & 0.002 \\
Sc & 0.146 & 0.965 & 2.999 & 0.010 & 0.000 & 0.006 \\
Ti & 0.553 & 0.697 & 0.247 & 0.942 & 0.000 & 0.002 \\
V & 3.763 & 0.005 & 0.827 & 0.530 & 0.000 & 0.006 \\
Cr & 1.189 & 0.314 & 3.361 & 0.005 & 0.000 & 0.002 \\
Mn & 0.454 & 0.769 & 1.264 & 0.276 & 0.000 & 0.002 \\
Fe & 0.159 & 0.959 & 1.178 & 0.317 & 0.000 & 0.002 \\
Co & 0.590 & 0.670 & 1.462 & 0.199 & 0.001 & 0.013 \\
Ni & 0.853 & 0.492 & 1.219 & 0.297 & 0.000 & 0.002 \\
Zn & 0.806 & 0.521 & 0.946 & 0.450 & 0.002 & 0.015 \\
Sr & 3.958 & 0.003 & 8.231 & $<$0.001 & 0.012 & 0.184 \\
Y & 3.481 & 0.008 & 0.686 & 0.634 & 0.034 & 0.097 \\
Zr & 3.334 & 0.010 & 1.542 & 0.173 & 0.415 & 0.387 \\
Ba & 11.406 & $<$0.001 & 0.651 & 0.661 & 0.643 & 0.561 \\
La & 6.574 & $<$0.001 & 1.295 & 0.263 & 0.878 & 0.522 \\
Nd & 8.760 & $<$0.001 & 3.062 & 0.009 & 0.946 & 0.768 \\
Eu & 15.139 & $<$0.001 & 0.916 & 0.470 & 0.649 & 0.576 \\
Dy & 6.388 & $<$0.001 & 0.817 & 0.537 & 0.974 & 0.846 \\
\hline
\end{tabular}
\end{table}

\clearpage

\bibliography{m92_nature}

@ARTICLE{and18,
       author = {{Andrae}, Ren{\'e} and {Fouesneau}, Morgan and {Creevey}, Orlagh and {Ordenovic}, Christophe and {Mary}, Nicolas and {Burlacu}, Alexandru and {Chaoul}, Laurence and {Jean-Antoine-Piccolo}, Anne and {Kordopatis}, Georges and {Korn}, Andreas and {Lebreton}, Yveline and {Panem}, Chantal and {Pichon}, Bernard and {Th{\'e}venin}, Fr{\'e}d{\'e}ric and {Walmsley}, Gavin and {Bailer-Jones}, Coryn A.~L.},
        title = "{Gaia Data Release 2. First stellar parameters from Apsis}",
      journal = {\aap},
         year = 2018,
        month = aug,
       volume = {616},
          eid = {A8},
        pages = {A8},
          doi = {10.1051/0004-6361/201732516},
archivePrefix = {arXiv},
       eprint = {1804.09374},
 primaryClass = {astro-ph.SR},
       adsurl = {https://ui.adsabs.harvard.edu/abs/2018A&A...616A...8A}
}

@ARTICLE{asp09,
   author = {{Asplund}, M. and {Grevesse}, N. and {Sauval}, A.~J. and {Scott}, P.
	},
    title = "{The Chemical Composition of the Sun}",
  journal = {\araa},
archivePrefix = "arXiv",
   eprint = {0909.0948},
 primaryClass = "astro-ph.SR",
     year = 2009,
    month = sep,
   volume = 47,
    pages = {481-522},
      doi = {10.1146/annurev.astro.46.060407.145222},
   adsurl = {http://adsabs.harvard.edu/abs/2009ARA%26A..47..481A}
}

@ARTICLE{bai09,
       author = {{Bailin}, Jeremy and {Harris}, William E.},
        title = "{Stochastic Self-Enrichment, Pre-Enrichment, and the Formation of Globular Clusters}",
      journal = {\apj},
         year = 2009,
        month = apr,
       volume = {695},
       number = {2},
        pages = {1082-1093},
          doi = {10.1088/0004-637X/695/2/1082},
archivePrefix = {arXiv},
       eprint = {0901.2302},
 primaryClass = {astro-ph.GA},
       adsurl = {https://ui.adsabs.harvard.edu/abs/2009ApJ...695.1082B}
}

@ARTICLE{bai18,
       author = {{Bailin}, Jeremy},
        title = "{A Model for Clumpy Self-enrichment in Globular Clusters}",
      journal = {\apj},
         year = 2018,
        month = aug,
       volume = {863},
       number = {1},
          eid = {99},
        pages = {99},
          doi = {10.3847/1538-4357/aad178},
archivePrefix = {arXiv},
       eprint = {1807.01447},
 primaryClass = {astro-ph.GA},
       adsurl = {https://ui.adsabs.harvard.edu/abs/2018ApJ...863...99B}
}

@ARTICLE{ban25,
       author = {{Bandyopadhyay}, Avrajit and {Ezzeddine}, Rana and {Placco}, Vinicius M. and {Frebel}, Anna and {Aguado}, David S. and {Roederer}, Ian U.},
        title = "{Probing Abundance Variations Among Multiple Stellar Populations in the Metal-poor Globular Cluster NGC 2298 Using Gemini-South/GHOST}",
      journal = {\aj},
         year = 2025,
        month = jul,
       volume = {170},
       number = {1},
          eid = {37},
        pages = {37},
          doi = {10.3847/1538-3881/add52a},
archivePrefix = {arXiv},
       eprint = {2505.04001},
 primaryClass = {astro-ph.SR},
       adsurl = {https://ui.adsabs.harvard.edu/abs/2025AJ....170...37B}
}

@software{bar24,
       author = {{Barlow}, Tom},
        title = "{MAKEE: MAuna Kea Echelle Extraction}",
 howpublished = {Astrophysics Source Code Library, record ascl:2407.001},
         year = 2024,
        month = jul,
          eid = {ascl:2407.001},
       adsurl = {https://ui.adsabs.harvard.edu/abs/2024ascl.soft07001B}
}

@ARTICLE{bas18,
       author = {{Bastian}, Nate and {Lardo}, Carmela},
        title = "{Multiple Stellar Populations in Globular Clusters}",
      journal = {\araa},
         year = "2018",
        month = "Sep",
       volume = {56},
        pages = {83-136},
          doi = {10.1146/annurev-astro-081817-051839},
archivePrefix = {arXiv},
       eprint = {1712.01286},
 primaryClass = {astro-ph.SR},
       adsurl = {https://ui.adsabs.harvard.edu/abs/2018ARA&A..56...83B}
}

@ARTICLE{ber26,
       author = {{Bergemann}, Maria and {Hoppe}, Richard},
        title = "{3D Non-LTE radiation transfer: theory and applications to stars, exoplanets, and kilonovae}",
      journal = {Living Reviews in Computational Astrophysics, accepted},
         year = 2025,
        month = nov,
          eid = {arXiv:2511.04254},
        pages = {arXiv:2511.04254},
          doi = {10.48550/arXiv.2511.04254},
archivePrefix = {arXiv},
       eprint = {2511.04254},
 primaryClass = {astro-ph.SR},
       adsurl = {https://ui.adsabs.harvard.edu/abs/2025arXiv251104254B}
}

@ARTICLE{cab24,
       author = {{Cabrera Garcia}, Jonathan and {Sakari}, Charli M. and {Roederer}, Ian U. and {Evans}, Donavon W. and {Silva}, Pedro and {Mateo}, Mario and {Song}, Ying-Yi and {Kremin}, Anthony and {Bailey}, John I. and {Walker}, Matthew G.},
        title = "{Abundances of Neutron-capture Elements in 62 Stars in the Globular Cluster Messier 15}",
      journal = {\apj},
         year = 2024,
        month = jun,
       volume = {967},
       number = {2},
          eid = {101},
        pages = {101},
          doi = {10.3847/1538-4357/ad380b},
archivePrefix = {arXiv},
       eprint = {2403.00063},
 primaryClass = {astro-ph.SR},
       adsurl = {https://ui.adsabs.harvard.edu/abs/2024ApJ...967..101C}
}

@ARTICLE{cam71,
       author = {{Cameron}, A.~G.~W. and {Fowler}, W.~A.},
        title = "{Lithium and the s-PROCESS in Red-Giant Stars}",
      journal = {\apj},
         year = 1971,
        month = feb,
       volume = {164},
        pages = {111},
          doi = {10.1086/150821},
       adsurl = {https://ui.adsabs.harvard.edu/abs/1971ApJ...164..111C}
}

@ARTICLE{car09a,
       author = {{Carretta}, E. and {Bragaglia}, A. and {Gratton}, R.~G. and {Lucatello}, S. and {Catanzaro}, G. and {Leone}, F. and {Bellazzini}, M. and {Claudi}, R. and {D'Orazi}, V. and {Momany}, Y. and {Ortolani}, S. and {Pancino}, E. and {Piotto}, G. and {Recio-Blanco}, A. and {Sabbi}, E.},
        title = "{Na-O anticorrelation and HB. VII. The chemical composition of first and second-generation stars in 15 globular clusters from GIRAFFE spectra}",
      journal = {\aap},
         year = 2009,
        month = oct,
       volume = {505},
       number = {1},
        pages = {117-138},
          doi = {10.1051/0004-6361/200912096},
archivePrefix = {arXiv},
       eprint = {0909.2938},
 primaryClass = {astro-ph.GA},
       adsurl = {https://ui.adsabs.harvard.edu/abs/2009A&A...505..117C}
}

@ARTICLE{car10,
       author = {{Carretta}, E. and {Bragaglia}, A. and {Gratton}, R.~G. and {Recio-Blanco}, A. and {Lucatello}, S. and {D'Orazi}, V. and {Cassisi}, S.},
        title = "{Properties of stellar generations in globular clusters and relations with global parameters}",
      journal = {\aap},
         year = 2010,
        month = jun,
       volume = {516},
          eid = {A55},
        pages = {A55},
          doi = {10.1051/0004-6361/200913451},
archivePrefix = {arXiv},
       eprint = {1003.1723},
 primaryClass = {astro-ph.GA},
       adsurl = {https://ui.adsabs.harvard.edu/abs/2010A&A...516A..55C}
}

@ARTICLE{car13_ngc362,
       author = {{Carretta}, E. and {Bragaglia}, A. and {Gratton}, R.~G. and {Lucatello}, S. and {D'Orazi}, V. and {Bellazzini}, M. and {Catanzaro}, G. and {Leone}, F. and {Momany}, Y. and {Sollima}, A.},
        title = "{NGC 362: another globular cluster with a split red giant branch}",
      journal = {\aap},
         year = 2013,
        month = sep,
       volume = {557},
          eid = {A138},
        pages = {A138},
          doi = {10.1051/0004-6361/201321905},
archivePrefix = {arXiv},
       eprint = {1307.4085},
 primaryClass = {astro-ph.GA},
       adsurl = {https://ui.adsabs.harvard.edu/abs/2013A&A...557A.138C}
}

@ARTICLE{car15,
       author = {{Carretta}, Eugenio},
        title = "{Five Groups of Red Giants with Distinct Chemical Composition in the Globular Cluster NGC 2808}",
      journal = {\apj},
         year = 2015,
        month = sep,
       volume = {810},
       number = {2},
          eid = {148},
        pages = {148},
          doi = {10.1088/0004-637X/810/2/148},
archivePrefix = {arXiv},
       eprint = {1507.07553},
 primaryClass = {astro-ph.SR},
       adsurl = {https://ui.adsabs.harvard.edu/abs/2015ApJ...810..148C}
}

@ARTICLE{cha16,
       author = {{Chantereau}, W. and {Charbonnel}, C. and {Meynet}, G.},
        title = "{Evolution of long-lived globular cluster stars. III. Effect of the initial helium spread on the position of stars in a synthetic Hertzsprung-Russell diagram}",
      journal = {\aap},
         year = 2016,
        month = aug,
       volume = {592},
          eid = {A111},
        pages = {A111},
          doi = {10.1051/0004-6361/201628418},
archivePrefix = {arXiv},
       eprint = {1606.01899},
 primaryClass = {astro-ph.SR},
       adsurl = {https://ui.adsabs.harvard.edu/abs/2016A&A...592A.111C}
}

@ARTICLE{coh11b,
   author = {{Cohen}, J.~G.},
    title = "{No Heavy-element Dispersion in the Globular Cluster M92}",
  journal = {\apjl},
archivePrefix = "arXiv",
   eprint = {1109.2876},
 primaryClass = "astro-ph.SR",
     year = 2011,
    month = oct,
   volume = 740,
      eid = {L38},
    pages = {L38},
      doi = {10.1088/2041-8205/740/2/L38},
   adsurl = {http://adsabs.harvard.edu/abs/2011ApJ...740L..38C}
}

@ARTICLE{dan01,
       author = {{Ventura}, Paolo and {D'Antona}, Francesca and {Mazzitelli}, Italo and {Gratton}, Raffaele},
        title = "{Predictions for Self-Pollution in Globular Cluster Stars}",
      journal = {\apjl},
         year = 2001,
        month = mar,
       volume = {550},
       number = {1},
        pages = {L65-L69},
          doi = {10.1086/319496},
archivePrefix = {arXiv},
       eprint = {astro-ph/0103337},
 primaryClass = {astro-ph},
       adsurl = {https://ui.adsabs.harvard.edu/abs/2001ApJ...550L..65V}
}

@ARTICLE{dec07,
       author = {{Decressin}, T. and {Meynet}, G. and {Charbonnel}, C. and {Prantzos}, N. and {Ekstr{\"o}m}, S.},
        title = "{Fast rotating massive stars and the origin of the abundance patterns in galactic globular clusters}",
      journal = {\aap},
         year = 2007,
        month = mar,
       volume = {464},
       number = {3},
        pages = {1029-1044},
          doi = {10.1051/0004-6361:20066013},
archivePrefix = {arXiv},
       eprint = {astro-ph/0611379},
 primaryClass = {astro-ph},
       adsurl = {https://ui.adsabs.harvard.edu/abs/2007A&A...464.1029D}
}

@ARTICLE{der10,
       author = {{D'Ercole}, Annibale and {D'Antona}, Francesca and {Ventura}, Paolo and {Vesperini}, Enrico and {McMillan}, Stephen L.~W.},
        title = "{Abundance patterns of multiple populations in globular clusters: a chemical evolution model based on yields from AGB ejecta}",
      journal = {\mnras},
         year = 2010,
        month = sep,
       volume = {407},
       number = {2},
        pages = {854-869},
          doi = {10.1111/j.1365-2966.2010.16996.x},
archivePrefix = {arXiv},
       eprint = {1005.1892},
 primaryClass = {astro-ph.GA},
       adsurl = {https://ui.adsabs.harvard.edu/abs/2010MNRAS.407..854D}
}

@ARTICLE{doh14,
       author = {{Doherty}, Carolyn L. and {Gil-Pons}, Pilar and {Lau}, Herbert H.~B. and {Lattanzio}, John C. and {Siess}, Lionel and {Campbell}, Simon W.},
        title = "{Super and massive AGB stars - III. Nucleosynthesis in metal-poor and very metal-poor stars - Z = 0.001 and 0.0001}",
      journal = {\mnras},
         year = 2014,
        month = jun,
       volume = {441},
       number = {1},
        pages = {582-598},
          doi = {10.1093/mnras/stu571},
archivePrefix = {arXiv},
       eprint = {1403.5054},
 primaryClass = {astro-ph.SR},
       adsurl = {https://ui.adsabs.harvard.edu/abs/2014MNRAS.441..582D}
}

@ARTICLE{dup13,
       author = {{Dupree}, A.~K. and {Avrett}, E.~H.},
        title = "{Direct Evaluation of the Helium Abundances in Omega Centauri}",
      journal = {\apjl},
         year = 2013,
        month = aug,
       volume = {773},
       number = {2},
          eid = {L28},
        pages = {L28},
          doi = {10.1088/2041-8205/773/2/L28},
archivePrefix = {arXiv},
       eprint = {1307.5860},
 primaryClass = {astro-ph.SR},
       adsurl = {https://ui.adsabs.harvard.edu/abs/2013ApJ...773L..28D}
}

@ARTICLE{for13,
       author = {{Foreman-Mackey}, Daniel and {Hogg}, David W. and {Lang}, Dustin and {Goodman}, Jonathan},
        title = "{emcee: The MCMC Hammer}",
      journal = {\pasp},
         year = 2013,
        month = mar,
       volume = {125},
       number = {925},
        pages = {306},
          doi = {10.1086/670067},
archivePrefix = {arXiv},
       eprint = {1202.3665},
 primaryClass = {astro-ph.IM},
       adsurl = {https://ui.adsabs.harvard.edu/abs/2013PASP..125..306F}
}

@ARTICLE{fro06,
       author = {{Fr{\"o}hlich}, C. and {Mart{\'\i}nez-Pinedo}, G. and {Liebend{\"o}rfer}, M. and {Thielemann}, F. -K. and {Bravo}, E. and {Hix}, W.~R. and {Langanke}, K. and {Zinner}, N.~T.},
        title = "{Neutrino-Induced Nucleosynthesis of A>64 Nuclei: The {\ensuremath{\nu}}p Process}",
      journal = {\prl},
         year = 2006,
        month = apr,
       volume = {96},
       number = {14},
          eid = {142502},
        pages = {142502},
          doi = {10.1103/PhysRevLett.96.142502},
archivePrefix = {arXiv},
       eprint = {astro-ph/0511376},
 primaryClass = {astro-ph},
       adsurl = {https://ui.adsabs.harvard.edu/abs/2006PhRvL..96n2502F}
}

@article{gaiadr3,
	author = {{Gaia Collaboration} and {Vallenari, A.} and {Brown, A.G.A.} and {Prusti, T.} and {et al.}},
	title = {Gaia Data Release 3. Summary of the content and survey properties},
	DOI= "10.1051/0004-6361/202243940",
	url= "https://doi.org/10.1051/0004-6361/202243940",
	journal = {A\&A},
	year = 2022,
}

@ARTICLE{gaia18,
       author = {{Gaia Collaboration} and {Babusiaux}, C. and {van Leeuwen}, F. and {Barstow}, M.~A. and {Jordi}, C. and {Vallenari}, A. and {Bossini}, D. and {Bressan}, A. and {Cantat-Gaudin}, T. and {van Leeuwen}, M. and {Brown}, A.~G.~A. and {Prusti}, T. and {de Bruijne}, J.~H.~J. and {Bailer-Jones}, C.~A.~L. and {Biermann}, M. and {Evans}, D.~W. and {Eyer}, L. and {Jansen}, F. and {Klioner}, S.~A. and {Lammers}, U. and {Lindegren}, L. and {Luri}, X. and {Mignard}, F. and {Panem}, C. and {Pourbaix}, D. and {Randich}, S. and {Sartoretti}, P. and {Siddiqui}, H.~I. and {Soubiran}, C. and {Walton}, N.~A. and {Arenou}, F. and {Bastian}, U. and {Cropper}, M. and {Drimmel}, R. and {Katz}, D. and {Lattanzi}, M.~G. and {Bakker}, J. and {Cacciari}, C. and {Casta{\~n}eda}, J. and {Chaoul}, L. and {Cheek}, N. and {De Angeli}, F. and {Fabricius}, C. and {Guerra}, R. and {Holl}, B. and {Masana}, E. and {Messineo}, R. and {Mowlavi}, N. and {Nienartowicz}, K. and {Panuzzo}, P. and {Portell}, J. and {Riello}, M. and {Seabroke}, G.~M. and {Tanga}, P. and {Th{\'e}venin}, F. and {Gracia-Abril}, G. and {Comoretto}, G. and {Garcia-Reinaldos}, M. and {Teyssier}, D. and {Altmann}, M. and {Andrae}, R. and {Audard}, M. and {Bellas-Velidis}, I. and {Benson}, K. and {Berthier}, J. and {Blomme}, R. and {Burgess}, P. and {Busso}, G. and {Carry}, B. and {Cellino}, A. and {Clementini}, G. and {Clotet}, M. and {Creevey}, O. and {Davidson}, M. and {De Ridder}, J. and {Delchambre}, L. and {Dell'Oro}, A. and {Ducourant}, C. and {Fern{\'a}ndez-Hern{\'a}ndez}, J. and {Fouesneau}, M. and {Fr{\'e}mat}, Y. and {Galluccio}, L. and {Garc{\'\i}a-Torres}, M. and {Gonz{\'a}lez-N{\'u}{\~n}ez}, J. and {Gonz{\'a}lez-Vidal}, J.~J. and {Gosset}, E. and {Guy}, L.~P. and {Halbwachs}, J. -L. and {Hambly}, N.~C. and {Harrison}, D.~L. and {Hern{\'a}ndez}, J. and {Hestroffer}, D. and {Hodgkin}, S.~T. and {Hutton}, A. and {Jasniewicz}, G. and {Jean-Antoine-Piccolo}, A. and {Jordan}, S. and {Korn}, A.~J. and {Krone-Martins}, A. and {Lanzafame}, A.~C. and {Lebzelter}, T. and {L{\"o}ffler}, W. and {Manteiga}, M. and {Marrese}, P.~M. and {Mart{\'\i}n-Fleitas}, J.~M. and {Moitinho}, A. and {Mora}, A. and {Muinonen}, K. and {Osinde}, J. and {Pancino}, E. and {Pauwels}, T. and {Petit}, J. -M. and {Recio-Blanco}, A. and {Richards}, P.~J. and {Rimoldini}, L. and {Robin}, A.~C. and {Sarro}, L.~M. and {Siopis}, C. and {Smith}, M. and {Sozzetti}, A. and {S{\"u}veges}, M. and {Torra}, J. and {van Reeven}, W. and {Abbas}, U. and {Abreu Aramburu}, A. and {Accart}, S. and {Aerts}, C. and {Altavilla}, G. and {{\'A}lvarez}, M.~A. and {Alvarez}, R. and {Alves}, J. and {Anderson}, R.~I. and {Andrei}, A.~H. and {Anglada Varela}, E. and {Antiche}, E. and {Antoja}, T. and {Arcay}, B. and {Astraatmadja}, T.~L. and {Bach}, N. and {Baker}, S.~G. and {Balaguer-N{\'u}{\~n}ez}, L. and {Balm}, P. and {Barache}, C. and {Barata}, C. and {Barbato}, D. and {Barblan}, F. and {Barklem}, P.~S. and {Barrado}, D. and {Barros}, M. and {Bartholom{\'e} Mu{\~n}oz}, L. and {Bassilana}, J. -L. and {Becciani}, U. and {Bellazzini}, M. and {Berihuete}, A. and {Bertone}, S. and {Bianchi}, L. and {Bienaym{\'e}}, O. and {Blanco-Cuaresma}, S. and {Boch}, T. and {Boeche}, C. and {Bombrun}, A. and {Borrachero}, R. and {Bouquillon}, S. and {Bourda}, G. and {Bragaglia}, A. and {Bramante}, L. and {Breddels}, M.~A. and {Brouillet}, N. and {Br{\"u}semeister}, T. and {Brugaletta}, E. and {Bucciarelli}, B. and {Burlacu}, A. and {Busonero}, D. and {Butkevich}, A.~G. and {Buzzi}, R. and {Caffau}, E. and {Cancelliere}, R. and {Cannizzaro}, G. and {Carballo}, R. and {Carlucci}, T. and {Carrasco}, J.~M. and {Casamiquela}, L. and {Castellani}, M. and {Castro-Ginard}, A. and {Charlot}, P. and {Chemin}, L. and {Chiavassa}, A. and {Cocozza}, G. and {Costigan}, G. and {Cowell}, S. and {Crifo}, F. and {Crosta}, M. and {Crowley}, C. and {Cuypers}, J. and {Dafonte}, C. and {Damerdji}, Y. and {Dapergolas}, A. and {David}, P. and {David}, M. and {de Laverny}, P. and {De Luise}, F. and {De March}, R. and {de Martino}, D. and {de Souza}, R. and {de Torres}, A. and {Debosscher}, J. and {del Pozo}, E. and {Delbo}, M. and {Delgado}, A. and {Delgado}, H.~E. and {Diakite}, S. and {Diener}, C. and {Distefano}, E. and {Dolding}, C. and {Drazinos}, P. and {Dur{\'a}n}, J. and {Edvardsson}, B. and {Enke}, H. and {Eriksson}, K. and {Esquej}, P. and {Eynard Bontemps}, G. and {Fabre}, C. and {Fabrizio}, M. and {Faigler}, S. and {Falc{\~a}o}, A.~J. and {Farr{\`a}s Casas}, M. and {Federici}, L. and {Fedorets}, G. and {Fernique}, P. and {Figueras}, F. and {Filippi}, F. and {Findeisen}, K. and {Fonti}, A. and {Fraile}, E. and {Fraser}, M. and {Fr{\'e}zouls}, B. and {Gai}, M. and {Galleti}, S. and {Garabato}, D. and {Garc{\'\i}a-Sedano}, F. and {Garofalo}, A. and {Garralda}, N. and {Gavel}, A. and {Gavras}, P. and {Gerssen}, J. and {Geyer}, R. and {Giacobbe}, P. and {Gilmore}, G. and {Girona}, S. and {Giuffrida}, G. and {Glass}, F. and {Gomes}, M. and {Granvik}, M. and {Gueguen}, A. and {Guerrier}, A. and {Guiraud}, J. and {Guti{\'e}}, R. and {Haigron}, R. and {Hatzidimitriou}, D. and {Hauser}, M. and {Haywood}, M. and {Heiter}, U. and {Helmi}, A. and {Heu}, J. and {Hilger}, T. and {Hobbs}, D. and {Hofmann}, W. and {Holland}, G. and {Huckle}, H.~E. and {Hypki}, A. and {Icardi}, V. and {Jan{\ss}en}, K. and {Jevardat de Fombelle}, G. and {Jonker}, P.~G. and {Juh{\'a}sz}, {\'A}. L. and {Julbe}, F. and {Karampelas}, A. and {Kewley}, A. and {Klar}, J. and {Kochoska}, A. and {Kohley}, R. and {Kolenberg}, K. and {Kontizas}, M. and {Kontizas}, E. and {Koposov}, S.~E. and {Kordopatis}, G. and {Kostrzewa-Rutkowska}, Z. and {Koubsky}, P. and {Lambert}, S. and {Lanza}, A.~F. and {Lasne}, Y. and {Lavigne}, J. -B. and {Le Fustec}, Y. and {Le Poncin-Lafitte}, C. and {Lebreton}, Y. and {Leccia}, S. and {Leclerc}, N. and {Lecoeur-Taibi}, I. and {Lenhardt}, H. and {Leroux}, F. and {Liao}, S. and {Licata}, E. and {Lindstr{\o}m}, H.~E.~P. and {Lister}, T.~A. and {Livanou}, E. and {Lobel}, A. and {L{\'o}pez}, M. and {Managau}, S. and {Mann}, R.~G. and {Mantelet}, G. and {Marchal}, O. and {Marchant}, J.~M. and {Marconi}, M. and {Marinoni}, S. and {Marschalk{\'o}}, G. and {Marshall}, D.~J. and {Martino}, M. and {Marton}, G. and {Mary}, N. and {Massari}, D. and {Matijevi{\v{c}}}, G. and {Mazeh}, T. and {McMillan}, P.~J. and {Messina}, S. and {Michalik}, D. and {Millar}, N.~R. and {Molina}, D. and {Molinaro}, R. and {Moln{\'a}r}, L. and {Montegriffo}, P. and {Mor}, R. and {Morbidelli}, R. and {Morel}, T. and {Morris}, D. and {Mulone}, A.~F. and {Muraveva}, T. and {Musella}, I. and {Nelemans}, G. and {Nicastro}, L. and {Noval}, L. and {O'Mullane}, W. and {Ord{\'e}novic}, C. and {Ord{\'o}{\~n}ez-Blanco}, D. and {Osborne}, P. and {Pagani}, C. and {Pagano}, I. and {Pailler}, F. and {Palacin}, H. and {Palaversa}, L. and {Panahi}, A. and {Pawlak}, M. and {Piersimoni}, A.~M. and {Pineau}, F. -X. and {Plachy}, E. and {Plum}, G. and {Poggio}, E. and {Poujoulet}, E. and {Pr{\v{s}}a}, A. and {Pulone}, L. and {Racero}, E. and {Ragaini}, S. and {Rambaux}, N. and {Ramos-Lerate}, M. and {Regibo}, S. and {Reyl{\'e}}, C. and {Riclet}, F. and {Ripepi}, V. and {Riva}, A. and {Rivard}, A. and {Rixon}, G. and {Roegiers}, T. and {Roelens}, M. and {Romero-G{\'o}mez}, M. and {Rowell}, N. and {Royer}, F. and {Ruiz-Dern}, L. and {Sadowski}, G. and {Sagrist{\`a} Sell{\'e}s}, T. and {Sahlmann}, J. and {Salgado}, J. and {Salguero}, E. and {Sanna}, N. and {Santana-Ros}, T. and {Sarasso}, M. and {Savietto}, H. and {Schultheis}, M. and {Sciacca}, E. and {Segol}, M. and {Segovia}, J.~C. and {S{\'e}gransan}, D. and {Shih}, I. -C. and {Siltala}, L. and {Silva}, A.~F. and {Smart}, R.~L. and {Smith}, K.~W. and {Solano}, E. and {Solitro}, F. and {Sordo}, R. and {Soria Nieto}, S. and {Souchay}, J. and {Spagna}, A. and {Spoto}, F. and {Stampa}, U. and {Steele}, I.~A. and {Steidelm{\"u}ller}, H. and {Stephenson}, C.~A. and {Stoev}, H. and {Suess}, F.~F. and {Surdej}, J. and {Szabados}, L. and {Szegedi-Elek}, E. and {Tapiador}, D. and {Taris}, F. and {Tauran}, G. and {Taylor}, M.~B. and {Teixeira}, R. and {Terrett}, D. and {Teyssandier}, P. and {Thuillot}, W. and {Titarenko}, A. and {Torra Clotet}, F. and {Turon}, C. and {Ulla}, A. and {Utrilla}, E. and {Uzzi}, S. and {Vaillant}, M. and {Valentini}, G. and {Valette}, V. and {van Elteren}, A. and {Van Hemelryck}, E. and {Vaschetto}, M. and {Vecchiato}, A. and {Veljanoski}, J. and {Viala}, Y. and {Vicente}, D. and {Vogt}, S. and {von Essen}, C. and {Voss}, H. and {Votruba}, V. and {Voutsinas}, S. and {Walmsley}, G. and {Weiler}, M. and {Wertz}, O. and {Wevers}, T. and {Wyrzykowski}, {\L}. and {Yoldas}, A. and {{\v{Z}}erjal}, M. and {Ziaeepour}, H. and {Zorec}, J. and {Zschocke}, S. and {Zucker}, S. and {Zurbach}, C. and {Zwitter}, T.},
        title = "{Gaia Data Release 2. Observational Hertzsprung-Russell diagrams}",
      journal = {\aap},
         year = 2018,
        month = aug,
       volume = {616},
          eid = {A10},
        pages = {A10},
          doi = {10.1051/0004-6361/201832843},
archivePrefix = {arXiv},
       eprint = {1804.09378},
 primaryClass = {astro-ph.SR},
       adsurl = {https://ui.adsabs.harvard.edu/abs/2018A&A...616A..10G}
}

@ARTICLE{gie11,
       author = {{Gieles}, Mark and {Heggie}, Douglas C. and {Zhao}, Hongsheng},
        title = "{The life cycle of star clusters in a tidal field}",
      journal = {\mnras},
         year = 2011,
        month = jun,
       volume = {413},
       number = {4},
        pages = {2509-2524},
          doi = {10.1111/j.1365-2966.2011.18320.x},
archivePrefix = {arXiv},
       eprint = {1101.1821},
 primaryClass = {astro-ph.GA},
       adsurl = {https://ui.adsabs.harvard.edu/abs/2011MNRAS.413.2509G}
}

@ARTICLE{gie25,
       author = {{Gieles}, Mark and {Padoan}, Paolo and {Charbonnel}, Corinne and {Vink}, Jorick S. and {Ram{\'\i}rez-Galeano}, Laura},
        title = "{Globular cluster formation from inertial inflows: accreting extremely massive stars as the origin of abundance anomalies}",
      journal = {\mnras},
         year = 2025,
        month = aug,
          doi = {10.1093/mnras/staf1314},
archivePrefix = {arXiv},
       eprint = {2501.12138},
 primaryClass = {astro-ph.GA},
       adsurl = {https://ui.adsabs.harvard.edu/abs/2025MNRAS.tmp.1257G}
}

@ARTICLE{gra19,
       author = {{Gratton}, Raffaele and {Bragaglia}, Angela and {Carretta}, Eugenio and {D'Orazi}, Valentina and {Lucatello}, Sara and {Sollima}, Antonio},
        title = "{What is a globular cluster? An observational perspective}",
      journal = {\aapr},
         year = 2019,
        month = nov,
       volume = {27},
       number = {1},
          eid = {8},
        pages = {8},
          doi = {10.1007/s00159-019-0119-3},
archivePrefix = {arXiv},
       eprint = {1911.02835},
 primaryClass = {astro-ph.SR},
       adsurl = {https://ui.adsabs.harvard.edu/abs/2019A&ARv..27....8G}
}

@ARTICLE{hen25,
       author = {{Henderson}, Lauren E. and {Gerasimov}, Roman and {Kirby}, Evan N.},
        title = "{Population-Dependent r-process Scatter in the Globular Cluster M15}",
      journal = {\apjl, in press},
         year = 2025,
        month = sep,
          eid = {arXiv:2509.16840},
        pages = {arXiv:2509.16840},
          doi = {10.48550/arXiv.2509.16840},
archivePrefix = {arXiv},
       eprint = {2509.16840},
 primaryClass = {astro-ph.SR},
       adsurl = {https://ui.adsabs.harvard.edu/abs/2025arXiv250916840H}
}

@ARTICLE{hig23,
       author = {{Higgins}, Erin R. and {Vink}, Jorick S. and {Hirschi}, Raphael and {Laird}, Alison M. and {Sabhahit}, Gautham N.},
        title = "{Stellar wind yields of very massive stars}",
      journal = {\mnras},
         year = 2023,
        month = nov,
       volume = {526},
       number = {1},
        pages = {534-547},
          doi = {10.1093/mnras/stad2537},
archivePrefix = {arXiv},
       eprint = {2308.10941},
 primaryClass = {astro-ph.SR},
       adsurl = {https://ui.adsabs.harvard.edu/abs/2023MNRAS.526..534H}
}

@ARTICLE{ji20,
       author = {{Ji}, Alexander P. and {Li}, Ting S. and {Hansen}, Terese T. and {Casey}, Andrew R. and {Koposov}, Sergey E. and {Pace}, Andrew B. and {Mackey}, Dougal and {Lewis}, Geraint F. and {Simpson}, Jeffrey D. and {Bland-Hawthorn}, Joss and {Cullinane}, Lara R. and {Da Costa}, Gary. S. and {Hattori}, Kohei and {Martell}, Sarah L. and {Kuehn}, Kyler and {Erkal}, Denis and {Shipp}, Nora and {Wan}, Zhen and {Zucker}, Daniel B.},
        title = "{The Southern Stellar Stream Spectroscopic Survey (S$^{5}$): Chemical Abundances of Seven Stellar Streams}",
      journal = {\aj},
         year = 2020,
        month = oct,
       volume = {160},
       number = {4},
          eid = {181},
        pages = {181},
          doi = {10.3847/1538-3881/abacb6},
archivePrefix = {arXiv},
       eprint = {2008.07568},
 primaryClass = {astro-ph.SR},
       adsurl = {https://ui.adsabs.harvard.edu/abs/2020AJ....160..181J}
}

@ARTICLE{joh10,
       author = {{Johnson}, Christian I. and {Pilachowski}, Catherine A.},
        title = "{Chemical Abundances for 855 Giants in the Globular Cluster Omega Centauri (NGC 5139)}",
      journal = {\apj},
         year = 2010,
        month = oct,
       volume = {722},
       number = {2},
        pages = {1373-1410},
          doi = {10.1088/0004-637X/722/2/1373},
archivePrefix = {arXiv},
       eprint = {1008.2232},
 primaryClass = {astro-ph.SR},
       adsurl = {https://ui.adsabs.harvard.edu/abs/2010ApJ...722.1373J}
}

@ARTICLE{joh17,
       author = {{Johnson}, Christian I. and {Caldwell}, Nelson and {Rich}, R. Michael and {Mateo}, Mario and {Bailey}, III, John I. and {Clarkson}, William I. and {Olszewski}, Edward W. and {Walker}, Matthew G.},
        title = "{A Chemical Composition Survey of the Iron-complex Globular Cluster NGC 6273 (M19)}",
      journal = {\apj},
         year = 2017,
        month = feb,
       volume = {836},
       number = {2},
          eid = {168},
        pages = {168},
          doi = {10.3847/1538-4357/836/2/168},
archivePrefix = {arXiv},
       eprint = {1611.05830},
 primaryClass = {astro-ph.SR},
       adsurl = {https://ui.adsabs.harvard.edu/abs/2017ApJ...836..168J}
}

@ARTICLE{kar07,
       author = {{Karakas}, Amanda and {Lattanzio}, John C.},
        title = "{Stellar Models and Yields of Asymptotic Giant Branch Stars}",
      journal = {\pasa},
         year = 2007,
        month = oct,
       volume = {24},
       number = {3},
        pages = {103-117},
          doi = {10.1071/AS07021},
archivePrefix = {arXiv},
       eprint = {0708.4385},
 primaryClass = {astro-ph},
       adsurl = {https://ui.adsabs.harvard.edu/abs/2007PASA...24..103K}
}

@ARTICLE{kar14,
       author = {{Karakas}, Amanda I. and {Lattanzio}, John C.},
        title = "{The Dawes Review 2: Nucleosynthesis and Stellar Yields of Low- and Intermediate-Mass Single Stars}",
      journal = {\pasa},
         year = "2014",
        month = "Jul",
       volume = {31},
          eid = {e030},
        pages = {e030},
          doi = {10.1017/pasa.2014.21},
archivePrefix = {arXiv},
       eprint = {1405.0062},
 primaryClass = {astro-ph.SR},
       adsurl = {https://ui.adsabs.harvard.edu/abs/2014PASA...31...30K}
}

@ARTICLE{kir11d,
   author = {{Kirby}, E.~N.},
    title = "{Grids of ATLAS9 Model Atmospheres and MOOG Synthetic Spectra}",
  journal = {\pasp},
archivePrefix = "arXiv",
   eprint = {1103.1385},
 primaryClass = "astro-ph.SR",
     year = 2011,
    month = may,
   volume = 123,
    pages = {531-535},
      doi = {10.1086/660019},
   adsurl = {http://adsabs.harvard.edu/abs/2011PASP..123..531K}
}

@ARTICLE{kir09,
   author = {{Kirby}, E.~N. and {Guhathakurta}, P. and {Bolte}, M. and {Sneden}, C. and 
	{Geha}, M.~C.},
    title = "{Multi-element Abundance Measurements from Medium-resolution Spectra. I. The Sculptor Dwarf Spheroidal Galaxy}",
  journal = {\apj},
archivePrefix = "arXiv",
   eprint = {0909.3092},
 primaryClass = "astro-ph.GA",
     year = 2009,
    month = nov,
   volume = 705,
    pages = {328-346},
      doi = {10.1088/0004-637X/705/1/328},
   adsurl = {http://adsabs.harvard.edu/abs/2009ApJ...705..328K}
}

@ARTICLE{kir20,
       author = {{Kirby}, Evan N. and {Duggan}, Gina and {Ramirez-Ruiz}, Enrico and {Macias}, Phillip},
        title = "{The Stars in M15 Were Born with the r-process}",
      journal = {\apjl},
         year = 2020,
        month = mar,
       volume = {891},
       number = {1},
          eid = {L13},
        pages = {L13},
          doi = {10.3847/2041-8213/ab78a1},
archivePrefix = {arXiv},
       eprint = {2002.09495},
 primaryClass = {astro-ph.SR},
       adsurl = {https://ui.adsabs.harvard.edu/abs/2020ApJ...891L..13K}
}

@ARTICLE{kur93,
   author = {{Kurucz}, R.},
    title = "{ATLAS9 Stellar Atmosphere Programs and 2 km/s grid.}",
  journal = {ATLAS9 Stellar Atmosphere Programs and 2 km/s grid.~Kurucz CD-ROM No.~13.~ Cambridge, Mass.: Smithsonian Astrophysical Observatory, 1993.},
     year = 1993,
   volume = 13,
   adsurl = {http://adsabs.harvard.edu/abs/1993KurCD..13.....K}
}

@ARTICLE{lat74,
       author = {{Lattimer}, J.~M. and {Schramm}, D.~N.},
        title = "{Black-Hole-Neutron-Star Collisions}",
      journal = {\apjl},
         year = "1974",
        month = "Sep",
       volume = {192},
        pages = {L145},
          doi = {10.1086/181612},
       adsurl = {https://ui.adsabs.harvard.edu/abs/1974ApJ...192L.145L}
}

@ARTICLE{lar22,
       author = {{Lardo}, Carmela and {Salaris}, Maurizio and {Cassisi}, Santi and {Bastian}, Nate},
        title = "{Confirmation of a metallicity spread amongst first population stars in globular clusters}",
      journal = {\aap},
         year = 2022,
        month = jun,
       volume = {662},
          eid = {A117},
        pages = {A117},
          doi = {10.1051/0004-6361/202243843},
archivePrefix = {arXiv},
       eprint = {2205.03323},
 primaryClass = {astro-ph.GA},
       adsurl = {https://ui.adsabs.harvard.edu/abs/2022A&A...662A.117L}
}

@ARTICLE{lar23,
       author = {{Lardo}, C. and {Salaris}, M. and {Cassisi}, S. and {Bastian}, N. and {Mucciarelli}, A. and {Cabrera-Ziri}, I. and {Dalessandro}, E.},
        title = "{High-precision abundances of first-population stars in NGC 2808: confirmation of a metallicity spread}",
      journal = {\aap},
         year = 2023,
        month = jan,
       volume = {669},
          eid = {A19},
        pages = {A19},
          doi = {10.1051/0004-6361/202245090},
archivePrefix = {arXiv},
       eprint = {2210.13369},
 primaryClass = {astro-ph.GA},
       adsurl = {https://ui.adsabs.harvard.edu/abs/2023A&A...669A..19L}
}

@ARTICLE{kir23,
       author = {{Kirby}, Evan N. and {Ji}, Alexander P. and {Kovalev}, Mikhail},
        title = "{r-process Abundance Patterns in the Globular Cluster M92}",
      journal = {\apj},
         year = 2023,
        month = nov,
       volume = {958},
       number = {1},
          eid = {45},
        pages = {45},
          doi = {10.3847/1538-4357/acf309},
archivePrefix = {arXiv},
       eprint = {2308.10980},
 primaryClass = {astro-ph.SR},
       adsurl = {https://ui.adsabs.harvard.edu/abs/2023ApJ...958...45K}
}

@ARTICLE{kra12,
       author = {{Krause}, M. and {Charbonnel}, C. and {Decressin}, T. and {Meynet}, G. and {Prantzos}, N. and {Diehl}, R.},
        title = "{Superbubble dynamics in globular cluster infancy. I. How do globular clusters first lose their cold gas?}",
      journal = {\aap},
         year = 2012,
        month = oct,
       volume = {546},
          eid = {L5},
        pages = {L5},
          doi = {10.1051/0004-6361/201220244},
archivePrefix = {arXiv},
       eprint = {1209.4518},
 primaryClass = {astro-ph.GA},
       adsurl = {https://ui.adsabs.harvard.edu/abs/2012A&A...546L...5K}
}

@ARTICLE{lam10,
       author = {{Lamers}, Henny J.~G.~L.~M. and {Baumgardt}, Holger and {Gieles}, Mark},
        title = "{Mass-loss rates and the mass evolution of star clusters}",
      journal = {\mnras},
         year = 2010,
        month = nov,
       volume = {409},
       number = {1},
        pages = {305-328},
          doi = {10.1111/j.1365-2966.2010.17309.x},
archivePrefix = {arXiv},
       eprint = {1007.1078},
 primaryClass = {astro-ph.GA},
       adsurl = {https://ui.adsabs.harvard.edu/abs/2010MNRAS.409..305L}
}

@ARTICLE{lat25,
       author = {{Latour}, M. and {Kamann}, S. and {Martocchia}, S. and {Husser}, T.-O. and {Saracino}, S. and {Dreizler}, S.},
        title = "{A stellar census in globular clusters with MUSE: Metallicity spread and dispersion among first-population stars}",
      journal = {\aap},
         year = 2025,
        month = feb,
       volume = {694},
          eid = {A248},
        pages = {A248},
          doi = {10.1051/0004-6361/202452420},
archivePrefix = {arXiv},
       eprint = {2501.09558},
 primaryClass = {astro-ph.GA},
       adsurl = {https://ui.adsabs.harvard.edu/abs/2025A&A...694A.248L}
}

@ARTICLE{lee24,
       author = {{Lee}, Jae-Woo},
        title = "{A Comparative Study between M30 and M92: M92 is a Merger Remnant with a Large Helium Enhancement}",
      journal = {\apj},
         year = 2024,
        month = feb,
       volume = {961},
       number = {2},
          eid = {227},
        pages = {227},
          doi = {10.3847/1538-4357/ad12ca},
archivePrefix = {arXiv},
       eprint = {2312.02442},
 primaryClass = {astro-ph.GA},
       adsurl = {https://ui.adsabs.harvard.edu/abs/2024ApJ...961..227L}
}

@ARTICLE{leg22,
       author = {{Legnardi}, M.~V. and {Milone}, A.~P. and {Armillotta}, L. and {Marino}, A.~F. and {Cordoni}, G. and {Renzini}, A. and {Vesperini}, E. and {D'Antona}, F. and {McKenzie}, M. and {Yong}, D. and {Dondoglio}, E. and {Lagioia}, E.~P. and {Carlos}, M. and {Tailo}, M. and {Jang}, S. and {Mohandasan}, A.},
        title = "{Constraining the original composition of the gas forming first-generation stars in globular clusters}",
      journal = {\mnras},
         year = 2022,
        month = jun,
       volume = {513},
       number = {1},
        pages = {735-751},
          doi = {10.1093/mnras/stac734},
archivePrefix = {arXiv},
       eprint = {2203.07571},
 primaryClass = {astro-ph.GA},
       adsurl = {https://ui.adsabs.harvard.edu/abs/2022MNRAS.513..735L}
}

@ARTICLE{leg24,
       author = {{Legnardi}, M.~V. and {Milone}, A.~P. and {Cordoni}, G. and {Marino}, A.~F. and {Dondoglio}, E. and {Jang}, S. and {Lagioia}, E.~P. and {Muratore}, F. and {Ziliotto}, T. and {Bortolan}, E. and {Mohandasan}, A.},
        title = "{The original composition of the gas forming first-generation stars in clusters: Insights from HST and JWST}",
      journal = {\aap},
         year = 2024,
        month = jul,
       volume = {687},
          eid = {A160},
        pages = {A160},
          doi = {10.1051/0004-6361/202449533},
archivePrefix = {arXiv},
       eprint = {2405.02006},
 primaryClass = {astro-ph.GA},
       adsurl = {https://ui.adsabs.harvard.edu/abs/2024A&A...687A.160L}
}

@ARTICLE{lin11,
   author = {{Lind}, K. and {Charbonnel}, C. and {Decressin}, T. and {Primas}, F. and 
	{Grundahl}, F. and {Asplund}, M.},
    title = "{Tracing the evolution of NGC 6397 through the chemical composition of its stellar populations}",
  journal = {\aap},
archivePrefix = "arXiv",
   eprint = {1012.0477},
 primaryClass = "astro-ph.SR",
     year = 2011,
    month = mar,
   volume = 527,
      eid = {A148},
    pages = {A148},
      doi = {10.1051/0004-6361/201015356},
   adsurl = {http://adsabs.harvard.edu/abs/2011A%26A...527A.148L}
}

@ARTICLE{lin22,
       author = {{Lind}, K. and {Nordlander}, T. and {Wehrhahn}, A. and {Montelius}, M. and {Osorio}, Y. and {Barklem}, P.~S. and {Af{\textcommabelow s}ar}, M. and {Sneden}, C. and {Kobayashi}, C.},
        title = "{Non-LTE abundance corrections for late-type stars from 2000 {\r{A}} to 3 {\textmu}m. I. Na, Mg, and Al}",
      journal = {\aap},
         year = 2022,
        month = sep,
       volume = {665},
          eid = {A33},
        pages = {A33},
          doi = {10.1051/0004-6361/202142195},
archivePrefix = {arXiv},
       eprint = {2206.11070},
 primaryClass = {astro-ph.SR},
       adsurl = {https://ui.adsabs.harvard.edu/abs/2022A&A...665A..33L}
}

@ARTICLE{mar19,
       author = {{Marino}, A.~F. and {Milone}, A.~P. and {Sills}, A. and {Yong}, D. and {Renzini}, A. and {Bedin}, L.~R. and {Cordoni}, G. and {D'Antona}, F. and {Jerjen}, H. and {Karakas}, A. and {Lagioia}, E. and {Piotto}, G. and {Tailo}, M.},
        title = "{Chemical Abundances along the 1G Sequence of the Chromosome Maps: The Globular Cluster NGC 3201}",
      journal = {\apj},
         year = 2019,
        month = dec,
       volume = {887},
       number = {1},
          eid = {91},
        pages = {91},
          doi = {10.3847/1538-4357/ab53d9},
archivePrefix = {arXiv},
       eprint = {1910.02892},
 primaryClass = {astro-ph.SR},
       adsurl = {https://ui.adsabs.harvard.edu/abs/2019ApJ...887...91M}
}

@MISC{mar12,
   author = {{Markwardt}, C.},
    title = "{MPFIT: Robust non-linear least squares curve fitting}",
     note = {Astrophysics Source Code Library},
     year = 2012,
archivePrefix = "ascl",
   eprint = {1208.019},
    month = aug,
   adsurl = {http://adsabs.harvard.edu/abs/2012ascl.soft08019M}
}

@ARTICLE{massari14,
       author = {{Massari}, D. and {Mucciarelli}, A. and {Ferraro}, F.~R. and {Origlia}, L. and {Rich}, R.~M. and {Lanzoni}, B. and {Dalessandro}, E. and {Valenti}, E. and {Ibata}, R. and {Lovisi}, L. and {Bellazzini}, M. and {Reitzel}, D.},
        title = "{Ceci N'est Pas a Globular Cluster: The Metallicity Distribution of the Stellar System Terzan 5}",
      journal = {\apj},
         year = 2014,
        month = nov,
       volume = {795},
       number = {1},
          eid = {22},
        pages = {22},
          doi = {10.1088/0004-637X/795/1/22},
archivePrefix = {arXiv},
       eprint = {1409.1682},
 primaryClass = {astro-ph.SR},
       adsurl = {https://ui.adsabs.harvard.edu/abs/2014ApJ...795...22M}
}

@ARTICLE{mas19,
       author = {{Masseron}, T. and {Garc{\'\i}a-Hern{\'a}ndez}, D.~A. and {M{\'e}sz{\'a}ros}, Sz. and {Zamora}, O. and {Dell'Agli}, F. and {Allende Prieto}, C. and {Edvardsson}, B. and {Shetrone}, M. and {Plez}, B. and {Fern{\'a}ndez-Trincado}, J.~G. and {Cunha}, K. and {J{\"o}nsson}, H. and {Geisler}, D. and {Beers}, T.~C. and {Cohen}, R.~E.},
        title = "{Homogeneous analysis of globular clusters from the APOGEE survey with the BACCHUS code. I. The northern clusters}",
      journal = {\aap},
         year = 2019,
        month = feb,
       volume = {622},
          eid = {A191},
        pages = {A191},
          doi = {10.1051/0004-6361/201834550},
archivePrefix = {arXiv},
       eprint = {1812.08817},
 primaryClass = {astro-ph.SR},
       adsurl = {https://ui.adsabs.harvard.edu/abs/2019A&A...622A.191M}
}

@ARTICLE{mck22,
       author = {{McKenzie}, M. and {Yong}, D. and {Marino}, A.~F. and {Monty}, S. and {Wang}, E. and {Karakas}, A.~I. and {Milone}, A.~P. and {Legnardi}, M.~V. and {Roederer}, I.~U. and {Martell}, S. and {Horta}, D.},
        title = "{The complex stellar system M 22: confirming abundance variations with high precision differential measurements}",
      journal = {\mnras},
         year = 2022,
        month = nov,
       volume = {516},
       number = {3},
        pages = {3515-3531},
          doi = {10.1093/mnras/stac2254},
       adsurl = {https://ui.adsabs.harvard.edu/abs/2022MNRAS.516.3515M}
}

@ARTICLE{mel09,
       author = {{Mel{\'e}ndez}, J. and {Asplund}, M. and {Gustafsson}, B. and {Yong}, D.},
        title = "{The Peculiar Solar Composition and Its Possible Relation to Planet Formation}",
      journal = {\apjl},
         year = 2009,
        month = oct,
       volume = {704},
       number = {1},
        pages = {L66-L70},
          doi = {10.1088/0004-637X/704/1/L66},
archivePrefix = {arXiv},
       eprint = {0909.2299},
 primaryClass = {astro-ph.SR},
       adsurl = {https://ui.adsabs.harvard.edu/abs/2009ApJ...704L..66M}
}

@ARTICLE{mil15,
       author = {{Milone}, A.~P.},
        title = "{Helium and multiple populations in the massive globular cluster NGC 6266 (M 62)}",
      journal = {\mnras},
         year = 2015,
        month = jan,
       volume = {446},
       number = {2},
        pages = {1672-1684},
          doi = {10.1093/mnras/stu2198},
archivePrefix = {arXiv},
       eprint = {1409.7230},
 primaryClass = {astro-ph.SR},
       adsurl = {https://ui.adsabs.harvard.edu/abs/2015MNRAS.446.1672M}
}

@ARTICLE{mil17,
       author = {{Milone}, A.~P. and {Piotto}, G. and {Renzini}, A. and {Marino}, A.~F. and {Bedin}, L.~R. and {Vesperini}, E. and {D'Antona}, F. and {Nardiello}, D. and {Anderson}, J. and {King}, I.~R. and {Yong}, D. and {Bellini}, A. and {Aparicio}, A. and {Barbuy}, B. and {Brown}, T.~M. and {Cassisi}, S. and {Ortolani}, S. and {Salaris}, M. and {Sarajedini}, A. and {van der Marel}, R.~P.},
        title = "{The Hubble Space Telescope UV Legacy Survey of Galactic globular clusters - IX. The Atlas of multiple stellar populations}",
      journal = {\mnras},
         year = 2017,
        month = jan,
       volume = {464},
       number = {3},
        pages = {3636-3656},
          doi = {10.1093/mnras/stw2531},
archivePrefix = {arXiv},
       eprint = {1610.00451},
 primaryClass = {astro-ph.SR},
       adsurl = {https://ui.adsabs.harvard.edu/abs/2017MNRAS.464.3636M}
}

@ARTICLE{mil18,
       author = {{Milone}, A.~P. and {Marino}, A.~F. and {Renzini}, A. and {D'Antona}, F. and {Anderson}, J. and {Barbuy}, B. and {Bedin}, L.~R. and {Bellini}, A. and {Brown}, T.~M. and {Cassisi}, S. and {Cordoni}, G. and {Lagioia}, E.~P. and {Nardiello}, D. and {Ortolani}, S. and {Piotto}, G. and {Sarajedini}, A. and {Tailo}, M. and {van der Marel}, R.~P. and {Vesperini}, E.},
        title = "{The Hubble Space Telescope UV legacy survey of galactic globular clusters - XVI. The helium abundance of multiple populations}",
      journal = {\mnras},
         year = 2018,
        month = dec,
       volume = {481},
       number = {4},
        pages = {5098-5122},
          doi = {10.1093/mnras/sty2573},
archivePrefix = {arXiv},
       eprint = {1809.05006},
 primaryClass = {astro-ph.SR},
       adsurl = {https://ui.adsabs.harvard.edu/abs/2018MNRAS.481.5098M}
}

@ARTICLE{mon23a,
       author = {{Monty}, Stephanie and {Yong}, David and {Marino}, Anna F. and {Karakas}, Amanda I. and {McKenzie}, Madeleine and {Grundahl}, Frank and {Mura-Guzm{\'a}n}, Aldo},
        title = "{Peeking beneath the precision floor - I. Metallicity spreads and multiple elemental dispersions in the globular clusters NGC 288 and NGC 362}",
      journal = {\mnras},
         year = 2023,
        month = jan,
       volume = {518},
       number = {1},
        pages = {965-986},
          doi = {10.1093/mnras/stac3040},
archivePrefix = {arXiv},
       eprint = {2210.15061},
 primaryClass = {astro-ph.GA},
       adsurl = {https://ui.adsabs.harvard.edu/abs/2023MNRAS.518..965M}
}

@ARTICLE{muc11,
       author = {{Mucciarelli}, A. and {Salaris}, M. and {Lovisi}, L. and {Ferraro}, F.~R. and {Lanzoni}, B. and {Lucatello}, S. and {Gratton}, R.~G.},
        title = "{Lithium abundance in the globular cluster M4: from the turn-off to the red giant branch bump}",
      journal = {\mnras},
         year = 2011,
        month = mar,
       volume = {412},
       number = {1},
        pages = {81-94},
          doi = {10.1111/j.1365-2966.2010.17884.x},
archivePrefix = {arXiv},
       eprint = {1010.3879},
 primaryClass = {astro-ph.SR},
       adsurl = {https://ui.adsabs.harvard.edu/abs/2011MNRAS.412...81M}
}

@ARTICLE{muc21,
       author = {{Mucciarelli}, A. and {Bellazzini}, M. and {Massari}, D.},
        title = "{Exploiting the Gaia EDR3 photometry to derive stellar temperatures}",
      journal = {\aap},
         year = 2021,
        month = sep,
       volume = {653},
          eid = {A90},
        pages = {A90},
          doi = {10.1051/0004-6361/202140979},
archivePrefix = {arXiv},
       eprint = {2106.03882},
 primaryClass = {astro-ph.SR},
       adsurl = {https://ui.adsabs.harvard.edu/abs/2021A&A...653A..90M}
}

@ARTICLE{mur90,
       author = {{Murray}, Stephen D. and {Lin}, Douglas N.~C.},
        title = "{On the Origin of Metal Homogeneities in Globular Clusters}",
      journal = {\apj},
         year = 1990,
        month = jul,
       volume = {357},
        pages = {105},
          doi = {10.1086/168895},
       adsurl = {https://ui.adsabs.harvard.edu/abs/1990ApJ...357..105M}
}

@ARTICLE{nal25,
       author = {{Nalamwar}, Pranav and {Kirby}, Evan N. and {Cai}, Alice},
        title = "{r-process Abundance Dispersion in the Globular Cluster M5 Using Keck Archival Data}",
      journal = {\apj},
         year = 2025,
        month = sep,
       volume = {990},
       number = {2},
          eid = {132},
        pages = {132},
          doi = {10.3847/1538-4357/adf3ad},
archivePrefix = {arXiv},
       eprint = {2508.11001},
 primaryClass = {astro-ph.GA},
       adsurl = {https://ui.adsabs.harvard.edu/abs/2025ApJ...990..132N}
}

@ARTICLE{nis15,
       author = {{Nishimura}, Nobuya and {Takiwaki}, Tomoya and
         {Thielemann}, Friedrich-Karl},
        title = "{The r-process Nucleosynthesis in the Various Jet-like Explosions of Magnetorotational Core-collapse Supernovae}",
      journal = {\apj},
         year = "2015",
        month = "Sep",
       volume = {810},
       number = {2},
          eid = {109},
        pages = {109},
          doi = {10.1088/0004-637X/810/2/109},
archivePrefix = {arXiv},
       eprint = {1501.06567},
 primaryClass = {astro-ph.SR},
       adsurl = {https://ui.adsabs.harvard.edu/abs/2015ApJ...810..109N}
}

@ARTICLE{nor24,
       author = {{Nordlander}, T. and {Gruyters}, P. and {Richard}, O. and {Korn}, A.~J.},
        title = "{Atomic diffusion and mixing in old stars - VIII. Chemical abundance variations in the globular cluster M4 (NGC 6121)}",
      journal = {\mnras},
         year = 2024,
        month = feb,
       volume = {527},
       number = {4},
        pages = {12120-12139},
          doi = {10.1093/mnras/stad3973},
archivePrefix = {arXiv},
       eprint = {2312.09657},
 primaryClass = {astro-ph.SR},
       adsurl = {https://ui.adsabs.harvard.edu/abs/2024MNRAS.52712120N}
}

@ARTICLE{nor95,
   author = {{Norris}, J.~E. and {Da Costa}, G.~S.},
    title = "{The Giant Branch of omega Centauri. IV. Abundance Patterns Based on Echelle Spectra of 40 Red Giants}",
  journal = {\apj},
     year = 1995,
    month = jul,
   volume = 447,
    pages = {680},
      doi = {10.1086/175909},
   adsurl = {http://adsabs.harvard.edu/abs/1995ApJ...447..680N}
}

@ARTICLE{ori11,
       author = {{Origlia}, L. and {Rich}, R.~M. and {Ferraro}, F.~R. and {Lanzoni}, B. and {Bellazzini}, M. and {Dalessandro}, E. and {Mucciarelli}, A. and {Valenti}, E. and {Beccari}, G.},
        title = "{Spectroscopy Unveils the Complex Nature of Terzan 5}",
      journal = {\apjl},
         year = 2011,
        month = jan,
       volume = {726},
       number = {2},
          eid = {L20},
        pages = {L20},
          doi = {10.1088/2041-8205/726/2/L20},
archivePrefix = {arXiv},
       eprint = {1012.2047},
 primaryClass = {astro-ph.GA},
       adsurl = {https://ui.adsabs.harvard.edu/abs/2011ApJ...726L..20O}
}

@ARTICLE{oso15,
       author = {{Osorio}, Y. and {Barklem}, P.~S. and {Lind}, K. and {Belyaev}, A.~K. and {Spielfiedel}, A. and {Guitou}, M. and {Feautrier}, N.},
        title = "{Mg line formation in late-type stellar atmospheres. I. The model atom}",
      journal = {\aap},
         year = 2015,
        month = jul,
       volume = {579},
          eid = {A53},
        pages = {A53},
          doi = {10.1051/0004-6361/201525846},
archivePrefix = {arXiv},
       eprint = {1504.07593},
 primaryClass = {astro-ph.SR},
       adsurl = {https://ui.adsabs.harvard.edu/abs/2015A&A...579A..53O}
}

@ARTICLE{pla21a,
       author = {{Placco}, Vinicius M. and {Sneden}, Christopher and {Roederer}, Ian U. and {Lawler}, James E. and {Den Hartog}, Elizabeth A. and {Hejazi}, Neda and {Maas}, Zachary and {Bernath}, Peter},
        title = "{Linemake: An Atomic and Molecular Line List Generator}",
      journal = {Research Notes of the American Astronomical Society},
         year = 2021,
        month = apr,
       volume = {5},
       number = {4},
          eid = {92},
        pages = {92},
          doi = {10.3847/2515-5172/abf651},
archivePrefix = {arXiv},
       eprint = {2104.08286},
 primaryClass = {astro-ph.IM},
       adsurl = {https://ui.adsabs.harvard.edu/abs/2021RNAAS...5...92P}
}

@MISC{pla21b,
       author = {{Placco}, Vinicius M. and {Sneden}, Christopher and {Roederer}, Ian U. and {Lawler}, James E. and {Den Hartog}, Elizabeth A. and {Hejazi}, Neda and {Maas}, Zachary and {Bernath}, Peter},
        title = "{linemake: Line list generator}",
 howpublished = {Astrophysics Source Code Library, record ascl:2104.027},
         year = 2021,
        month = apr,
          eid = {ascl:2104.027},
        pages = {ascl:2104.027},
archivePrefix = {ascl},
       eprint = {2104.027},
       adsurl = {https://ui.adsabs.harvard.edu/abs/2021ascl.soft04027P}
}

@ARTICLE{pra06,
       author = {{Prantzos}, N. and {Charbonnel}, C.},
        title = "{On the self-enrichment scenario of galactic globular clusters: constraints on the IMF}",
      journal = {\aap},
         year = 2006,
        month = oct,
       volume = {458},
       number = {1},
        pages = {135-149},
          doi = {10.1051/0004-6361:20065374},
archivePrefix = {arXiv},
       eprint = {astro-ph/0606112},
 primaryClass = {astro-ph},
       adsurl = {https://ui.adsabs.harvard.edu/abs/2006A&A...458..135P}
}

@ARTICLE{pra17,
       author = {{Prantzos}, N. and {Charbonnel}, C. and {Iliadis}, C.},
        title = "{Revisiting nucleosynthesis in globular clusters. The case of NGC 2808 and the role of He and K}",
      journal = {\aap},
         year = 2017,
        month = dec,
       volume = {608},
          eid = {A28},
        pages = {A28},
          doi = {10.1051/0004-6361/201731528},
archivePrefix = {arXiv},
       eprint = {1709.05819},
 primaryClass = {astro-ph.GA},
       adsurl = {https://ui.adsabs.harvard.edu/abs/2017A&A...608A..28P}
}

@ARTICLE{pra20,
       author = {{Prantzos}, N. and {Abia}, C. and {Cristallo}, S. and {Limongi}, M. and {Chieffi}, A.},
        title = "{Chemical evolution with rotating massive star yields II. A new assessment of the solar s- and r-process components}",
      journal = {\mnras},
         year = 2020,
        month = jan,
       volume = {491},
       number = {2},
        pages = {1832-1850},
          doi = {10.1093/mnras/stz3154},
archivePrefix = {arXiv},
       eprint = {1911.02545},
 primaryClass = {astro-ph.GA},
       adsurl = {https://ui.adsabs.harvard.edu/abs/2020MNRAS.491.1832P}
}

@ARTICLE{ram09,
       author = {{Ram{\'\i}rez}, I. and {Mel{\'e}ndez}, J. and {Asplund}, M.},
        title = "{Accurate abundance patterns of solar twins and analogs. Does the anomalous solar chemical composition come from planet formation?}",
      journal = {\aap},
         year = 2009,
        month = dec,
       volume = {508},
       number = {1},
        pages = {L17-L20},
          doi = {10.1051/0004-6361/200913038},
       adsurl = {https://ui.adsabs.harvard.edu/abs/2009A&A...508L..17R}
}

@ARTICLE{ram14,
       author = {{Ram{\'\i}rez}, I. and {Mel{\'e}ndez}, J. and {Bean}, J. and {Asplund}, M. and {Bedell}, M. and {Monroe}, T. and {Casagrande}, L. and {Schirbel}, L. and {Dreizler}, S. and {Teske}, J. and {Tucci Maia}, M. and {Alves-Brito}, A. and {Baumann}, P.},
        title = "{The Solar Twin Planet Search. I. Fundamental parameters of the stellar sample}",
      journal = {\aap},
         year = 2014,
        month = dec,
       volume = {572},
          eid = {A48},
        pages = {A48},
          doi = {10.1051/0004-6361/201424244},
archivePrefix = {arXiv},
       eprint = {1408.4130},
 primaryClass = {astro-ph.SR},
       adsurl = {https://ui.adsabs.harvard.edu/abs/2014A&A...572A..48R}
}

@ARTICLE{roe11a,
       author = {{Roederer}, Ian U.},
        title = "{Primordial r-process Dispersion in Metal-poor Globular Clusters}",
      journal = {\apjl},
         year = "2011",
        month = "May",
       volume = {732},
       number = {1},
          eid = {L17},
        pages = {L17},
          doi = {10.1088/2041-8205/732/1/L17},
archivePrefix = {arXiv},
       eprint = {1104.5056},
 primaryClass = {astro-ph.GA},
       adsurl = {https://ui.adsabs.harvard.edu/abs/2011ApJ...732L..17R}
}

@ARTICLE{roe11b,
       author = {{Roederer}, Ian U. and {Sneden}, Christopher},
        title = "{Heavy-element Dispersion in the Metal-poor Globular Cluster M92}",
      journal = {\aj},
         year = 2011,
        month = jul,
       volume = {142},
       number = {1},
          eid = {22},
        pages = {22},
          doi = {10.1088/0004-6256/142/1/22},
archivePrefix = {arXiv},
       eprint = {1104.5055},
 primaryClass = {astro-ph.SR},
       adsurl = {https://ui.adsabs.harvard.edu/abs/2011AJ....142...22R}
}

@ARTICLE{roe24,
       author = {{Roederer}, Ian U. and {Beers}, Timothy C. and {Hattori}, Kohei and {Placco}, Vinicius M. and {Hansen}, Terese T. and {Ezzeddine}, Rana and {Frebel}, Anna and {Holmbeck}, Erika M. and {Sakari}, Charli M.},
        title = "{The R-Process Alliance: 2MASS J22132050{\textendash}5137385, the Star with the Highest-known r-process Enhancement at [Eu/Fe] = +2.45}",
      journal = {\apj},
         year = 2024,
        month = aug,
       volume = {971},
       number = {2},
          eid = {158},
        pages = {158},
          doi = {10.3847/1538-4357/ad57bf},
archivePrefix = {arXiv},
       eprint = {2406.02691},
 primaryClass = {astro-ph.SR},
       adsurl = {https://ui.adsabs.harvard.edu/abs/2024ApJ...971..158R}
}

@ARTICLE{skr06,
       author = {{Skrutskie}, M.~F. and {Cutri}, R.~M. and {Stiening}, R. and {Weinberg}, M.~D. and {Schneider}, S. and {Carpenter}, J.~M. and {Beichman}, C. and {Capps}, R. and {Chester}, T. and {Elias}, J. and {Huchra}, J. and {Liebert}, J. and {Lonsdale}, C. and {Monet}, D.~G. and {Price}, S. and {Seitzer}, P. and {Jarrett}, T. and {Kirkpatrick}, J.~D. and {Gizis}, J.~E. and {Howard}, E. and {Evans}, T. and {Fowler}, J. and {Fullmer}, L. and {Hurt}, R. and {Light}, R. and {Kopan}, E.~L. and {Marsh}, K.~A. and {McCallon}, H.~L. and {Tam}, R. and {Van Dyk}, S. and {Wheelock}, S.},
        title = "{The Two Micron All Sky Survey (2MASS)}",
      journal = {\aj},
         year = 2006,
        month = feb,
       volume = {131},
       number = {2},
        pages = {1163-1183},
          doi = {10.1086/498708},
       adsurl = {https://ui.adsabs.harvard.edu/abs/2006AJ....131.1163S}
}

@PHDTHESIS{sne73,
   author = {{Sneden}, C.~A.},
    title = "{Carbon and Nitrogen Abundances in Metal-Poor Stars.}",
   school = {University of Texas Austin.},
     year = 1973,
   adsurl = {http://adsabs.harvard.edu/abs/1973PhDT.......180S}
}

@ARTICLE{sne97,
   author = {{Sneden}, C. and {Kraft}, R.~P. and {Shetrone}, M.~D. and {Smith}, G.~H. and 
	{Langer}, G.~E. and {Prosser}, C.~F.},
    title = "{Star-To-Star Abundance Variations Among Bright Giants in the Metal-Poor Globular Cluster M15}",
  journal = {\aj},
     year = 1997,
    month = nov,
   volume = 114,
    pages = {1964},
      doi = {10.1086/118618},
   adsurl = {http://adsabs.harvard.edu/abs/1997AJ....114.1964S}
}

@ARTICLE{sne08,
       author = {{Sneden}, C. and {Cowan}, J.~J. and {Gallino}, R.},
        title = "{Neutron-capture elements in the early galaxy.}",
      journal = {\araa},
         year = 2008,
        month = sep,
       volume = {46},
        pages = {241-288},
          doi = {10.1146/annurev.astro.46.060407.145207},
       adsurl = {https://ui.adsabs.harvard.edu/abs/2008ARA&A..46..241S}
}

@ARTICLE{sob11,
   author = {{Sobeck}, J.~S. and {Kraft}, R.~P. and {Sneden}, C. and {Preston}, G.~W. and 
	{Cowan}, J.~J. and {Smith}, G.~H. and {Thompson}, I.~B. and 
	{Shectman}, S.~A. and {Burley}, G.~S.},
    title = "{The Abundances of Neutron-capture Species in the Very Metal-poor Globular Cluster M15: A Uniform Analysis of Red Giant Branch and Red Horizontal Branch Stars}",
  journal = {\aj},
archivePrefix = "arXiv",
   eprint = {1103.1008},
 primaryClass = "astro-ph.SR",
     year = 2011,
    month = jun,
   volume = 141,
      eid = {175},
    pages = {175},
      doi = {10.1088/0004-6256/141/6/175},
   adsurl = {http://adsabs.harvard.edu/abs/2011AJ....141..175S}
}

@ARTICLE{tra04,
       author = {{Travaglio}, Claudia and {Gallino}, Roberto and {Arnone}, Enrico and {Cowan}, John and {Jordan}, Faith and {Sneden}, Christopher},
        title = "{Galactic Evolution of Sr, Y, And Zr: A Multiplicity of Nucleosynthetic Processes}",
      journal = {\apj},
         year = 2004,
        month = feb,
       volume = {601},
       number = {2},
        pages = {864-884},
          doi = {10.1086/380507},
archivePrefix = {arXiv},
       eprint = {astro-ph/0310189},
 primaryClass = {astro-ph},
       adsurl = {https://ui.adsabs.harvard.edu/abs/2004ApJ...601..864T}
}

@ARTICLE{val16,
       author = {{Valcarce}, A.~A.~R. and {Catelan}, M. and {Alonso-Garc{\'\i}a}, J. and {Contreras Ramos}, R. and {Alves}, S.},
        title = "{Level of helium enhancement among M3's horizontal branch stars}",
      journal = {\aap},
         year = 2016,
        month = may,
       volume = {589},
          eid = {A126},
        pages = {A126},
          doi = {10.1051/0004-6361/201526387},
archivePrefix = {arXiv},
       eprint = {1601.06747},
 primaryClass = {astro-ph.SR},
       adsurl = {https://ui.adsabs.harvard.edu/abs/2016A&A...589A.126V}
}

@ARTICLE{van16,
       author = {{VandenBerg}, Don A. and {Denissenkov}, P.~A. and {Catelan}, M{\'a}rcio},
        title = "{Constraints on the Distance Moduli, Helium and Metal Abundances, and Ages of Globular Clusters from their RR Lyrae and Non-variable Horizontal-branch Stars. I. M3, M15, and M92}",
      journal = {\apj},
         year = 2016,
        month = aug,
       volume = {827},
       number = {1},
          eid = {2},
        pages = {2},
          doi = {10.3847/0004-637X/827/1/2},
archivePrefix = {arXiv},
       eprint = {1607.02088},
 primaryClass = {astro-ph.SR},
       adsurl = {https://ui.adsabs.harvard.edu/abs/2016ApJ...827....2V}
}

@ARTICLE{ventura12,
       author = {{Ventura}, Paolo and {D'Antona}, Francesca and {Di Criscienzo}, Marcella and {Carini}, Roberta and {D'Ercole}, Annibale and {vesperini}, Enrico},
        title = "{Super-AGB-AGB Evolution and the Chemical Inventory in NGC 2419}",
      journal = {\apjl},
         year = 2012,
        month = dec,
       volume = {761},
       number = {2},
          eid = {L30},
        pages = {L30},
          doi = {10.1088/2041-8205/761/2/L30},
archivePrefix = {arXiv},
       eprint = {1211.3857},
 primaryClass = {astro-ph.SR},
       adsurl = {https://ui.adsabs.harvard.edu/abs/2012ApJ...761L..30V}
}

@INPROCEEDINGS{vog94,
       author = {{Vogt}, S.~S. and {Allen}, S.~L. and {Bigelow}, B.~C. and {Bresee}, L. and {Brown}, B. and {Cantrall}, T. and {Conrad}, A. and {Couture}, M. and {Delaney}, C. and {Epps}, H.~W. and {Hilyard}, D. and {Hilyard}, D.~F. and {Horn}, E. and {Jern}, N. and {Kanto}, D. and {Keane}, M.~J. and {Kibrick}, R.~I. and {Lewis}, J.~W. and {Osborne}, J. and {Pardeilhan}, G.~H. and {Pfister}, T. and {Ricketts}, T. and {Robinson}, L.~B. and {Stover}, R.~J. and {Tucker}, D. and {Ward}, J. and {Wei}, M.~Z.},
        title = "{HIRES: the high-resolution echelle spectrometer on the Keck 10-m Telescope}",
    booktitle = {Instrumentation in Astronomy VIII},
         year = 1994,
       editor = {{Crawford}, David L. and {Craine}, Eric R.},
       series = {Society of Photo-Optical Instrumentation Engineers (SPIE) Conference Series},
       volume = {2198},
        month = jun,
        pages = {362},
          doi = {10.1117/12.176725},
       adsurl = {https://ui.adsabs.harvard.edu/abs/1994SPIE.2198..362V}
}

@ARTICLE{wil12,
   author = {{Willman}, B. and {Strader}, J.},
    title = "{''Galaxy,'' Defined}",
  journal = {\aj},
archivePrefix = "arXiv",
   eprint = {1203.2608},
 primaryClass = "astro-ph.CO",
     year = 2012,
    month = sep,
   volume = 144,
      eid = {76},
    pages = {76},
      doi = {10.1088/0004-6256/144/3/76},
   adsurl = {http://adsabs.harvard.edu/abs/2012AJ....144...76W}
}

@ARTICLE{yon13_differential,
       author = {{Yong}, David and {Mel{\'e}ndez}, Jorge and {Grundahl}, Frank and {Roederer}, Ian U. and {Norris}, John E. and {Milone}, A.~P. and {Marino}, A.~F. and {Coelho}, P. and {McArthur}, Barbara E. and {Lind}, K. and {Collet}, R. and {Asplund}, Martin},
        title = "{High precision differential abundance measurements in globular clusters: chemical inhomogeneities in NGC 6752}",
      journal = {\mnras},
         year = 2013,
        month = oct,
       volume = {434},
       number = {4},
        pages = {3542-3565},
          doi = {10.1093/mnras/stt1276},
archivePrefix = {arXiv},
       eprint = {1307.4486},
 primaryClass = {astro-ph.GA},
       adsurl = {https://ui.adsabs.harvard.edu/abs/2013MNRAS.434.3542Y}
}

@ARTICLE{zen19,
       author = {{Zennaro}, M. and {Milone}, A.~P. and {Marino}, A.~F. and {Cordoni}, G. and {Lagioia}, E.~P. and {Tailo}, M.},
        title = "{Four stellar populations and extreme helium variation in the massive outer-halo globular cluster NGC 2419}",
      journal = {\mnras},
         year = 2019,
        month = aug,
       volume = {487},
       number = {3},
        pages = {3239-3251},
          doi = {10.1093/mnras/stz1477},
archivePrefix = {arXiv},
       eprint = {1902.02178},
 primaryClass = {astro-ph.SR},
       adsurl = {https://ui.adsabs.harvard.edu/abs/2019MNRAS.487.3239Z}
}

@ARTICLE{BasicATLAS,
       author = {{Larkin}, Mikaela M. and {Gerasimov}, Roman and {Burgasser}, Adam J.},
        title = "{Characterization of Population III Stars with Stellar Atmosphere and Evolutionary Modeling and Predictions of their Observability with the JWST}",
      journal = {\aj},
         year = 2023,
        month = jan,
       volume = {165},
       number = {1},
          eid = {2},
        pages = {2},
          doi = {10.3847/1538-3881/ac9b43},
archivePrefix = {arXiv},
       eprint = {2210.09185},
 primaryClass = {astro-ph.SR},
       adsurl = {https://ui.adsabs.harvard.edu/abs/2023AJ....165....2L}
}

@ARTICLE{SYNTHE,
       author = {{Kurucz}, Robert L. and {Avrett}, Eugene H.},
        title = "{Solar Spectrum Synthesis. I. A Sample Atlas from 224 to 300 nm}",
      journal = {SAO Special Report},
         year = 1981,
        month = may,
       volume = {391},
       adsurl = {https://ui.adsabs.harvard.edu/abs/1981SAOSR.391.....K}
}

@ARTICLE{Baumueller_Na_NLTE,
       author = {{Baumueller}, D. and {Butler}, K. and {Gehren}, T.},
        title = "{Sodium in the Sun and in metal-poor stars}",
      journal = {\aap},
         year = 1998,
        month = oct,
       volume = {338},
        pages = {637-650},
       adsurl = {https://ui.adsabs.harvard.edu/abs/1998A&A...338..637B}
}

@ARTICLE{Lind_Na_NLTE,
       author = {{Lind}, K. and {Asplund}, M. and {Barklem}, P.~S. and {Belyaev}, A.~K.},
        title = "{Non-LTE calculations for neutral Na in late-type stars using improved atomic data}",
      journal = {\aap},
         year = 2011,
        month = apr,
       volume = {528},
          eid = {A103},
        pages = {A103},
          doi = {10.1051/0004-6361/201016095},
archivePrefix = {arXiv},
       eprint = {1102.2160},
 primaryClass = {astro-ph.SR},
       adsurl = {https://ui.adsabs.harvard.edu/abs/2011A&A...528A.103L}
}

\end{document}